\documentclass{emulateapj}

\usepackage{psfig}
\usepackage{amsmath}
\usepackage{amssymb}
\usepackage{graphicx}
\usepackage{appendix}
\usepackage{color}
\usepackage{multirow}
\usepackage[normalem]{ulem}

\newcommand{\lSect}[1]{{\label{sec:#1}}}
\newcommand{\lFig}[1]{{\label{fig:#1}}}
\newcommand{\lEq}[1]{{\label{eq:#1}}}
\newcommand{\lTab}[1]{{\label{tab:#1}}}
\def\gtaprx {\lower .1ex\hbox{\rlap{\raise .6ex\hbox{\hskip .3ex
	{\ifmmode{\scriptscriptstyle >}\else
		{$\scriptscriptstyle >$}\fi}}}
	\kern -.4ex{\ifmmode{\scriptscriptstyle \sim}\else
		{$\scriptscriptstyle\sim$}\fi}}}
\def\ltaprx {\lower .1ex\hbox{\rlap{\raise .6ex\hbox{\hskip .3ex
	{\ifmmode{\scriptscriptstyle <}\else
		{$\scriptscriptstyle <$}\fi}}}
	\kern -.4ex{\ifmmode{\scriptscriptstyle \sim}\else
		{$\scriptscriptstyle\sim$}\fi}}}
\newcommand{\FIGFF}[2]{{\ref{fig:#2}{#1}}}

\newcommand{\FIG}[2]{{Fig.~\FIGFF{#1}{#2}}}
\newcommand{\Fig}[1]{{\FIG{}{#1}}}

\newcommand{\Sectff}[1]{{\ref{sec:#1}}}
\newcommand{\Sect}[1]{{\S\Sectff{#1}}}

\newcommand{\Eqref}[1]{{\ref{eq:#1}}}
\newcommand{\Eqff}[1]{{(\Eqref{#1})}}
\newcommand{\eqff}[1]{{\Eqref{#1}}}

\newcommand{\Eq}[1]{{eq.~\Eqff{#1}}}
\newcommand{\eq}[1]{{equation~\eqff{#1}}}

\newcommand{\Msun}{\ensuremath{\mathrm{M}_\odot}}
\newcommand{\Lsun}{\ensuremath{\mathrm{L}_\odot}}
\newcommand{\Rsun}{\ensuremath{\mathrm{R}_\odot}}
\newcommand{\Zsun}{\ensuremath{\mathrm{Z}_\odot}}
\newcommand{\Tab}[1]{{Table \ref{tab:#1}}}

\def\gtaprx {\lower .1ex\hbox{\rlap{\raise .6ex\hbox{\hskip .3ex
	{\ifmmode{\scriptscriptstyle >}\else
		{$\scriptscriptstyle >$}\fi}}}
	\kern -.4ex{\ifmmode{\scriptscriptstyle \sim}\else
		{$\scriptscriptstyle\sim$}\fi}}}
\def\ltaprx {\lower .1ex\hbox{\rlap{\raise .6ex\hbox{\hskip .3ex
	{\ifmmode{\scriptscriptstyle <}\else
		{$\scriptscriptstyle <$}\fi}}}
	\kern -.4ex{\ifmmode{\scriptscriptstyle \sim}\else
		{$\scriptscriptstyle\sim$}\fi}}}

\begin{document}


\title{The Evolution of Massive Helium Stars Including Mass Loss}

\author{S. E. Woosley\altaffilmark{1}}
\altaffiltext{1}{Department of Astronomy and Astrophysics, University
  of California, Santa Cruz, CA 95064, woosley@ucolick.org}

\begin{abstract} 
The evolution of helium stars with initial masses in the range 1.6 to
120 \Msun \ is studied, including the effects of mass loss by
winds. These stars are assumed to form in binary systems when their
expanding hydrogenic envelopes are promptly lost just after helium
ignition.  Significant differences are found with single star
evolution, chiefly because the helium core loses mass during helium
burning rather than gaining it from hydrogen shell burning.
Consequently presupernova stars for a given initial mass function have
considerably smaller mass when they die and will be easier to
explode. Even accounting for this difference, the helium stars with
mass loss develop more centrally condensed cores that should explode
more easily than their single-star counterparts. The production of
low mass black holes may be diminished. Helium stars with initial
masses below 3.2 \Msun \ experience significant radius expansion after
helium depletion, reaching blue supergiant proportions. This could
trigger additional mass exchange or affect the light curve of the
supernova.  The most common black hole masses produced in binaries is
estimated to be about 9 \Msun. A new maximum mass for black holes
derived from pulsational pair-instability supernovae is derived - 46
\Msun, and a new potential gap at 10 - 12 \Msun \ is noted. Models
pertinent to SN 2014ft are presented and a library of presupernova
models is generated.
\end{abstract}


\section{INTRODUCTION}
\lSect{intro}

Half or more of massive stars are found in binaries with such close
separations that the stars will interact when one of them becomes a
supergiant \citep{San11,San12}. This interaction will radically affect
the sorts of supernovae they produce
\citep{Pod92,Wel99,Lan12,Dem17}. Many of the supernovae will no longer
be Type II, but Type I. More subtle structural changes also happen to
the core structure that affect the explosion physics, nucleosynthesis,
and remnant properties.

In cases where the stars do not fully merge, binary interaction
preferentially removes the low density hydrogen envelope, so binary
evolution is often studied using helium stars. Doing so reduces the
computational expense trivially, but also isolates the evolution of
the central core from uncertainties in red giant mass loss,
semiconvection and overshoot mixing during hydrogen core and shell
burning, and rotationally-induced mixing on the main sequence. A
significant fraction of rapidly rotating massive stars might also
experience chemically homogeneous evolution \citep{Mae87} and
end up resembling helium stars. These could be the precursors of
merging black hole pairs in binary systems \citep{Man16} or gamma-ray
bursts \citep{Woo06}.

The study of massive helium stars has a rich history. Some of the
first explorations of advanced burning stages were carried out in
helium stars by \citet[e.g.][]{Arn74}. \citet{Nom88} systematically
explored massive stellar evolution and nucleosynthesis using helium
stars. \citet{Woo95} included mass loss as a way to get realistic 
progenitors for Type Ib supernovae. More recently, helium star
evolution has been considered specifically in the context of mass
exchanging binaries by \citet{Kru18,Mcc16,Tau15,Yoo10,Yoo17}, and
others. These studies paid close attention to important details
of the binary interaction, but did not usually follow the full
range of helium star masses expected to give supernovae, nor 
the evolution through core collapse and beyond. In contrast, the
present study includes all massive stars expected to explode or
produce compact remnants, and, except for the lightest members,
follows the evolution to its completion.

A simple, approximate approach is used to account for binary mass
exchange.  It is assumed that the dominant effect of binary
interaction is to remove the hydrogen envelope, revealing a bare
helium star. This could happen due to mass transfer through a Roche
lobe or by the formation and ejection of a common envelope. A similar
starting point might be generated by chemically homogeneous
evolution. The key quantities are then the mass of the initial helium
core when it is uncovered, the central abundance of helium then, and
the star's subsequent mass loss history. The simplest assumption, which is
adopted here, is to assume that the helium core is always revealed
when helium burning ignites. The initial central helium mass fraction
is thus near 1, and the final evolution is determined by the
progenitor star's initial mass on the main sequence and the mass loss
rate. Given a relation, to be derived, between ZAMS mass and helium
core mass at helium ignition, the outcome of stellar evolution for all
masses can be surveyed. The results should be useful both to
understanding, qualitatively, the outcome of binary evolution for a
large range of masses, and for calculating the distribution of compact
remnants in binaries \citep[e.g.][]{Fry12}.

For mass loss by winds, the recent prescription of \citet{Yoo17} is
used. This, in turn, is a restatement of previous estimates for
various kinds of Wolf-Rayet stars. It is neither a unique prescription
\citep[see e.g.][]{Vin17}, nor one unlikely to change, but it is an
improvement over what the KEPLER code has used in the past. The stars
studied lack rotation and have solar metallicity.  Rotation, even
rapid rotation, is not expected to alter the results appreciably.  The
carbon-oxygen (CO) core may be a bit larger, but post-helium burning
phases occur so rapidly that substantial mixing does not occur. The
ratio of centrifugal force to gravity is not large, except near the
surface. Except during core collapse, the central evolution is
unaffected. 

Similarly, except for nucleosynthesis, the metallicity does not
greatly alter the presupernova evolution, except as it affects the
mass loss. The mass loss is varied here to test that sensitivity.  To
some extent, varying the mass loss rate also accounts for the fact
that the helium core may not be uncovered until later in the evolution
after appreciable helium has burned.

As in previous studies, one finds domains where white dwarfs or
electron-capture supernovae are the likely outcome (\Sect{NeO}); a
heavier range where residual degeneracy leads to off-center burning
and possible thermonuclear flashes (\Sect{deflag}); a range of normal
Type Ib and Ic supernovae and a mixture of neutron stars and black
holes (\Sect{normal}); and some very heavy stars where the pair
instability is important (\Sect{ppisn}). These stars all have
counterparts in previous studies of single stars, but happen for
different initial masses and differ in outcome in subtle and important
ways. For example, the maximum mass for black holes as truncated by
the pulsational pair instability is smaller because the the expanding
core is not  decelerated by a massive hydrogen envelope. The light curves
for all models are different, and of course Type I, not II. In
\Sect{lite}, a brief survey is given of all the Type I supernova light
curves expected from binaries. Emphasis is on low mass models where
new phenomena are expected.

While a large network was carried in all models, studies of
nucleosynthesis, explosion kinematics and remnant masses are deferred
to future papers.

\section{Computation and Physics}
\lSect{compute}

Over 150 helium star models were evolved with initial masses from 1.6
\Msun \ to 120 \Msun. From 1.6 \Msun \ to 4.5 \Msun, the mass
increment was $\Delta M =$ 0.1 \Msun; from 4.5 to 23.0, $\Delta M =$
0.25 \Msun; from 23 to 28, $\Delta M =$ 0.5 \Msun; from 28 to 70,
$\Delta M =$ 2 \Msun; and from 70 to 120 \Msun, $\Delta M =$ 5
\Msun. Models were characterized by their initial mass, mass loss
rate (\Sect{mdot}), initial composition, and the physics used in their
study.

The initial composition was taken to be the products of hydrogen
burning in a massive star with solar metallicity. The particular star
used to generate the initial composition was the  13 \Msun \ model of
\citet{Woo15}. By mass fraction, the abundances of species that
constituted more than 0.01\% of the mass were $^4$He (0.9855);
$^{12}$C (2.2(-4)); $^{14}$N (8.98(-3)); $^{16}$O (2.07(-4));
$^{20}$Ne (1.14(-3)); $^{23}$Na (1.47(-4); $^{24}$Mg (5.65(-4);
$^{26}$Mg (1.55(-4)); $^{28}$Si (7.55(-4)); $^{32}$S (3.96(-4)); and
$^{56}$Fe (1.26(-3)). The total mass fraction of all isotopes of Mn,
Fe, Co and Ni was 1.46(-3). The abundance of nitrogen was sufficiently
large that each star experienced a brief stage of convective nitrogen
burning ($^{14}$N($\alpha,\gamma)^{18}$F$(e^+ \nu)^{18}$O) before
settling down to burn helium.

Nuclear physics was handled in different ways for the various burning
stages and star masses.  Four tools are available for tracking
nucleosynthesis and energy generation in the KEPLER code: a) an
adaptive network that includes all isotopes with any appreciable
abundances within user-specified bounds \citep{Rau02}, all coupled to
directly to the energy generation \citep{Woo04}; b) an approximation
network with 19 species from H to $^{56}$Ni with steady state assumed
for 10 other species \citep{Wea78}; c) a silicon quasi-equilibrium
network that assumes nuclear statistical equilibrium within two large
groups from silicon to scandium and titanium to nickel \citep{Wea78}
and contains 128 species; and d) a large network like in a), but used
in passive mode just to follow nucleosynthesis and gradual changes in
the electron mole number $Y_e$. Temperature and density-dependent weak
interaction rates were included in all cases but b)
\citep[e.g.,][]{Heg01}.  To save time the network in cases a) and d)
was truncated at molybdenum (Z = 42). That will not affect energy
generation, but means that the heavy s-process was not tracked in the
present study. Typically 300 to 400 isotopes from $^1$H to $^{114}$Mo
were carried in cases a) and d), depending on the burning stage.

The large network (case a) was used for the full evolution of all
models lighter than 4.5 \Msun.  This was necessary to follow weak
interactions on trace species during high-density carbon, neon, and
oxygen burning where the quasi-equilibrium approximation is not
valid. The energy generation can become quite complicated when, e.g.,
in silicon burning, the dominant species are not alpha-particle (Z =
N) nuclei and the most abundant species are $^{30}$Si and $^{34}$S.
The large network was also used to study a few pulsational-pair
instability supernova where the explosions left silicon-group and
iron-group species sitting at low temperature for a long time.  Good
agreement with runs using approximations b) and c) was found
(\Sect{ppisn}; \Tab{ppisntab}).

All other cases used a combination of the approximation network (case
b) and quasi-equilibrium network (case c) to follow energy
generation. The passive network (case d) was also used to track
nucleosynthesis and gradual changes in $Y_e$ which were fed back into
the equation of state and thus, indirectly, affected structure.

Zoning was similar to that employed in \citet{Suk18} and varied
approximately as $M^{0.5}$, with M the initial mass of star. The 2
\Msun \ helium star had about 900 zones; the 120 \Msun \ star, about
7000.  Surface zoning and boundary conditions were important in
resolving the photosphere.  Models up to 10 \Msun \ had surface zoning
approaching 10$^{25}$ gm and a surface boundary pressure of 10$^4$
dyne cm$^{-1}$.  From 10 to 20 \Msun \ the boundary pressure was
increased to 10$^6$ dyne cm$^{-2}$. Above 10 \Msun, severe density
inversions and pulsational instabilities developed in a tiny bit of
matter at the surface, $<10^{-6}$ \Msun, that made it difficult to
determine the photospheric radius exactly. These heavier stars did not
experience radius expansion and coarser zoning sufficed.  From 20 to
120 \Msun, surface zoning approached 10$^{27}$ gm \ and the surface
boundary pressure was 10$^8$ dyne cm$^{-2}$. 

Fortunately, the mass loss rate employed in this study depends only
upon the luminosity (\Sect{mdot}). Finer zoning was useful for
resolving the radius, but not the luminosity.  Several limited sets of
models were run to test the sensitivity of results to zoning and
little variation was found. For example, a 10.0 \Msun \ model with
surface zoning to 10$^{27}$ gm and boundary pressure 10$^8$ dyne
cm$^{-2}$ (the standard values for the heavy models) had a final mass,
luminosity, and radius of 6.737 \Msun, $6.82 \times 10^{38}$ erg
s$^{-1}$, and $3.96 \times 10^{10}$ cm. The same star run with surface
zoning of 10$^{21}$ gm and a boundary pressure of 10$^4$ dyne
cm$^{-2}$ had final mass, luminosity, and radius 6.750 \Msun, $7.37
\times 10^{38}$ erg s$^{-1}$, and $9.9 \times 10^{10}$ cm. In this
fine zoned model, a radius of $4 \times 10^{10}$ cm existed just
10$^{-9}$ \Msun \ deeper in. Such a small mass can affect the observed
temperature of the progenitor star, but not the internal evolution or
light curve. The differences were even less for lighter stars. A 4.0
\Msun \ model with its standard zoning had mass, luminosity and radius
3.155 \Msun, $2.33 \times 10^{38}$ erg s$^{-1}$, and $2.43 \times
10^{11}$ cm. With surface zoning of 10$^{22}$ gm instead of 10$^{25}$
gm, the mass, luminosity, and radius were 3.158 \Msun, $2.35 \times
10^{38}$ erg s$^{-1}$, and $2.27 \times 10^{11}$ cm.

Resolving temperature and density gradients in the thin helium shell
of the lower mass stars also took zoning that approached 10$^{-6}$ \Msun. For
models close to the Chandrasekhar mass, gradients in density were so
steep that the zoning would have been too fine for the star to evolve,
even in hundreds of thousands of steps. This limited the mass to which
degenerate core growth was followed for stars in the 1.6 to 2.4 \Msun
\ range, typically to around 1.26 \Msun where carbon flames ignited.

The treatment of convectively bounded flames (CBFs) during oxygen and
silicon burning was as in \citet{Woo15}.  Opacities, the treatment of
convection, the equation of state, reaction rates, simulations of
explosions, and radiation transport in supernovae were all the same as
in past works using the KEPLER code, \citep{Woo02,Suk18}.

\subsection{Mass Loss}
\lSect{mdot}

Once revealed, a helium core in a close binary system continues to
experience mass loss through winds. The uncertainty in mass loss rate
and its dependence on metallicity dominate other factors such as
exactly when and how the helium star was formed. Here we use the mass
loss rate compilation of \citet{Yoo17}, which is, in turn, an
amalgamation of the rates of \citet{Tra16} for Wolf-Rayet stars of
types WC and WO and \citet{Han14} for WNE stars. In particular, as
\citet{Yoo17} discusses, we take a mass loss rate for WC and WO stars,
defined as when the surface mass fraction of helium drops below $Y_s =
0.9$, of
\begin{multline}
\log \, \dot M_{\rm CO} \ = \  -9.2 \, + \, 0.85 \, \log
\left(\frac{L}{\Lsun} \right) \, + \cr 0.44 \, \log \, Y_s \,
  +  0.25 \, \log \, \left(\frac{X_{\rm Fe}}{X_{\rm Fe \, \sun}} \right).
\lEq{mdotwr}
\end{multline}
The metallicity in the form of iron is assumed to include all isotopes
of the iron group and is $X_{\rm Fe \, \sun} = 1.46 \times
10^{-3}$. All models calculated in this paper had Z = \Zsun, so the
metallicity term in these equations is zero. For further discussion
of the metallicity dependence see \citet{Eld06}. 

For WNE stars, nominally $Y_s > 0.98$,
\begin{multline}
\log \, \dot M_{\rm WNE} \ = \ -11.32 \, + \, 1.18 \, \log
\left(\frac{L}{\Lsun}\right) \cr + 0.6 \, \log \, \left(\frac{X_{\rm
    Fe}}{X_{\rm Fe \, \sun}} \right)
\lEq{mdotwne}
\end{multline}.
\noindent
For surface helium mass fractions between 0.9 and 0.98, the mass loss
rate is linearly interpolated 
\begin{equation}
\dot M_{\rm int} \ = \ f \, \dot M_{\rm CO} \, + \, (1 -f) \, \dot M_{\rm WNE},
\lEq{mdot}
\end{equation}
with $f = 12.5 \, (0.98 - Y_s$).

\begin{figure}
\includegraphics[width=0.48\textwidth]{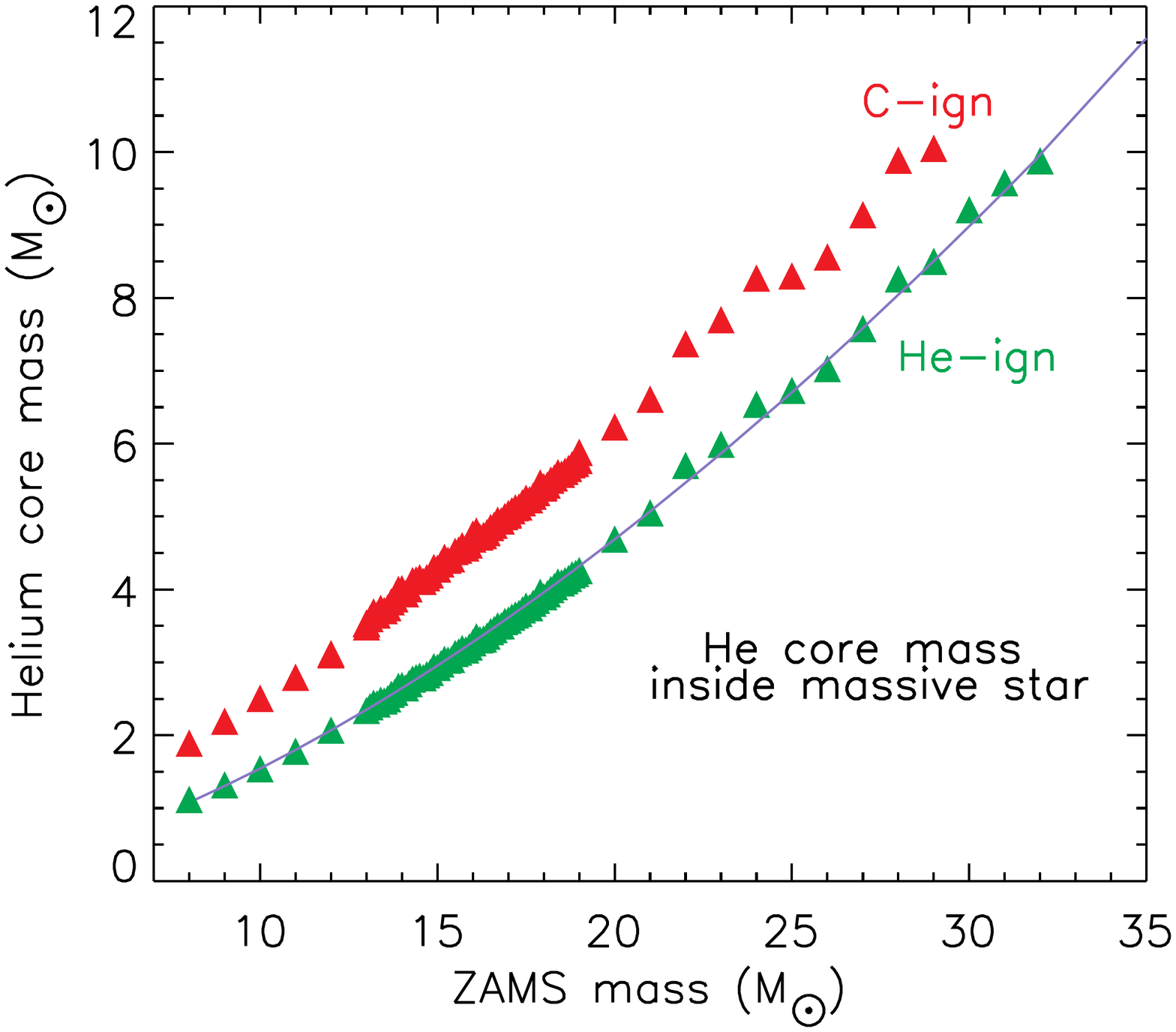}
\includegraphics[width=0.48\textwidth]{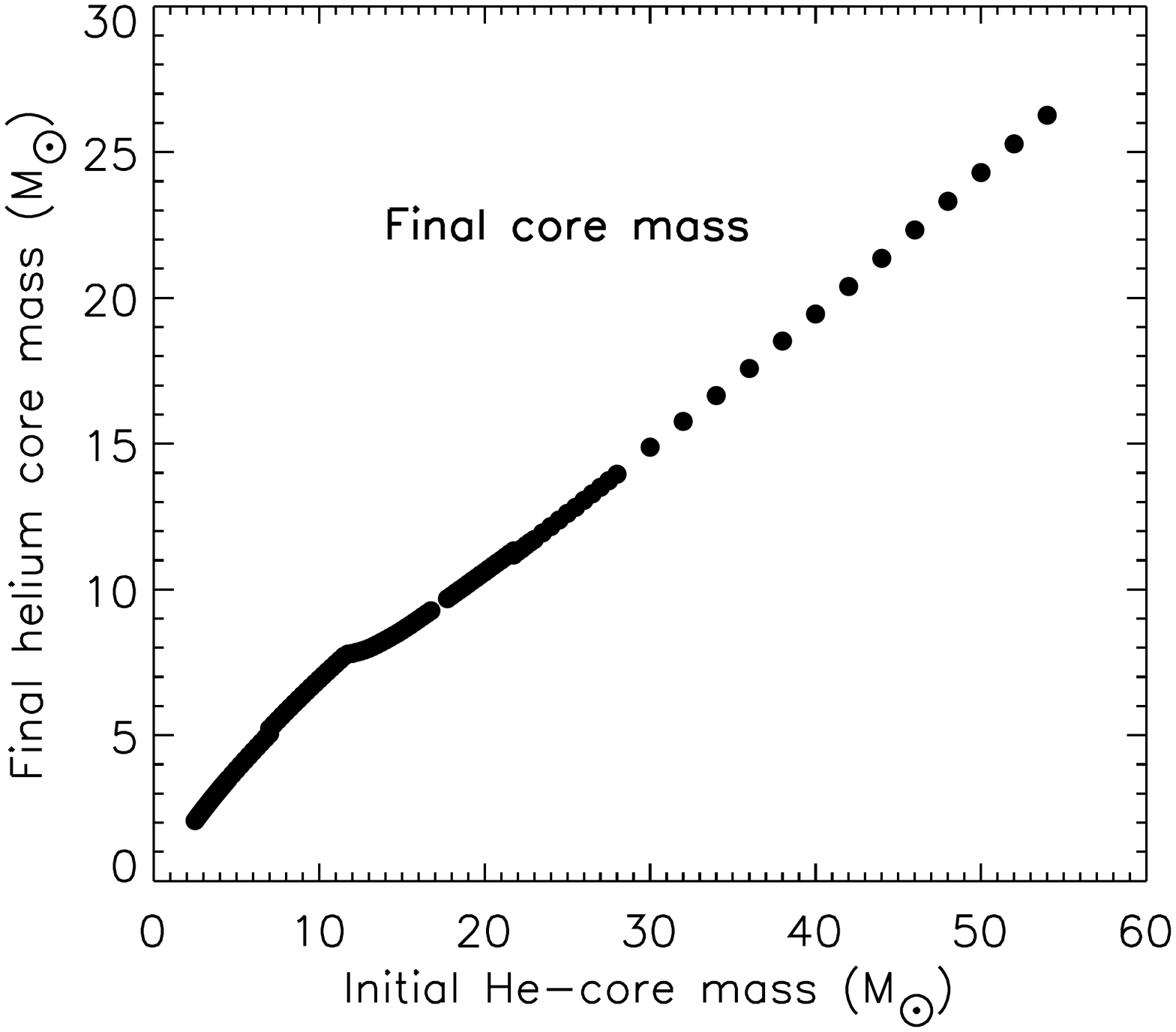}
\caption{The core mass of helium and heavy elements is given for
  single stars of solar metallicity that do not lose all their
  hydrogen envelope prior to their death, and for bare helium stars
  evolved with mass loss throughout their helium-burning
  evolution. (Top:) Masses at the time of central helium ignition in
  full stars, close to the time when the star first becomes a
  supergiant, are given as green points. Helium core masses in
  presupernova stars that do not lose their envelopes are in red. This
  mass does not change appreciably after carbon ignition. The
  blue line is a fit to the data below 30 \Msun \ (\Eq{mzams}). (Bottom:)
  The final presupernova mass as a function of initial helium core
  mass assuming the standard mass loss rate. The inflection around 11
  \Msun \ reflects the uncovering of the CO-core by mass loss.
  \lFig{cores}}
\end{figure}

It is assumed that the helium core is revealed at helium ignition and
experiences this mass loss throughout its lifetime. Based on the
models of \citet{Woo07} and \citet{Suk18}, a fit to the helium core
mass in single massive stars, {\sl at helium ignition}, as a function
of zero age main sequence mass (ZAMS mass, i.e., the original
hydrogenic star) is
\begin{equation}
M_{\rm He,i} \ \approx \ 0.0385 \, M_{\rm ZAMS}^{1.603}  \, \Msun \ \ \ 
(M_{\rm ZAMS} < 30 \Msun).
\lEq{mzams}
\end{equation}
This fit is valid for main sequence masses in the range 6 to 30 \Msun.
For example, a 6 \Msun \ initial helium core mass corresponds to a
23.3 \Msun \ main sequence star. This expression is plotted as a line
through the green points in the upper panel of \Fig{cores}.
Above 30 \Msun, and up to at least 140 \Msun, a better fit for stars
that do not lose their entire hydrogen envelope is 
\begin{equation}
M_{\rm He,i}
\ \approx \ 0.50 \, M_{\rm ZAMS} \, - 5.87 \, \Msun \ \ (M_{\rm ZAMS} \ge 30 \Msun).
\lEq{mzamsh}
\end{equation}

\section{Presupernova Evolution}
\lSect{evolution}

Depending on mass, a wide variety of outcomes is expected. In each
case, the range referred to in the section head is the zero age
helium core mass, i.e., the mass at helium ignition immediately
following the removal of any hydrogen envelope.

\subsection{$M_{\rm He,i}$ = 1.6 - 3.2 \Msun; White Dwarfs and Unusual Supernovae}

Helium stars with initial masses between 1.6 and 3.2 \Msun \ develop
degenerate cores of carbon and oxygen ($<1.6$ - 1.8 \Msun); neon,
oxygen, and magnesium (1.9 - 2.4 \Msun); or silicon (2.5 - 3.2
\Msun). Their final evolution is characterized by temperature
inversions, flashes, off-center ignition, and convectively bounded
flames. End states depend upon the mass loss during the final
burning stages which may be due to winds or a combination of winds and
resumed transfer to the binary companion. Radii exceeding
10$^{13}$ cm are common (\Tab{cign} and \Tab{finallow}).  Above about 3.2
\Msun, for the present choice of mass loss rates, this large expansion
does not occur (\Tab{siign}).  We consider first the case of mass loss
by winds alone, and then comment on how the results might be altered
by late stage mass loss to a companion (\Sect{binary2}).

\subsubsection{$M_{\rm He,i}$ = 1.6 - 2.4 \Msun; White Dwarfs and Electron-capture 
Supernovae}
\lSect{NeO}

After central helium depletion, helium stars from the least massive
considered here (1.6 \Msun) to 2.4 \Msun \ develop degenerate
carbon-oxygen or neon-oxygen-magnesium (NeO) cores surrounded by steep
density gradients, thin helium-burning shells, and low density,
convective helium envelopes. The radii of such stars can become quite
large (\Tab{cign}), and grows with time \Tab{finallow}). The
degenerate core also increases its mass as carbon and oxygen are added
by helium burning. These attributes are similar to Super Asymptotic
Giant Branch (SAGB) stars \citep[e.g.][]{Gar94,Sie06}, but differ in
that the envelope here consists entirely of helium. Nevertheless, the
term SAGB will be used to describe this structure.  Another key
difference is that the helium burning shell here, despite being thin
in mass, is not unstable. Perhaps a mild instability is suppressed by
the implicit hydrodynamics of the code and large time steps which are
typically 10$^{6.5}$ to 10$^{7.5}$ s during the thin shell
epoch. Resolving the helium burning shell requires very fine zoning
though, down to 10$^{-6}$ \Msun. With this resolution, the burning
breaks up into two regions, a nitrogen burning shell which lies just
beneath the base of the convective envelope, and a broader helium
burning region, which is also radiative.

Provided these stars retain sufficient mass, they have two
opportunities to ignite carbon burning, first as the helium burning
ashes contract and heat up and second, as the helium burning shell
builds up a critical mass and ignites carbon closer to the surface. The
critical mass for that second ignition is estimated here to be near
1.26 \Msun.  Consequently, the most massive white dwarf with a thick
carbon shell should be 1.26 \Msun.

\paragraph{Ignition of carbon within the helium-depleted core}

\begin{figure*}
\includegraphics[width=0.49\textwidth]{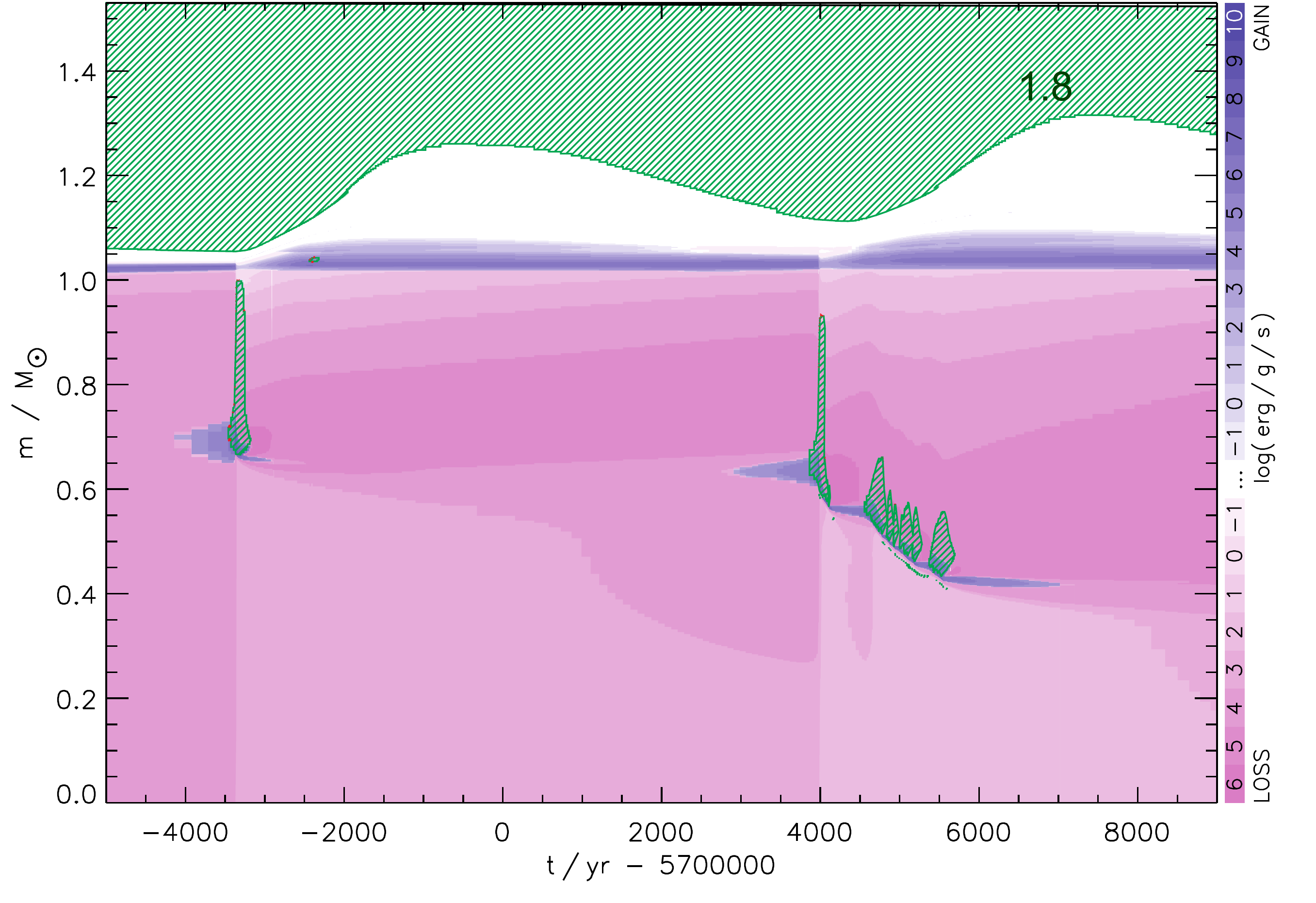}
\includegraphics[width=0.49\textwidth]{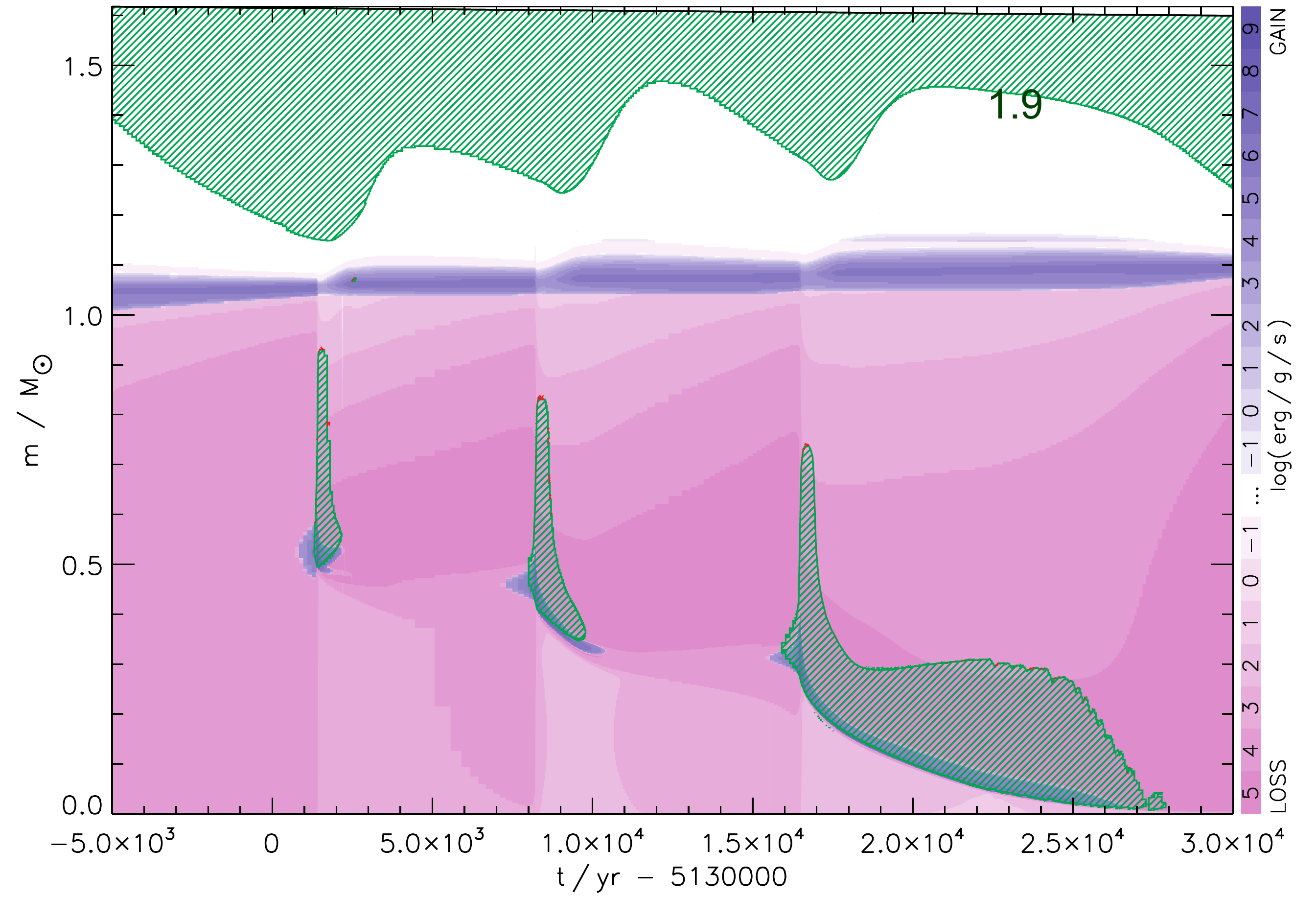}
\\
\includegraphics[width=0.49\textwidth]{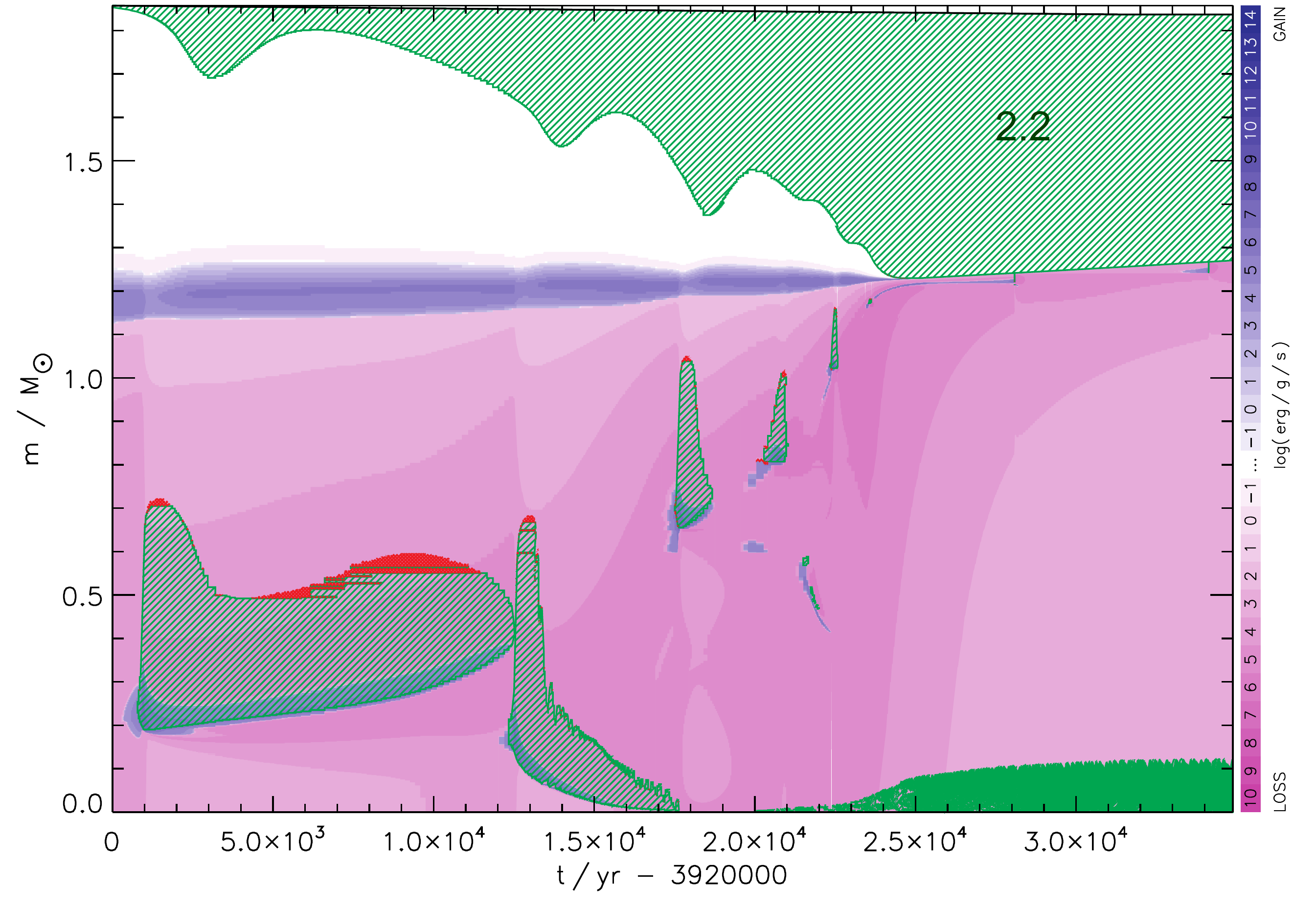}
\includegraphics[width=0.49\textwidth]{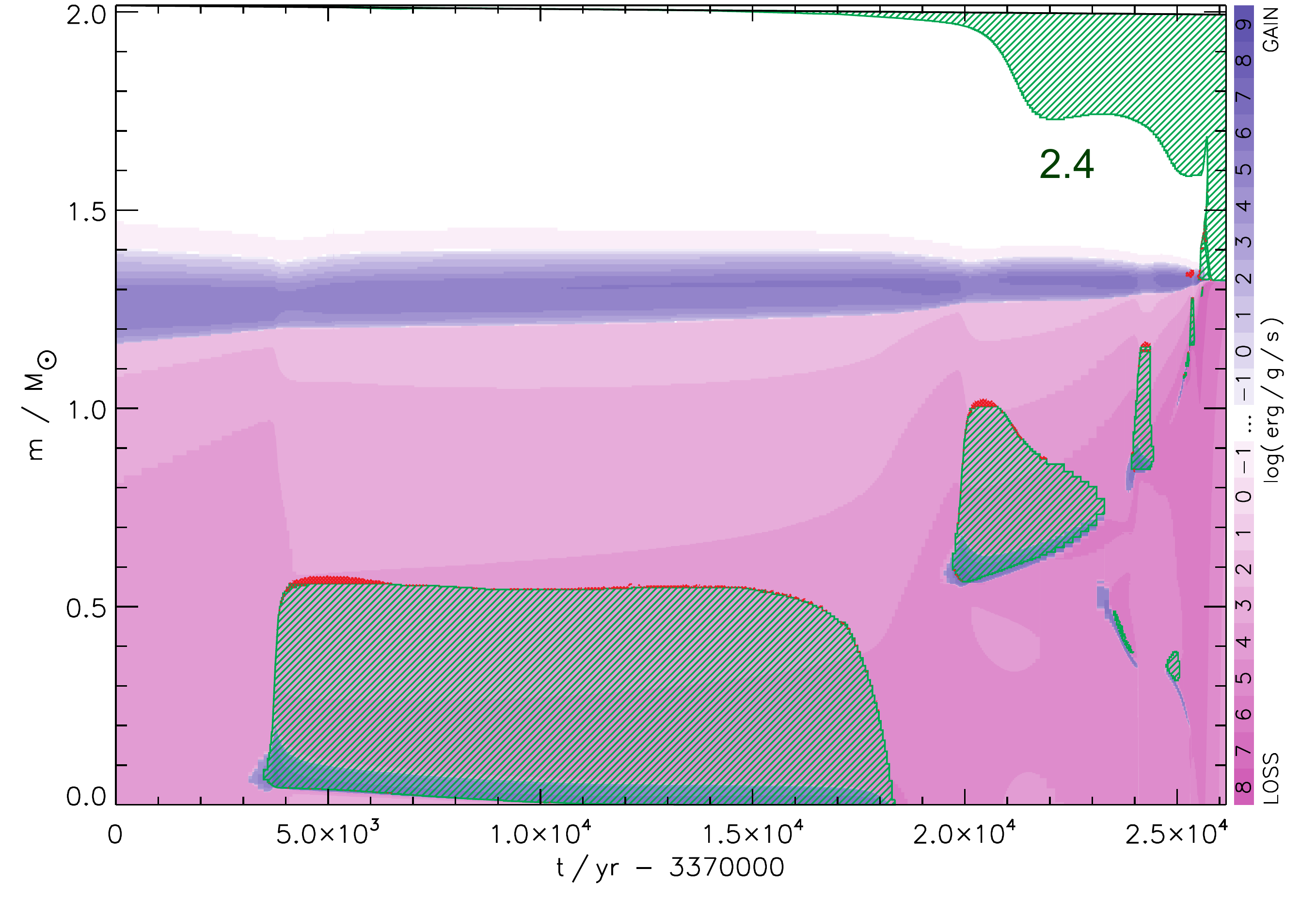}
\caption{Convection following carbon ignition in the 1.8 (top left),
  1.9 (top right), 2.2 (bottom left), and 2.4 (bottom right). Green
  cross hatching indicates regions that are convective; shades of blue
  indicate positive net nuclear energy generation and pink, negative
  due to neutrino losses. Semiconvective regions are red. The offset
  in time is indicated on the x-axis. Carbon ignites in
  the 1.8 \Msun \ model about 5.7 million years after helium 
  ignition. The top of the y-axis is the current mass of the star
  (1.53 \Msun \ for the 1.8 \Msun \ model). This model
  experiences several off-center carbon flashes but never ignites a
  convectively bounded flame. The 1.9 and 2.2 \Msun \ models experience
  3 and 2 flashes respectively, and do ignite a flame that burns to
  the center of the star. The 2.4 \Msun model \ ignites a CBF
  0.057 \Msun \ off center that burns quickly to the middle. Carbon
  flashes and flames cause the core to expand which diminishes the
  depth of the convective envelope and shrinks the stellar radius.
  Eventually all stars have fully convective envelopes.
  \lFig{convection}}.
\end{figure*}

\Fig{convection} and \Tab{cign} show some characteristics of ``first''
carbon ignition in this mass range. Below 1.8 \Msun, carbon burning
does not ignite prior to the formation of an SAGB-like structure.  The
CO-core is initially less than the 1.06 \Msun \ necessary for ignition
\citep{Nom86,Nom88}. The 1.8 \Msun \ model itself is a transition case
that ignites carbon burning in a flash 0.69 \Msun \ off center
(\Fig{convection}) with a CO-core mass of only 1.03 \Msun. A second
flash occurs 7500 years later. Both flashes are weak though, and fail
to exhaust the carbon at the ignition sites, or to ignite a sustained
CBF. The star at this point has not become become a fully developed
giant and the pressure of the overlying helium is not negligible,
hence the lower carbon ignition mass and incomplete burning. It
remains mostly a CO-white dwarf.

It is not uncommon to see separate flashes preceding the formation of
a CBF \citep{Sie06,Far15}, or cases where flashes do not lead to
flame formation. See, for example, the 7.0 \Msun \ model of
\citet{Far15}, especially their Figure 3. This sort of evolution is
thought to lead to ``hybrid CO+NeO white dwarfs'' and that is what
is also found here for the 1.8 \Msun \ model. Most of the core
remains CO, but there is a shell of partly burned carbon.

By comparing diffusion and burning time scales, \citet{Far15} estimate
a unique density for off-center carbon ignition, $2.1 \times 10^6$ g
cm$^{-3}$. In good agreement, the 1.8 \Msun \ model here ignited
carbon at $1.5 \times 10^6$ g cm$^{-3}$ when its central density was
$1.30 \times 10^7$ g cm$^{-3}$.

The 1.9 \Msun \ model was the lowest mass to clearly ignite a
sustained carbon CBF (\Tab{cign}) and did so with a core mass of 1.06
\Msun \ (\Tab{cign}).  It did so after three flashes
(\Fig{convection}), the third of which developed into a flame that
propagated to the center of the star. Even so the composition remained
contaminated by a substantial trace of residual, unburned carbon.

From 1.9 to 2.4 \Msun, all stars ignite carbon CBFs \citep{Nom85} that
burn to the center of the star \citep{Tim94} prior to becoming SAGB
stars. This turns the composition of the core to neon, oxygen, and
magnesium. As the mass of the core increases, a diminishing number of
flashes precedes flame formation (\Fig{convection}). The flash that
bounds the eventual flame ignites systematically deeper in the star at
a density that remains near $1.5 \times 10^6$ g cm$^{-3}$. The
decreasing central density of more massive stars causes the ignition
to move inwards \citep{Far15}. These flashes lead to an expansion of
the core that temporarily weakens the helium burning shell so that the
base of the convective envelope moves out in mass. The flashes are
less violent at higher mass where the degeneracy is less.  The 2.5
\Msun \ model was the last to ignite carbon off center, doing so at an
offset from the center of only 0.005 \Msun.

In all models with a CBF, zoning was sufficiently fine that the flame
moved by thermal diffusion. Rezoning was not allowed within the flame,
but the calculation included  standard overshooting as described
  in \citet{Suk18}. Thermohaline mixing was not included, though the
Ledoux criterion led to the mixing of regions with an inverted atomic
mass when the temperature gradient was not strongly inverted. The
flame speed is known to be sensitive to zoning, convective overshoot,
and thermohaline mixing \citep{Tim94,Sie06,Den13}. Sensitivity studies
were not carried out, but the speeds here were about five times slower
than \citet{Tim94} and their speeds should not be critical to the
subsequent evolution.

\paragraph{Ignition of carbon as an SAGB star}

The luminosity of models during their SAGB phase, which is due to
helium shell burning, can be fitted to a power of the core mass, i.e.,
there is a ``core luminosity relation'' for helium SAGB stars
\citep{Jef88}. Using the values in \Tab{cign} and \Tab{finallow} and
earlier data points for core masses in the range 1.0 to 1.3 \Msun,
\begin{equation}
L \ \approx 1.3 \times 10^{38} \, M_c^{4.1} \ {\rm erg \ s^{-1}},
\lEq{lsagb}
\end{equation}
where $M_c$ is the core mass (CO + NeO) in solar masses \citep[see
  also][their eq. 3.2]{Hav79}.

Burning one gram of helium to one gram of carbon and oxygen (50\% by
mass each) yields $7.3 \times 10^{17}$ erg. The core luminosity
relation thus implies that the compact core will increase in mass at a
rate
\begin{equation}
\dot M_c \ \approx 2.8 \times 10^{-6} \ M_c^{4.1} \ {\rm \Msun \ y^{-1}},
\lEq{cogrow}
\end{equation}
which is about 6 to 8 $\times 10^{-6}$ \Msun \ y$^{-1}$ for core
masses between 1.2 and 1.3 \Msun. Growth rates in this range imply
that carbon will ignite when the white dwarf reaches a mass of 1.25 -
1.3 \Msun \ \citep[Table 2][]{Sai04}.  Combining \Eq{lsagb} and
\Eq{mdotwne}, the mass loss to winds in the SAGB phase is
\begin{equation}
\dot M_{\rm wind} \ \approx 1.1 \times 10^{-6} \ M_c^{4.84} \ {\rm
  \Msun \ y^{-1}},
\lEq{mdotagb}
\end{equation}
so for $M_c \approx 1.3$ \Msun, the core gains about 2 gm for every 1
gm lost to winds.

\begin{deluxetable}{cccccc} 
\tablecaption{Carbon Ignition in Low Mass Models} 
\tablehead{ \colhead{${M_{\rm init}}$}           & 
            \colhead{${M_{\rm ign}}$}            &
            \colhead{${M_{\rm CO}}$}             &
            \colhead{${M_{\rm C-ign}}$}             &
            \colhead{${L_{38}}$}                & 
            \colhead{${R_{\rm 13}}$}        
            \\
            \colhead{[\Msun]}                   &  
            \colhead{[\Msun]}                   &
            \colhead{[\Msun]}                   &
            \colhead{[\Msun]}                   &
            \colhead{[10$^{38}$ erg \ s$^{-1}$]}  & 
            \colhead{[10$^{13}$ cm]}               
            }
\startdata
1.6 &   1.36 & 0.953 &  -    & 1.14  & 0.94    \\     
1.7 &   1.45 & 0.985 &  -    & 1.20  & 0.96    \\     
1.75 &  1.49 & 1.00  &  -    & 1.25  & 0.98    \\     
1.8  &  1.53 & 1.03  & 0.622 & 0.955  & 0.83    \\     
1.9  &  1.61 & 1.06  & 0.519 & 1.09  & 0.89    \\     
2.0  &  1.70 & 1.08  & 0.407 & 0.872 & 0.77    \\     
2.1  &  1.78 & 1.12  & 0.306 & 0.787 & 0.66    \\    
2.2  &  1.85 & 1.18  & 0.213 & 0.854 & 0.52    \\
2.3  &  1.94 & 1.20  & 0.127 &  0.93 & 0.15    \\    
2.4  &  2.01 & 1.24  & 0.057 &  0.99 & 0.063   \\
2.5  &  2.09 & 1.28  & 0.005 &  1.04 & 0.037   \\
2.6  &  2.17 & 1.32  &  0    &  1.09 & 0.028   \\
2.7  &  2.22 & 1.46  &  0    &  1.15 & 0.023   \\    
\enddata
\tablecomments{For M$_{\rm init} > 1.75$ \Msun, $M_{\rm ign}$ and
  $M_{\rm CO}$ are the masses of the star and its CO core when carbon
  first ignites and R$_{13}$ and L$_{38}$ are the star's radius
  and luminosity.  M$_{\rm C-ign}$ is the mass shell where carbon
  ignites. For the three lighter models, approximate conditions are
  given when the star first develops a thin helium burning
  shell. Carbon burning has not ignited.}  \lTab{cign}
\end{deluxetable}

Carbon did not burn continuously in the shell that grew beneath
the thin helium burning shell in the SAGB stars.  A critical mass of
unburned carbon and oxygen accumulated before igniting and
running away. Each runaway ignited a localized CBF, and the
combination of convection and flame consumed the accumulated carbon
layer turning it to neon, oxygen, and magnesium. Similar to the
hydrogen critical mass in classical novae, the critical mass of this
carbon layer became smaller as the mass of the core and the
gravitational potential at its edge increased.  For models lighter than
2.1 \Msun \ case, the critical mass, $\sim0.1 - 0.2$ \Msun \ was so
large that only one layer was observed to form and run away during the
time studied. For example, the 2.0 \Msun \ model ended carbon core
burning with a NeO core mass of 1.15 \Msun. Once a SAGB structure
developed, helium burning in a thin shell increased the mass of the CO
core to 1.26 before a runaway ignited at 1.22 \Msun. Burning in this
layer eventually converted the carbon and oxygen from 1.15 to 1.25
\Msun \ to NeO. Though the subsequent evolution was not followed, the
critical mass for later runaways on this larger core will be
smaller. That is, the critical mass depends on the current core mass,
not the original helium star mass.  For the 2.3 \Msun \ model where
the initial core mass was similar to the final mass in the 2.0 \Msun
\ model, multiple carbon flashes were followed (\Fig{cflashes}) and
the critical mass was determined to be $\sim$0.01 \Msun \ for a core
mass of 1.27 \Msun. For the 2.4 \Msun \ model carbon burning
approached a steady state with helium burning, lagging just a few
thousandths of a solar mass beneath, and no flashes were observed.

\begin{deluxetable}{cccccc} 
\tablecaption{Late Evolution Below 2.5 \Msun} 
\tablehead{ \colhead{${M_{\rm init}}$}           & 
            \colhead{${M_{\rm current}}$}            &
            \colhead{${M_{\rm CO}}$}             &
            \colhead{${L_{38}}$}                & 
            \colhead{${R_{\rm 13}}$}             &
            \colhead{}
            \\
            \colhead{[\Msun]}                   &  
            \colhead{[\Msun]}                   &
            \colhead{[\Msun]}                   &
            \colhead{[10$^{38}$ erg \ s$^{-1}$]}  & 
            \colhead{[10$^{13}$ cm]}             &               
            \colhead{}
            }
\startdata
1.6  & 1.21 & 1.21 &  -   &    -   &  CO WD   \\
1.7  & 1.31 & 1.22 & 2.85 &   1.34 &  CO      \\
1.75 & 1.33 & 1.26 & 3.27 &   1.39 &  CO      \\
1.8  & 1.39 & 1.24 & 3.03 &   1.33 &  CO/NeO  \\
1.9  & 1.48 & 1.26 & 3.18 &   1.41 &  NeO     \\
2.0  & 1.59 & 1.26 & 3.25 &   1.48 &  NeO     \\
2.1  & 1.73 & 1.23 & 3.19 &   1.53 &  NeO     \\ 
2.2  & 1.80 & 1.28 & 3.71 &   1.63 &  NeO     \\
2.3  & 1.87 & 1.32 & 4.16 &   1.74 &  NeO     \\
2.4  & 1.96 & 1.32 & 4.18 &   1.77 &  NeO     \\
2.5  & 2.07 & 1.37 & 0.96 &   0.72 &  Si flash        
\enddata
\tablecomments{For models from 1.7 to 2.4 \Msun, conditions are given
  at the last model calculated and are not the terminal values. CO and
  NeO indicate the major constituents of the core at that time. Were
  the envelope not lost, continued growth of the core to the
  Chandrasekhar mass would lead to electron-capture supernovae in all
  cases from 1.8 to 2.5 \Msun \ model. } \lTab{finallow}
\end{deluxetable}

\begin{figure}
\includegraphics[width=0.48\textwidth]{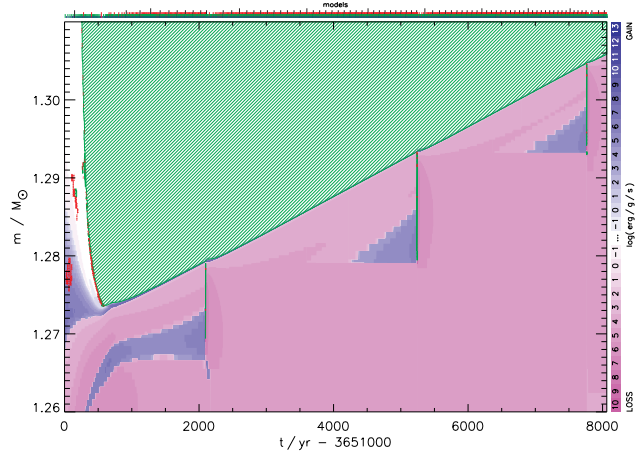}
\caption{Carbon flashes in the 2.3 \Msun \ model. Helium burning leads
  to the creation of carbon layers about 0.01 \Msun \ thick that then
  ignite somewhere near their center. Eventually the region becomes
  convective reaching almost to the helium shell and burns inwards as
  a flame. Three flashes are shown. The convective episode in each
  comes at the end and lasts about 10 years. Here purple is the CO
  plus NeO core which is losing energy to neutrinos, green cross
  hatching denotes convection and blue, positive energy
  generation. \lFig{cflashes}}
\end{figure}

No SAGB model was followed all the way to the Chandrasekhar mass. As
the core grew more massive and compact, its surface gravity
strengthened, making the helium and nitrogen burning shells thinner
and requiring more zones for their resolution. Burning helium to
carbon in zones of ever decreasing mass required more time steps to
maintain a given growth rate.  The envelope expanded to large radii
and partly recombined, causing additional numerical instability. The
future evolution of the models can be estimated though assuming: 1)
that winds dominate the mass loss; 2) the core grows at twice the rate
at which the total star loses mass (\Eq{mdotagb}), 3) all cores over
1.26 \Msun \ ignite carbon burning in the accumulated layer; and 4)
electron-capture supernovae happen when the core reaches 1.38
\Msun. This growth rate is consistent with the numerical models
evaluated during the last 0.1 \Msun \ of core growth
(\Tab{finallow}). It is estimated that the core of the 1.7 \Msun
\ model will grow to 1.27 \Msun \ while the star loses 0.03 to end up
as a white dwarf of 1.27 \Msun, probably composed of NeO. Similarly,
the 1.8 \Msun \ model will end up as a 1.34 \Msun \ NeO white
dwarf. The 1.9 \Msun \ model will become an electron-capture supernova
when its core mass reaches 1.38 \Msun \ and the envelope's mass is
only 0.05 \Msun; the 2.0 \Msun \ model will make a similar supernova with
an envelope mass of 0.16 \Msun; and so on.

\subsubsection{$M_{\rm He,i}$ = 2.5 - 3.2 \Msun; Silicon Flashes}
\lSect{deflag}

\begin{deluxetable*}{ccccccccc} 
\tablecaption{Silicon Ignition in Low Mass Models} 
\tablehead{ \colhead{${M_{\rm init}}$}        & 
            \colhead{${M_{\rm fin}}$}         &
            \colhead{${M_{\rm CO}}$}         &
            \colhead{${R_{13}}$}           & 
            \colhead{${\rho_c}$}           & 
            \colhead{${M_{\rm ign}}$}        & 
            \colhead{${\eta}$}             &
            \colhead{${T_{\rm ign}}$}        & 
            \colhead{Si-Ign} 
            \\
            \colhead{[\Msun]}               &  
            \colhead{[\Msun]}               &
            \colhead{[\Msun]}               &
            \colhead{[10$^{13}$ cm]}         &
            \colhead{[10$^9$ g \ cm$^{-3}$]}  & 
            \colhead{[\Msun]}               &
            \colhead{}                      & 
            \colhead{[10$^9$ K]}            &
            \colhead{}                       
            }
\startdata
2.5 & 2.07 & 1.367 & 0.719 & 1.76  &  0.504  &  9.39 & 3.23 &  deflagration \\
2.6 & 2.15 & 1.414 & 0.778 & 1.36  &  0.414  &  9.13 & 3.23 &  weak pulse \\
2.7 & 2.22 & 1.459 & 0.760 & 1.17  &  0.38   &  9.09 & 3.18 &  weak pulse \\ 
2.8 & 2.30 & 1.507 & 0.724 & 0.974 &  0.30   &  9.05 & 3.21 &  weak pulse \\
2.9 & 2.37 & 1.556 & 0.647 & 0.772 &  0.18   &  9.04 & 3.18 &  weak pulse \\
3.0 & 2.45 & 1.604 & 0.476 & 0.610 &  0.090  &  9.16 & 3.19 &  deflagration \\
3.1 & 2.52 & 1.656 & 0.191 & 0.495 &  0.011  &  9.27 & 3.18 &  deflagration \\
3.2 & 2.59 & 1.709 & 0.104 & 0.419 &   0     &  8.87 & 3.20 &  deflagration \\
3.3 & 2.67 & 1.761 & 0.070 & 0.341 &   0     &  8.13 & 3.20 &  normal       \\
3.4 & 2.74 & 1.809 & 0.053 & 0.299 &   0     &  7.59 & 3.23 &  \ \ normal 
\enddata
\tablecomments{$M_{\rm init}$, $M_{\rm fin}$, $M_{\rm CO}$, and
  $M_{\rm ign}$ are the masses of the initial helium core, the final
  presupernova mass, the CO core at silicon ignition, and
  the shell where silicon burning ignites. $T_{\rm ign}$ and $\eta$
  are the temperature (in 10$^9$ K) and electron degeneracy parameter
  at the location where silicon ignites, i.e., at $M_{\rm
    ign}$  and $R_{13}$ is the radius of the star then.}  \lTab{siign}
\end{deluxetable*}

Initial helium core masses of 2.5 to 3.2 \Msun \ correspond to stars
with main sequence masses 13.5 to 15.8 \Msun \ (\Eq{mzams},
\Fig{cores}). At death, the mass ranges from 2.07 to 2.59 \Msun \ and
the CO-core mass from 1.37 to 1.71 \Msun \ (\Tab{siign}). This range of
final helium and CO core masses is known to be characterized by the
lingering effects of off-center ignition and degenerate silicon
flashes \citep[e.g.][Table 1]{Woo15}.

As with carbon ignition, the displacement of oxygen ignition from the
center shrinks with increasing mass. In the 3.5 \Msun \ model, both
neon and oxygen burning ignite only 0.004 \Msun \ off center. The
oxygen-burning CBF propagates to the center, leaving behind
composition and entropy profiles that set the stage for silicon
ignition. How these flames progress may be different in a more
realistic three-dimensional simulation and the assumption of a
spherically symmetric flame propagation by conduction rather than
turbulent undershoot mixing is questionable \citep{Woo15}.

Silicon burning also ignites off center for helium stars up to 3.2
\Msun \ (\Tab{siign}) with the displacement, again, a declining
function of mass. Silicon ignition is defined as a region with
exoergic energy generation capable of driving convection with a base
temperature exceeding $3.2 \times 10^9$ K. The core's temperature
profile at silicon ignition is inverted by neutrino losses so that
higher temperatures occur farther out in the star. Lower mass stars
retain their high degeneracy farther out, and this accounts for the
violent runaway in the outer regions of the 2.5 \Msun \ model, but not
the 2.6 \Msun \ model. At still greater masses, higher temperature
migrates deeper into the star allowing another island of instability
from 3.0 to 3.2 \Msun.

The silicon flash in stars with violent runaways begins at a
degeneracy parameter $\eta \gtaprx 9$ and reaches, locally, very
high temperatures, around $6 \times 10^{9}$ K, leading either to
detonation or deflagration \citep{Woo15}. If convection is left on
until the energy transport exceeds 10$^{49}$ erg s$^{-1}$, the ``sonic
limit'', detonation forms that decays a short time later to a
deflagration. If convection is more restricted, a hot, less dense layer
of iron forms that is buoyant and also seeds a deflagration. 

The strength of these explosions is poorly determined,
especially in a 1D study, but the kinetic energy of any ejecta
probably does not exceed a few $\times 10^{49}$ erg.  Typically the
runaways burn 0.1 to 0.5 \Msun \ of silicon to iron
(\Sect{sidefl}). The fusion of a gram of silicon releases $1.7 \times
10^{17}$ erg g$^{-1}$. This yield is evaluated from the existing
composition, but typically the fuel is rich in $^{30}$Si and the ash
in $^{56}$Fe.  The burning thus yields about $3.4 \times 10^{50}$ erg
per solar mass implying an explosive yield of order $0.3 - 1.7 \times
10^{50}$ erg, but the binding energy of the core at the time of
ignition is $4 - 5 \times 10^{50}$ erg (e.g., $4.55 \times 10^{50}$
erg in the 3.0 \Msun \ model).  It is only by transporting some
fraction of the explosion energy to the envelope by a shock wave that
any mass is ejected. The shock is produced by the large amplitude
oscillation of the core and the more burning, the larger the amplitude
and the stronger the shock.

Even stars where the runaway was not clearly supersonic, i.e., the
2.6 - 2.9 \Msun \ models, experienced mildly degenerate silicon
flashes that gave weak shocks in the envelope. Typically these shocks
decayed by momentum conservation to less than 100 km s$^{-1}$ before
reaching the star's surface. A tiny bit of matter is puffed off at
low speeds, but that will be overtaken by the optically thick supernova
before it brightens appreciably.

The radii of the star at silicon ignition - the same as ``presupernova
radius'' in \Tab{models} - ranges from $7 \times 10^{12}$ cm (2.5
\Msun \ model) to $1 \times 10^{12}$ cm (3.2 \Msun \ model). Given the
low energy of the shock and small radius, the display resulting from
the silicon flash will not be bright, and would decline rapidly.  A
much brighter display could result when the remaining core completes
silicon burning and collapses to a neutron star. Even a
neutrino-powered wind would impacting a solar mass shell at roughly 1
- 10 AU would give a luminous supernova \citep{Woo15}. These
possibilities are explored in \Sect{sidefl}.  .

\begin{figure}
\includegraphics[width=0.48\textwidth]{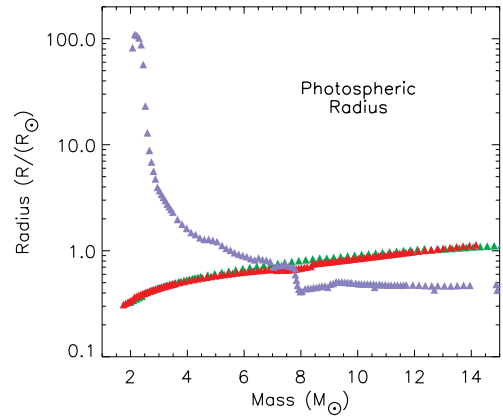}
\caption{Radii of the helium-star models with standard mass loss at
  various stages in their evolution. The mass is the mass of the star
  at the time of the plot. Green points are at the time of helium
  ignition; red points, the time of central helium depletion; and blue
  points, just prior to oxygen ignition. The dip near 8 \Msun \ for the
  blue points reflects the uncovering of the helium burning shell and
  a depletion of helium at the surface of the star. \lFig{radius}}
\end{figure}

\begin{figure}
\includegraphics[width=0.48\textwidth]{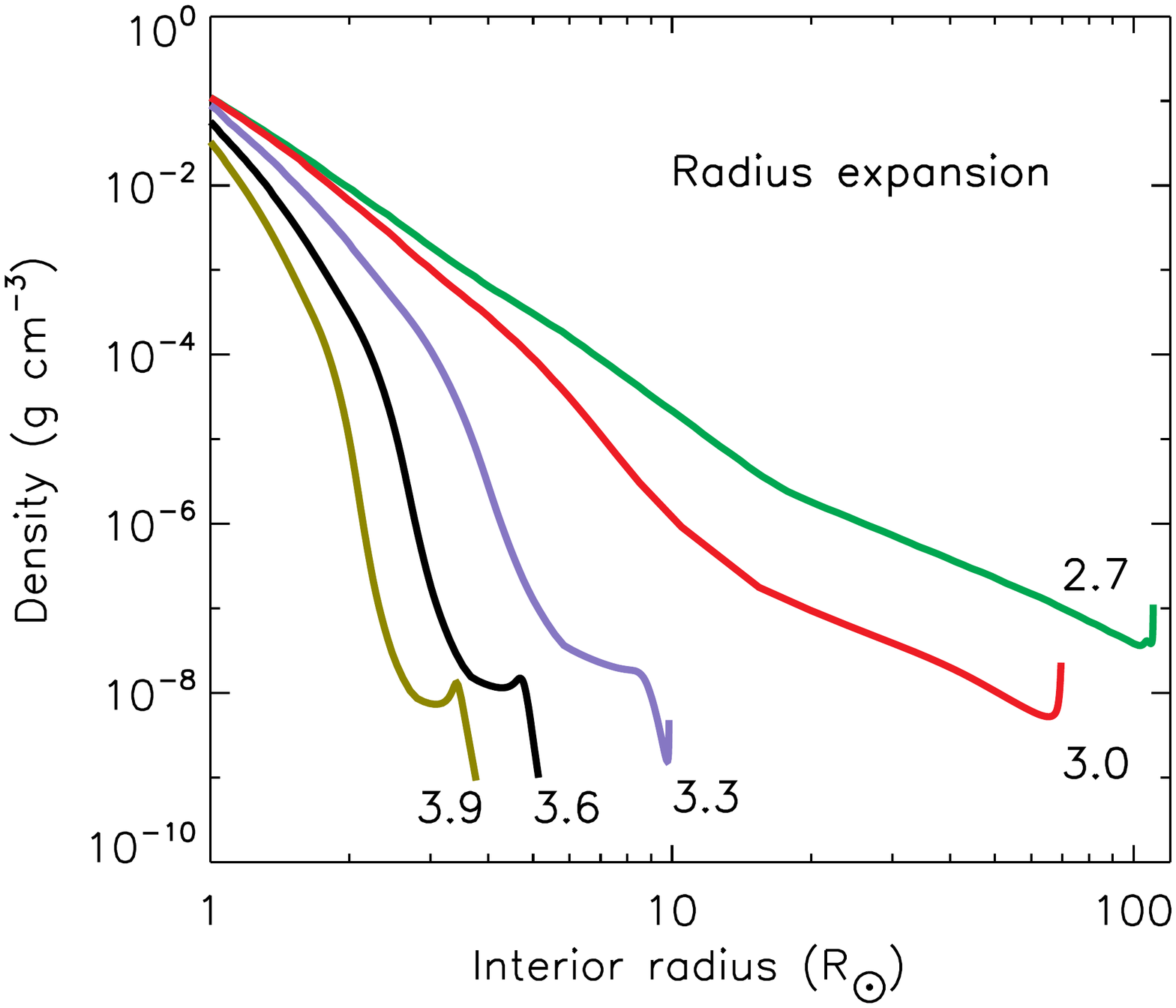}
\includegraphics[width=0.48\textwidth]{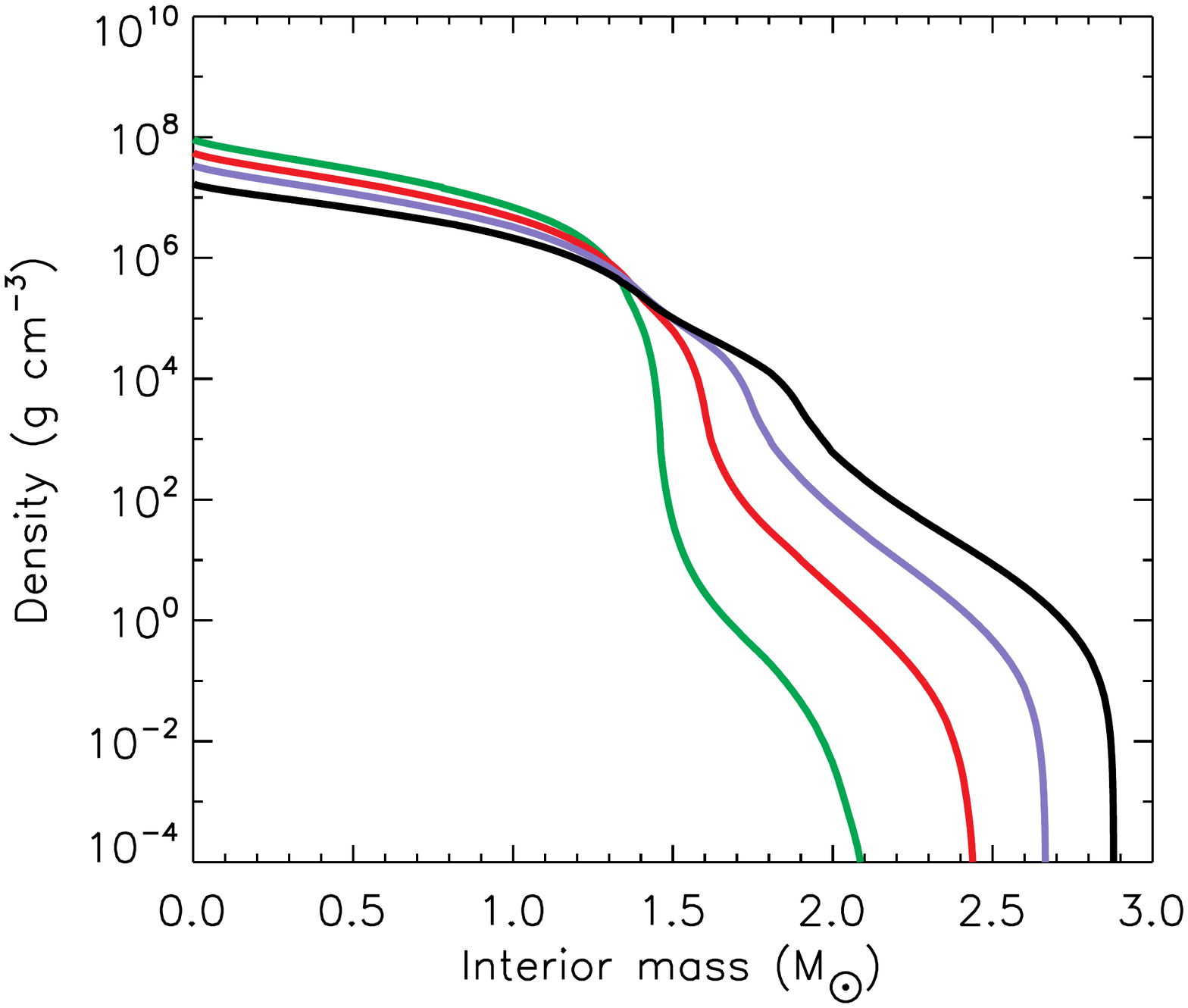}
\caption{Density profiles for stars experiencing radius expansion The
  density is evaluated at central oxygen depletion to avoid the
  complication of silicon flashes in some models.  (Top) The density
  in the outer layers of Models, 2.7, 3.0, 3.3, 3.6 and 3.9.  Below
  3.3 \Msun, helium stars experience significant radius expansion
  after helium core burning that may affect their shock break out and
  light curves \citep[see also][]{Pet06}. (Bottom) Density in the
  cores of the 2.7 to 3.6 \Msun \ models. (Final masses are 2.22,
  2.45, 2.67, and 2.88 \Msun). Only 0.12 \Msun \ is outside of 10 \Rsun
  \ in the 2.7 \Msun \ model, and 0.004 \Msun \ is outside 10 \Rsun
  \ in the 3.0 \Msun \ model.  \lFig{density}}
\end{figure}

\subsubsection{Modifications for late stage mass transfer}
\lSect{binary2}

Models below 2.5 \Msun \ develop thin helium shells and SAGB-like
structure after central helium depletion. This causes the surface to
expand to large radii. Heavier stars up to about 3.2 \Msun \ also
experience significant radius expansion during carbon burning, even
though their helium shells remains thick (\Tab{siign}).

For models below 1.8 \Msun, using \Eq{mdotwne}, a white dwarf results,
even if late stage binary transfer is neglected. If all stars that
expand beyond $1 \times 10^{13}$ cm are assumed to lose their
envelopes, then stars from 1.9 to 2.4 \Msun \ also become NeO white
dwarfs (\Tab{finallow}). From 2.4 to 3.2 \Msun \ things are
complicated. The mass of the CO core is already large enough when the
radius starts to expand that the central evolution will proceed pretty
much as already described. A supernova of some sort will result, but
how much helium remains and its radius will greatly affect the light
curves. Both these quantities are quite uncertain. For now, we adopt
an upper bound to the stars that make white dwarfs of 2.4 \Msun
\ \citep[see also][]{Tau15}.

Major adjustments to the star's outer structure happen during
the last roughly 10,000 years of its life so the
immediate progenitors of Type Ib and Ic supernovae might appear
different from ordinary Wolf-Rayet stars \citep[see also][]{Eld13}. For
example, assuming only loss by wind, the 2.5 \Msun \ model had a
radius of $1.2 \times 10^{11}$ cm at carbon ignition and a radius of
$4.8 \times 10^{12}$ cm when the central carbon mass fraction reached
1\%.

\subsection{$M_{\rm He,i}$ = 3.2 - 60 \Msun; ``Normal'' Evolution}
\lSect{normal}

Models with initial masses above 3.2 ignite all phases of nuclear
burning, including silicon burning, at their centers
(\Tab{siign}). Their radii also remain sufficiently small
that a second stage of binary mass transfer is probably
avoided.

The age, mass, luminosity, and effective temperature of the helium
stars on the (helium) main sequence, when convective burning has
reduced the central helium mass fraction to 0.50, are given in
\Tab{models} and \Fig{luminosity}. The luminosity as a function of
mass is given, approximately, for the entire distribution by $L/\Lsun
\approx 2.81 \times 10^3 M^{1.71}$ (Red curve), where $M$ is the {\sl
  current} mass in solar masses. A better fit can be obtained by
breaking the fitting interval into two mass ranges: 3 to 15 \Msun
\ (green line in \Fig{luminosity}, and above 15 \Msun \ (blue line)
\begin{equation}
\begin{split}
L/\Lsun & \approx 1.16 \times 10^3 M^{2.13}
 \ \ \ \ \ (3 \Msun \le M \le 15 \Msun) \cr
        & \approx 4.92 \times 10^4 \, M -4.57 \times 10^5 
 \ \ \ \ \ (M > 15 \Msun)
\end{split}
\lEq{lofm}
\end{equation}
This second equation has the correct form for extrapolating to masses
beyond the fit interval because the luminosity approaches the Eddington
value at high masses and is thus proportional to M.

Similarly, the lifetimes in \Tab{models} can be approximated, but are
not plotted.
\begin{equation}
\begin{split}
t_f/10^5 \ {\rm y} & \approx 3.0 + 75.3 M_{\rm He,i}^{-1.24}
 \ \ \ \ \ (3 \Msun \le M \le 75 \Msun) \cr
        & \approx 3.0 
 \ \ \ \ \ (M > 75 \Msun)
\end{split}
\lEq{tf}
\end{equation}
where again, at high masses the lifetime approaches the correct
(constant) Eddington value. Here $M_{\rm He,i}$ is the {\sl initial}
mass of the helium star.

A fit for the final mass as a function of starting helium core mass
(see second panel of \Fig{cores}) that is good to better than a few
percent for initial helium core masses up to 10 \Msun \ is
\begin{equation}
M_{\rm fin} \ \approx \ 0.939 \ M_{He,i}^{0.872},
\lEq{mfinhe}
\end{equation}
where all masses are in solar masses. For heavier helium cores,
\begin{equation}
M_{\rm fin} \ \approx \ 0.463 \ M_{He,i} \, + 1.49.
\end{equation}
These equations can be combined with \Eq{mzams} and \Eq{mzamsh} to
give
\begin{equation}
M_{\rm fin} \ \approx \ 0.0548 \ M_{\rm ZAMS}^{1.40}.
\lEq{mfinzams}
\end{equation}
for main sequence masses below 30 \Msun and 
\begin{equation}
M_{\rm fin} \ \approx \ 0.232 \ M_{\rm ZAMS} \, - 1.23.
\end{equation}
 for main sequence masses above 30 \Msun. For very high masses a good
 rule of thumb is that the final mass is one-fourth of the ZAMS mass.
 This is for the standard mass loss rate and would need to be
 modified for other values.

\begin{figure}
\includegraphics[width=0.48\textwidth]{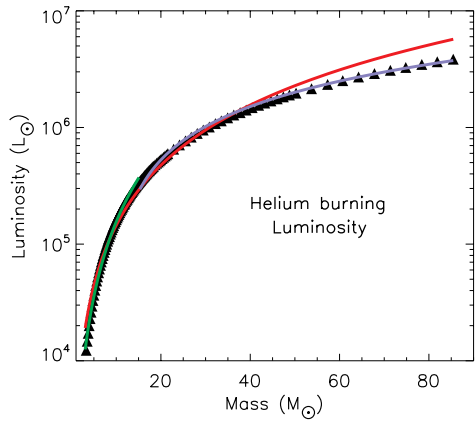}
\caption{The luminosity in solar units half way through helium
  burning as a function of the current mass of the core. The red solid
  line is a power law fit to the entire data set and the green and blue
  lines fits when the data is broken into two sets above and below 15
  \Msun \ (see text). \lFig{luminosity}}
\end{figure}


\begin{deluxetable*}{ccccccccccccc} 
\tablecaption{Properties of the model sets} 
\tablehead{ \colhead{${M_{\rm init}}$}           & 
            \colhead{${t_{\rm He/2}}$}           & 
            \colhead{${M_{\rm He/2}}$}           &
            \colhead{${\log \, L_{\rm He/2}}$}    & 
            \colhead{${\log \, T_{\rm eff,He/2}}$} & 
            \colhead{${t_{\rm fin}}$}            & 
            \colhead{${M_{\rm fin}}$}            &
            \colhead{${\log \, L_{\rm C/O \ ign}}$}   & 
            \colhead{${\log \, T_{\rm eff, CO-ign}}$} & 
            \colhead{${M_{\rm CO}}$}             &
            \colhead{${m_{\rm He}}$}             & 
            \colhead{${Y_{\rm s}}$} 
            \\
            \colhead{[\Msun]}                  &  
            \colhead{[10$^5$ y]}               & 
            \colhead{[\Msun]}                  &
            \colhead{[\Lsun]}                  & 
            \colhead{[K]}                      &
            \colhead{[10$^5$ y]}               & 
            \colhead{[\Msun]}                  &
            \colhead{[\Lsun]}                  & 
            \colhead{[K]}                      &
            \colhead{[\Msun]}                  &
            \colhead{[\Msun]}                  &
            \colhead{} 
            }\\
\startdata
\multicolumn{12}{c}{Nominal mass loss}\\
\multicolumn{12}{c}{}\\
\multicolumn{12}{l}{At helium burning and carbon ignition}\\
\multicolumn{12}{l}{}\\
1.8 & 19.66 & 1.73 & 3.27 & 4.82 & 57.03 & 1.53 & 4.37 & 3.82 & 1.03 & 0.49 & 0.99 \\
1.9 & 17.79 & 1.82 & 3.35 & 4.83 & 51.32 & 1.61 & 4.45 & 3.82 & 1.06 & 0.55 & 0.99 \\
2.0 & 16.04 & 1.92 & 3.41 & 4.84 & 46.59 & 1.70 & 4.35 & 3.83 & 1.08 & 0.59 & 0.99 \\
2.1 & 14.68 & 2.01 & 3.48 & 4.85 & 42.53 & 1.78 & 4.31 & 4.31 & 1.09 & 0.63 & 0.99 \\
2.2 & 13.56 & 2.10 & 3.53 & 4.86 & 39.10 & 1.86 & 4.31 & 4.59 & 1.10 & 0.67 & 0.99 \\
2.3 & 12.50 & 2.20 & 3.59 & 4.87 & 36.19 & 1.94 & 4.33 & 4.67 & 1.14 & 0.70 & 0.99 \\
2.4 & 11.76 & 2.29 & 3.64 & 4.87 & 33.63 & 2.02 & 4.35 & 4.71 & 1.18 & 0.73 & 0.99 \\
2.5 & 10.86 & 2.38 & 3.69 & 4.88 & 31.43 & 2.10 & 4.37 & 4.73 & 1.21 & 0.76 & 0.99 \\
\multicolumn{12}{c}{}\\
\multicolumn{12}{l}{At helium burning and oxygen ignition}\\
\multicolumn{12}{l}{}\\
      2.5 & 10.86 & 2.38 & 3.69 & 4.88 & 31.73 &  2.07 & 4.37 & 3.85 & 1.37 & 0.67 & 0.99 \\
      2.6 & 10.28 & 2.47 & 3.74 & 4.89 & 29.76 &  2.15 & 4.47 & 3.86 & 1.41 & 0.70 & 0.99 \\
      2.7 &  9.69 & 2.56 & 3.79 & 4.89 & 28.02 &  2.22 & 4.51 & 3.87 & 1.46 & 0.72 & 0.99 \\
      2.8 &  9.08 & 2.65 & 3.83 & 4.90 & 26.48 &  2.30 & 4.55 & 3.89 & 1.51 & 0.75 & 0.99 \\
      2.9 &  8.64 & 2.74 & 3.87 & 4.91 & 25.10 &  2.37 & 4.59 & 3.92 & 1.56 & 0.77 & 0.99 \\
      3.0 &  8.21 & 2.83 & 3.91 & 4.91 & 23.88 &  2.45 & 4.61 & 3.99 & 1.61 & 0.78 & 0.99 \\
      3.2 &  7.54 & 3.01 & 3.99 & 4.92 & 21.74 &  2.59 & 4.65 & 4.34 & 1.70 & 0.82 & 0.99 \\
      3.5 &  6.65 & 3.28 & 4.09 & 4.93 & 19.18 &  2.81 & 4.69 & 4.55 & 1.85 & 0.86 & 0.99 \\
      4.0 &  5.60 & 3.72 & 4.24 & 4.95 & 16.10 &  3.15 & 4.78 & 4.69 & 2.12 & 0.91 & 0.99 \\
     4.5 &  4.89 &  4.16 &  4.36 &  4.97 & 13.97 &  3.49 &  4.85 &  4.75 &  2.38 &  0.93 &  0.99 \\
     5.0 &  4.75 &  4.55 &  4.47 &  4.98 & 12.37 &  3.82 &  4.91 &  4.86 &  2.65 &  0.94 &  0.99 \\
     6.0 &  3.93 &  5.39 &  4.64 &  5.01 & 10.24 &  4.45 &  5.02 &  4.95 &  3.15 &  0.93 &  0.99 \\
     7.0 &  3.40 &  6.22 &  4.78 &  5.02 &  8.87 &  5.05 &  5.10 &  4.99 &  3.63 &  0.88 &  0.99 \\
     8.0 &  3.05 &  7.03 &  4.90 &  5.04 &  7.91 &  5.64 &  5.17 &  5.00 &  4.11 &  0.82 &  0.99 \\
     9.0 &  2.79 &  7.83 &  5.00 &  5.05 &  7.22 &  6.20 &  5.21 &  5.01 &  4.59 &  0.73 &  0.99 \\
    10.0 &  2.58 &  8.61 &  5.08 &  5.06 &  6.70 &  6.75 &  5.27 &  5.02 &  5.07 &  0.61 &  0.99 \\
    11.0 &  2.43 &  9.39 &  5.16 &  5.05 &  6.30 &  7.05 &  5.27 &  5.24 &  5.45 &  0.35 &  0.64 \\
    12.0 &  2.30 & 10.17 &  5.22 &  5.05 &  5.99 &  7.27 &  5.28 &  5.22 &  5.55 &  0.20 &  0.38 \\
    14.0 &  2.10 & 11.70 &  5.34 &  5.06 &  5.52 &  8.04 &  5.36 &  5.24 &  6.11 &  0.21 &  0.19 \\
    16.0 &  1.95 & 13.20 &  5.43 &  5.06 &  5.18 &  8.84 &  5.39 &  5.25 &  6.79 &  0.24 &  0.22 \\
    18.0 &  1.84 & 14.69 &  5.51 &  5.07 &  4.92 &  9.62 &  5.42 &  5.25 &  7.42 &  0.25 &  0.23 \\
    20.0 &  1.70 & 16.23 &  5.58 &  5.12 &  4.72 & 10.39 &  5.47 &  5.30 &  8.01 &  0.26 &  0.25 \\
    25.0 &  1.59 & 19.68 &  5.71 &  5.12 &  4.35 & 12.53 &  5.61 &  5.34 &  9.76 &  0.25 &  0.22 \\
    30.0 &  1.49 & 22.85 &  5.81 &  5.12 &  4.08 & 14.81 &  5.72 &  5.37 & 11.63 &  0.25 &  0.21 \\
    40.0 &  1.37 & 29.48 &  5.98 &  5.13 &  3.74 & 19.55 &  5.89 &  5.40 & 15.67 &  0.27 &  0.19 \\
    50.0 &  1.29 & 36.32 &  6.11 &  5.13 &  3.52 & 24.48 &  6.02 &  5.42 & 20.00 &  0.32 &  0.18 \\
    60.0 &  1.24 & 43.25 &  6.21 &  5.14 &  3.36 & 29.53 &  6.13 &  5.42 & 24.31 &  0.38 &  0.17 \\
    70.0 &  1.20 & 50.26 &  6.30 &  5.14 &  3.25 & 34.66 &  6.21 &  5.42 & 28.81 &  0.46 &  0.16 \\
    80.0 &  1.17 & 57.32 &  6.37 &  5.14 &  3.16 & 39.83 &  6.29 &  5.40 & 33.38 &  0.53 &  0.16 \\
    90.0 &  1.15 & 64.25 &  6.43 &  5.14 &  3.09 & 45.04 &  6.35 &  5.34 & 38.04 &  0.60 &  0.16 \\
   100.0 &  1.13 & 71.37 &  6.49 &  5.14 &  3.04 & 50.27 &  6.40 &  5.29 & 42.70 &  0.66 &  0.16 \\
   110.0 &  1.12 & 78.41 &  6.54 &  5.15 &  2.99 & 55.51 &  6.45 &  5.22 & 47.45 &  0.71 &  0.16 \\
   120.0 &  1.11 & 85.45 &  6.59 &  5.15 &  2.95 & 60.78 &  6.49 &  5.16 & 52.05 &  0.76 &  0.16 \\
\multicolumn{12}{c}{}\\
\multicolumn{12}{c}{1.5*Nominal mass loss}\\
\multicolumn{12}{c}{}\\
     3.5 &  6.89 &  3.18 &  4.05 &  4.93 &  19.60 & 2.57 &  4.65 &  4.69 &  1.73 &  0.77 &  0.99 \\
     4.0 &  5.80 &  3.59 &  4.20 &  4.95 &  16.45 & 2.87 &  4.73 &  4.80 &  1.95 &  0.82 &  0.99 \\
     4.5 &  5.08 &  4.00 &  4.32 &  4.97 &  14.26 & 3.16 &  4.79 &  4.86 &  2.16 &  0.84 &  0.99 \\
     5.0 &  4.44 &  4.41 &  4.43 &  4.98 & 12.66 &  3.43 &  4.85 &  4.89 &  2.37 &  0.83 &  0.99 \\
     6.0 &  3.79 &  5.17 &  4.60 &  5.00 & 10.48 &  3.96 &  4.95 &  4.98 &  2.77 &  0.80 &  0.99 \\
     7.0 &  3.35 &  5.92 &  4.73 &  5.02 &  9.09 &  4.45 &  5.03 &  5.05 &  3.17 &  0.72 &  0.99 \\
     8.0 &  2.94 &  6.69 &  4.85 &  5.04 &  8.15 &  4.92 &  5.09 &  5.09 &  3.56 &  0.60 &  0.98 \\
     9.0 &  2.72 &  7.40 &  4.94 &  5.05 &  7.50 &  4.87 &  5.07 &  5.05 &  3.76 &  0.22 &  0.49 \\
    10.0 &  2.52 &  8.13 &  5.03 &  5.06 &  7.05 &  4.96 &  5.08 &  5.06 &  3.73 &  0.13 &  0.22 \\
    12.0 &  2.28 &  9.49 &  5.16 &  5.07 &  6.40 &  5.43 &  5.12 &  5.07 &  4.04 &  0.18 &  0.21 \\
    14.0 &  2.08 & 10.86 &  5.27 &  5.09 &  5.97 &  5.86 &  5.18 &  5.15 &  4.40 &  0.19 &  0.22 \\
    16.0 &  1.96 & 12.14 &  5.36 &  5.10 &  5.64 &  6.34 &  5.20 &  5.15 &  4.79 &  0.18 &  0.21 \\
    18.0 &  1.86 & 13.43 &  5.44 &  5.11 &  5.38 &  6.84 &  5.26 &  5.16 &  5.19 &  0.18 &  0.20 \\
    20.0 &  1.80 & 14.36 &  5.49 &  5.11 &  5.16 &  7.39 &  5.29 &  5.17 &  5.62 &  0.18 &  0.19 \\
    25.0 &  1.65 & 16.82 &  5.60 &  5.12 &  4.74 &  8.85 &  5.38 &  5.18 &  6.78 &  0.18 &  0.18 \\
    30.0 &  1.55 & 19.45 &  5.70 &  5.12 &  4.43 & 10.41 &  5.48 &  5.18 &  8.01 &  0.18 &  0.17 \\
    35.0 &  1.48 & 22.21 &  5.79 &  5.13 &  4.20 & 12.04 &  5.58 &  5.18 &  9.34 &  0.18 &  0.16 
\enddata
\tablecomments{
Ages, masses, luminosities, and effective temperatures when helium has
half burned (``He/2'') and when carbon (first 8 lines) or oxygen (the
rest of the table) ignites. The oxygen ignition values are also
essentially the presupernova star properties. $M_{CO}$ is the carbon
oxygen core mass; $M_{\rm He}$, the remaining mass of helium in the
star; and $Y_{\rm s}$ the surface helium mass fraction.}
\lTab{models}
\end{deluxetable*}

\subsection{$M_{\rm He,i}$ = 60 - 120 \Msun; Pulsational Pair-Instability 
Supernovae and Black Holes}
\lSect{ppisn}

For the standard choice of mass loss rates (\eq{mdot}), the pair
instability is first encountered for the 60 \Msun \ model. At the time
of the instability (carbon depletion) the core has shrunk to 29.53
\Msun \ and consists mostly of oxygen and heavy elements. Helium has a
small surface abundance, 0.16 by mass fraction, and the total mass of
the star is small (\Tab{models}). As with models having similar core
mass in the study of \citet{Woo17} (see his Table 1), the instability
in these lighter cores is a weak one, occurring at late times in the
oxygen burning shell, and resulting in the low energy ejection of only
a few hundredths of a solar mass. The onset of the instability occurs
at smaller masses here than in the earlier study because of the larger
effective CO-core mass (\Fig{cocore}). As \citet{Woo17} pointed out,
the strength of the pair instability is most sensitive to the CO-core
mass.

\begin{figure}
\includegraphics[width=0.48\textwidth]{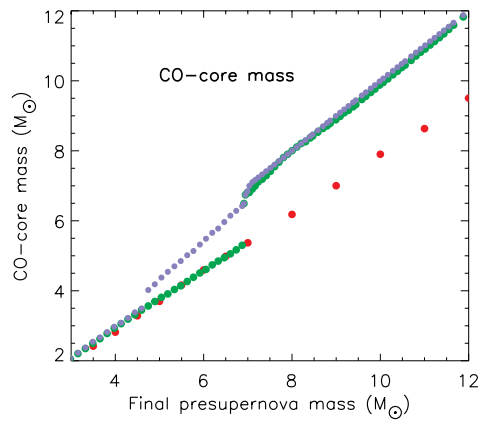}
\caption{CO core masses in presupernova helium stars evolved with and
  without mass loss. The CO-core mass is plotted against the final
  mass, which is the same as the original helium core mass in the case
  of no mass loss (red points), but equals the presupernova helium
  core mass after mass loss in the other models (green and blue
  points). For the green points, CO-core masses are evaluated where
  the helium mass fraction, moving in from the surface, declines below
  0.2. This is usually the location of the outer boundary of the
  helium burning shell. The blue points use a fiducial helium mass
  fraction of 0.5 which helps separate regions that have had
  substantial helium shell burning. Below 4.7 \Msun \ and above 7.0
  \Msun \ the green and blue points are almost identical. For final
  masses above 7.0 \Msun \ the presupernova mass essentially equals
  the CO-core mass. Green and red points agree below 7 \Msun \ which
  is where mass loss uncovers the CO core
  (\Tab{models}). \lFig{cocore}}
\end{figure}

The instability continues to be weak and brief for helium stars up to
about 75 \Msun \ (final mass 37 \Msun). The instability happens at
late times in the oxygen burning shell after a silicon core has
already formed in hydrostatic equilibrium and the total energy of the
ejecta is less than 10$^{50}$ erg.  These low energy, small ejecta
mass, and small radii models will not make luminous supernovae, but if
the core collapses uneventfully to a black hole, the small amount of
mass ejected may be the only optical display of their death. Their
transients will be brief, faint, and very blue (see \citet{Woo17} and
\Sect{lite}). If, on the other hand, collapse leads to an explosive
event, these small masses at radii 10$^{12}$ to 10$^{13}$ cm may
enable a bright collisionally-powered supernova. The light curves for
the heavier more energetic models would be similar to those
already published by \citet{Woo17}.

\begin{deluxetable}{ccccccc} 
\tablecaption{Pulsational Pair-Instability Supernovae} 
\tablehead{ \colhead{${M_{\rm init}}$}       & 
            \colhead{${M_{\rm f}}$}         &
            \colhead{${\tau}$}            &
            \colhead{E}                   &
            \colhead{${M_{ej}}$}           & 
            \colhead{${M_{\rm rem}}$}       &        
            \colhead{${M_{\rm Fe}}$}        
            \\
            \colhead{[\Msun]}              &  
            \colhead{[\Msun]}              &
            \colhead{[sec]}                &
            \colhead{[10$^{50}$ erg ]}      & 
            \colhead{[\Msun]}              &
            \colhead{[\Msun]}              &
            \colhead{[\Msun]}               
            }
\startdata
60  &  29.53  &  6.4(4)  &  0.015 &   0.02 & 29.51 & 2.32 \\
62  &  30.56  &  6.1(4)  &  0.020 &   0.05 & 30.51 & 2.36 \\
64  &  31.57  &  5.8(4)  &  0.022 &   0.07 & 31.50 & 2.49 \\
66  &  32.60  &  5.7(4)  &  0.030 &   0.10 & 32.50 & 2.55 \\    
68  &  33.63  &  5.8(4)  &  0.038 &   0.18 & 33.45 & 2.67 \\
70  &  34.66  &  6.7(4)  &  0.060 &   0.43 & 34.23 & 2.84 \\
75  &  36.83  &  8.0(4)  &  0.36  &   0.66 & 36.17 & 2.95 \\
80  &  39.38  &  4.1(5)  &    1.5 &   1.95 & 37.43 & 3.22  \\
85  &  41.95  &  9.4(5)  &    4.6 &   3.78 & 38.17 & 2.99   \\
90  &  44.54  &  2.3(6)  &    5.6 &   5.78 & 38.76 & 2.62   \\
95  &  47.13  &  1.1(7)  &    5.8 &   6.37 & 40.76 & 2.52   \\
100 &  49.75  &  1.1(8)  &    5.7 &   7.15 & 42.60 & 2.73   \\
100* &  50.22  &  8.6(7)  &   4.1 &   7.29 & 42.93 & 2.22   \\
105 &  52.24  &  9.1(9)  &    5.2 &   7.21 & 45.03 & 1.73  \\
105* &  52.82  &  7.6(9)  &   3.1 &   6.83 & 45.99 & 2.04  \\
110 &  54.79  &  5.8(10) &    5.2 &  10.57 & 44.22 & 2.58  \\
110* &  55.43  &  7.3(10) &   5.9 &  10.63 & 44.80 & 2.14  \\
115 &  57.42  &  1.5(11) &   14.8 &  16.07 & 41.35 & 2.63  \\
120 &  60.12  &  1.2(12) &   35.6 &  56.61 & 3.51  & 1.76   \\
\enddata

\tablecomments{$\tau$ is the time between the onset of the first pulse
  and core collapse. $E_{50}$ is the total kinetic energy of all mass
  ejected in units of 10$^{50}$ erg. Models 105*, 110*, and 115* were
  run using large networks coupled directly to the burning,  but
    are otherwise like Models 100, 105, and 110. These and all other
  unstarred models used the approximation network and
  quasi-equilibrium assumption.}  \lTab{ppisntab}
\end{deluxetable}

Moving on up in mass, one encounters PPISN of increasing duration and
energy that resemble the helium stars of constant mass studied
by \citet{Woo17}. The final mass is offset by about 5\% due to the
larger CO-core. For example, \citet{Woo17} found the maximum helium
core mass that made a PPISN rather than experiencing complete
disruption in a single pulse (PISN) was about 63 \Msun. Here the
equivalent value is 60 \Msun \ (\Tab{ppisntab}). The maximum remnant
mass, 46 \Msun, is also lighter. This would be the best current
estimate for the onset of the black hole mass gap that might be
detected by gravitational radiation experiments. It is considerably
smaller than the 52 \Msun \ value given by \citet{Woo17} and 50 \Msun
\ given by \citet{Bel16}. The larger values were for single
stars. For helium stars, there is no hydrogen envelope to tamp the
explosion and cause more material to fall back. The CO core is also
larger in the new models.

\begin{figure}
\includegraphics[width=0.48\textwidth]{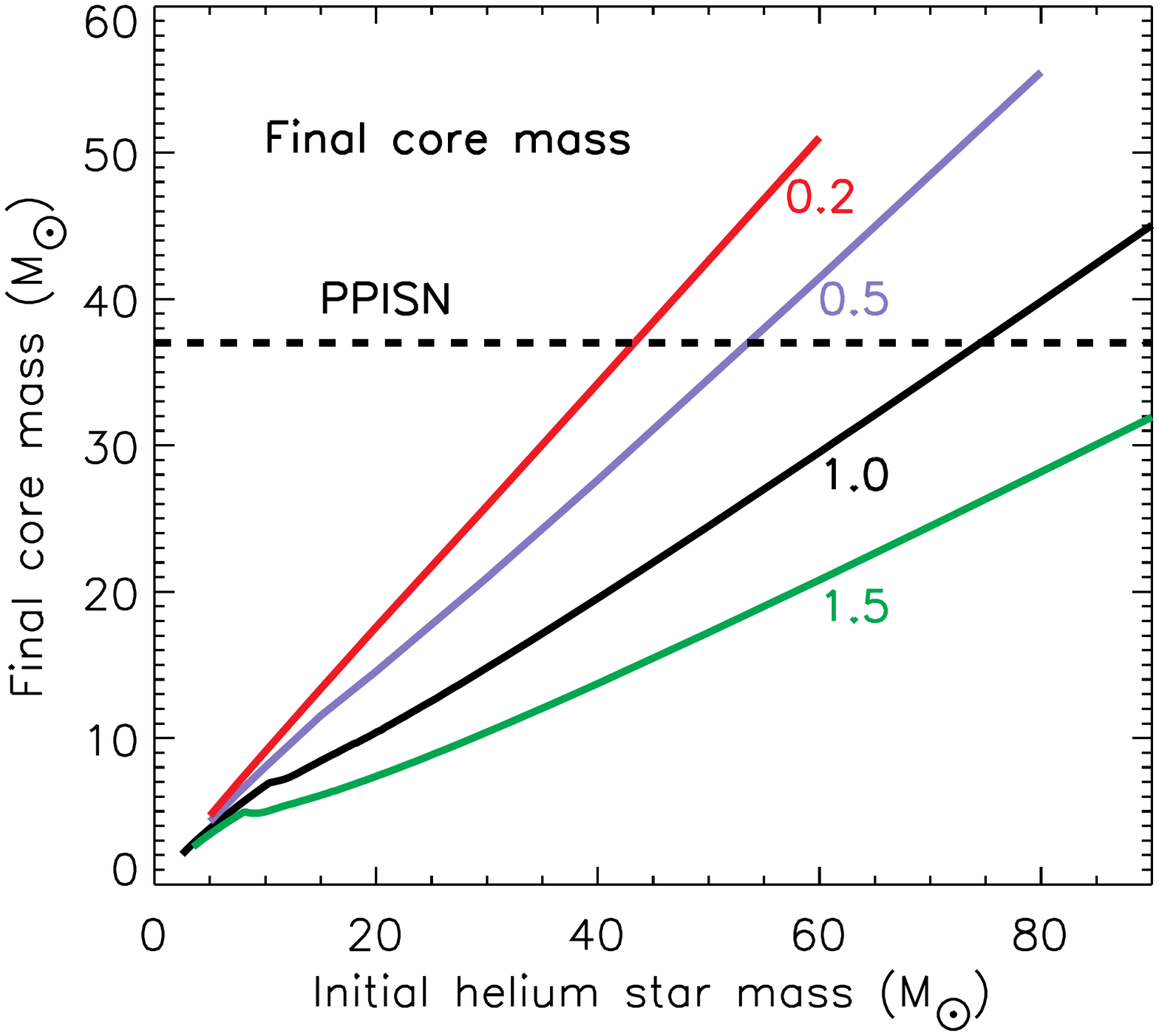}
\includegraphics[width=0.48\textwidth]{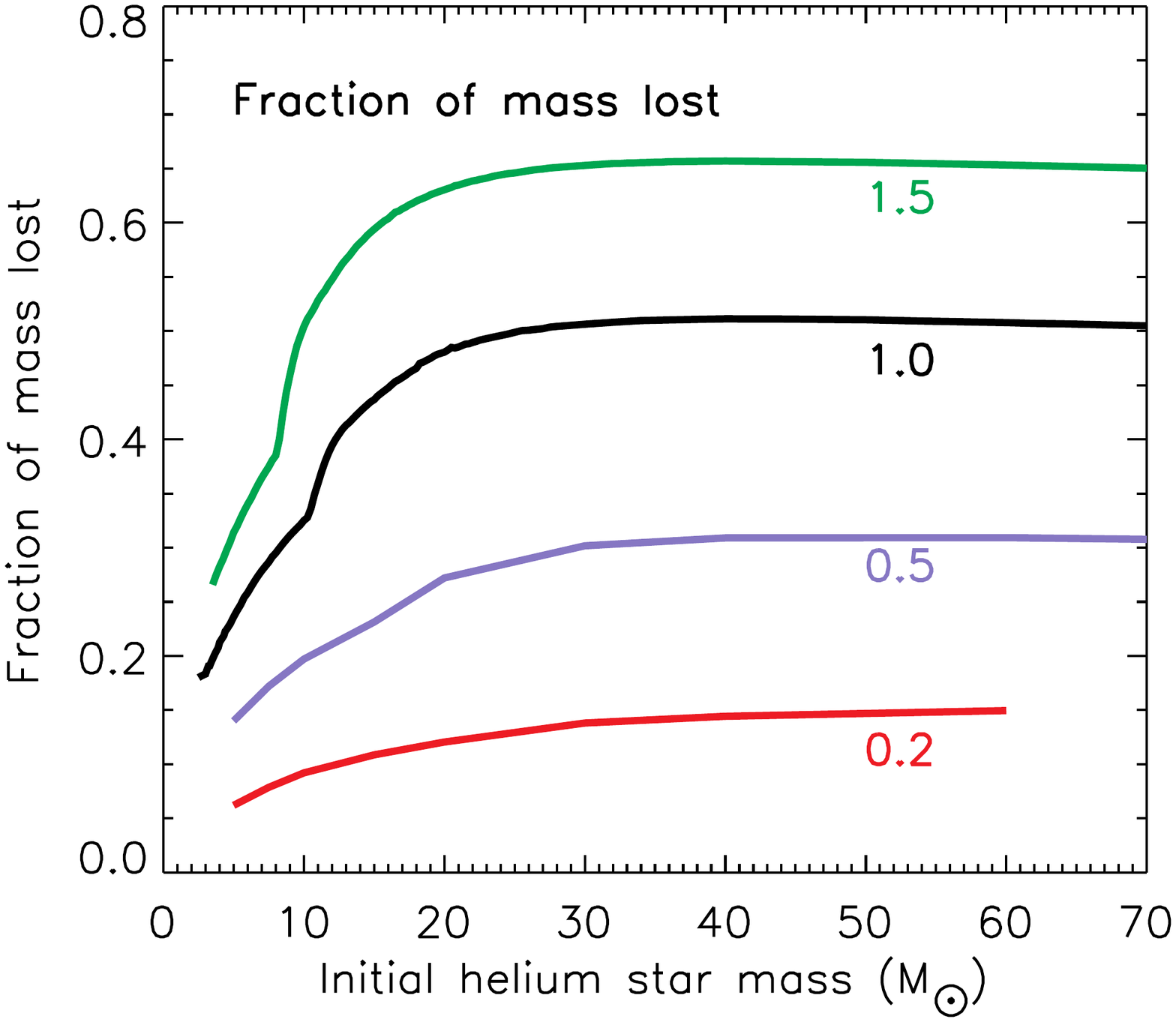}
\caption{(Top:)Final helium and CO-core masses as a function of
  initial helium core mass for several choices of mass loss rate. The
  number on each plot is the factor by which the overall mass loss
  rate throughout the evolution was multiplied relative to
  \eq{mdot}. Final cores over 37 \Msun \ encounter a strong
  pulsational pair-instability. (bottom:) The fraction of total mass
  lost during the entire evolution is shown for the same choices of
  mass loss rates.  \lFig{mloss}}
\end{figure}

For the standard mass loss rate, the strong pulsational pair
instability is first encountered for a final core mass of 37 \Msun,
(\Tab{ppisntab}) which corresponds to an initial helium core mass near
75 \Msun. Using \eq{mzamsh}, this corresponds approximately to a star
whose main sequence mass was 160 \Msun. Reductions in the mass loss
rate result in the instability being encountered at lower masses. The
reduction may either be because the standard rate assumed for solar
metallicity stars here is too large \citep{Vin17}, or because of
reduced metallicity. Though metallicity scaling is given in
\eq{mdotwr} and \eq{mdotwne}, the exponent is uncertain \citep{Eld06}.

An additional 18 models were calculated with initial masses between 5
and 80 \Msun. Half of these used a mass loss rate 50\% of standard and
half used 20\%. The results from this sparse grid are displayed along
with the more densely sampled grids with normal and 1.5 times normal
mass loss in \Fig{mloss}. Reducing the mass loss by a factor of two
results in a strong pulsational pair-instability (final mass over 37
\Msun; \Tab{ppisntab}) being encountered for an initial helium core mass
of only 52 \Msun.  Reducing the standard mass loss by
only a factor of two would allow a 110 \Msun \ star, either single or
in a binary system, to become a PPISN.

The fraction of mass lost for high mass helium stars where the
luminosity is almost linear in the mass and the lifetime approaches $3
\times 10^5$ years (i.e., the Eddington limit) is
\begin{equation}
\frac{\Delta M}{M_{\rm init}} \ \approx \ (1 \, - \, \exp (-0.7 F)
\end{equation}
where F is the factor by which the overall mass loss is
multiplied. For example, at F = 1 the star loses half
its mass. 

\section{Presupernova Properties}
\lSect{presn}

\subsection{Surface Composition}
\lSect{surfacey}

Winds remove the surface layers of the star ultimately revealing what
lies beneath. Initially that material consists of almost pure helium,
but as mass loss reveals the products of helium burning, a variable
amount of carbon and oxygen appear. None of the stars here were so
extreme as to lose all their helium. The mass loss occurs mostly
during helium core burning, and the outer edge of the convective
core recedes as mass is lost, leaving behind a gradient of helium.
\Tab{models} and \Fig{he} show that the surface consists of helium and
nitrogen until some critical mass is reached, about 11 \Msun \ (final
mass 7 \Msun) for the nominal mass loss case and 9 (final mass 4.9
\Msun) for a loss rate that is 1.5 times higher.  \Fig{he} here is
very similar to Figure 7 of \citet{Yoo17}.

As \citet{Yoo17} points out, the higher mass limit is inconsistent
with the observed luminosity of the faintest WC/WO stars and this
argues for mass loss rates greater than the standard value used here.
On the other hand, \citet{Yoo17} and \citet{Mcc16}, argue that the
temperatures of observed Wolf-Rayet stars are cooler than the models
unless a low mass loss rate is used. \citet{Vin17} also predicts mass
loss rates that are substantially smaller even than the standard one
used here. He gives
\begin{equation}
\log \dot M_{\rm Vink} \ = \ −13.3 \ + \ 1.36 \, \log
\left(\frac{L}{\Lsun}\right) + 0.61 \log \left(\frac{Z}{\Zsun}\right).
\end{equation}
For solar metallicity and a 6 \Msun \ star, for example, with log
L/\Lsun = 4.64 (\Tab{models}), this expression gives 10$^{-7}$ \Msun
\ y$^{-1}$, whereas \Eq{mdotwne} gives a value 10 times larger.

In this paper, most models use the standard value of \citet{Yoo17}. As
we shall see (\Sect{sn1bc}), however, a larger value may also be
needed to produce Type Ic supernovae if their progenitors must have
mass total less than 6 \Msun, but lose their helium-rich layers before
exploding.  There too is an issue of whether mixing or mass loss
is the essential ingredient in making a Type Ic.

\begin{figure}
\includegraphics[width=0.48\textwidth]{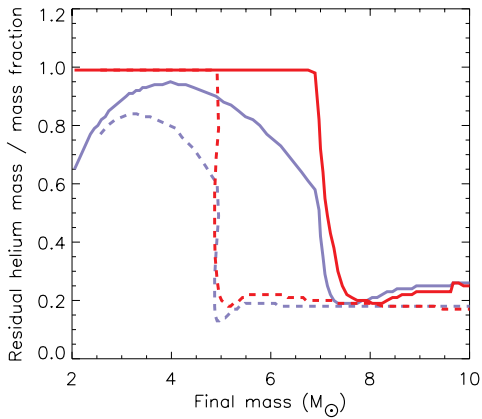}
\caption{Total remaining helium mass in solar masses (blue) and surface
  helium mass fraction (red), both in the presupernova
  star. Quantities are plotted as a function of the presupernova mass
  for normal mass loss (solid lines) and 1.5 times the nominal rate.
  \citep[See also][]{Yoo17}. \lFig{he}}
\end{figure}

\begin{figure*}
\includegraphics[width=0.48\textwidth]{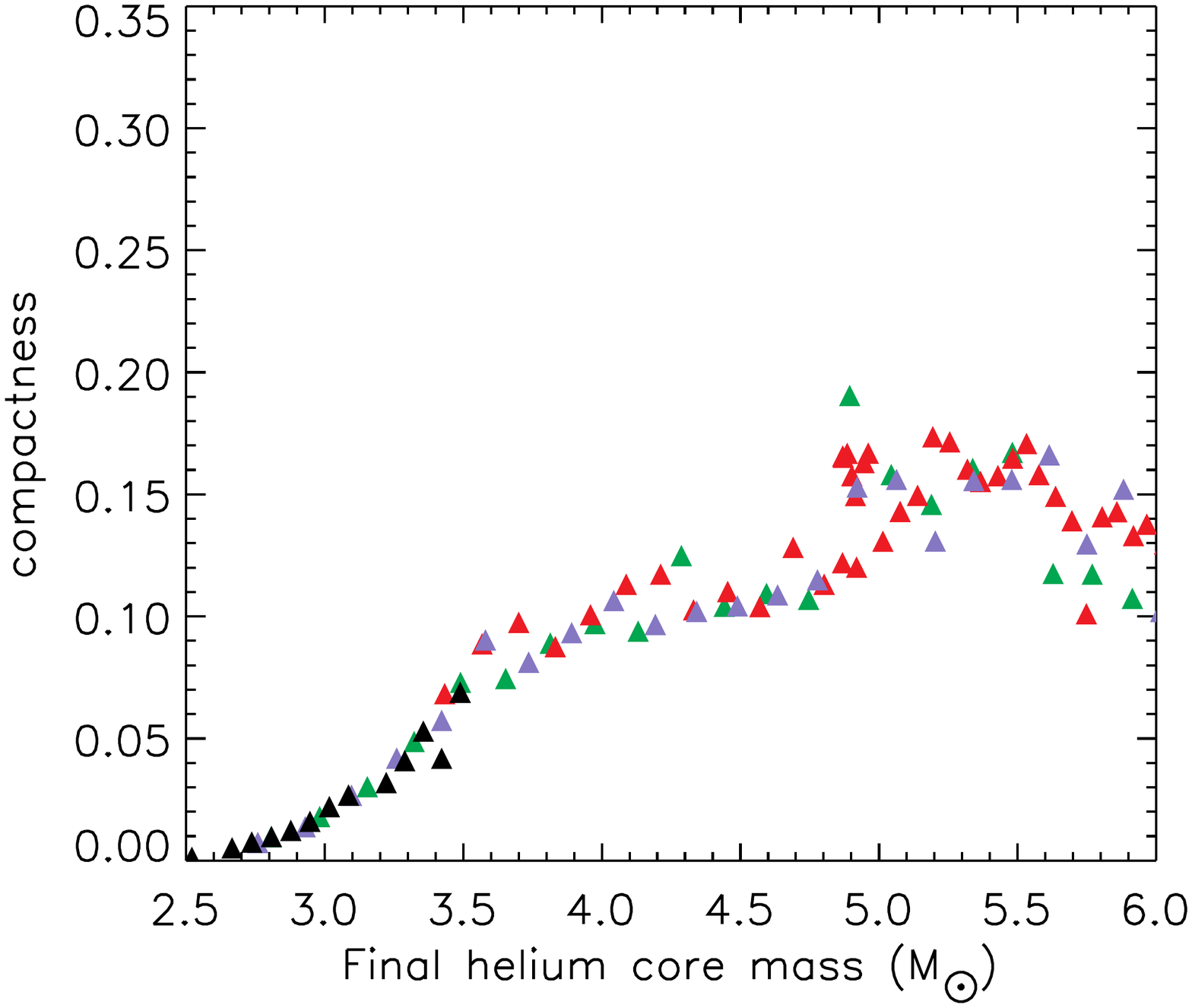}
\includegraphics[width=0.48\textwidth]{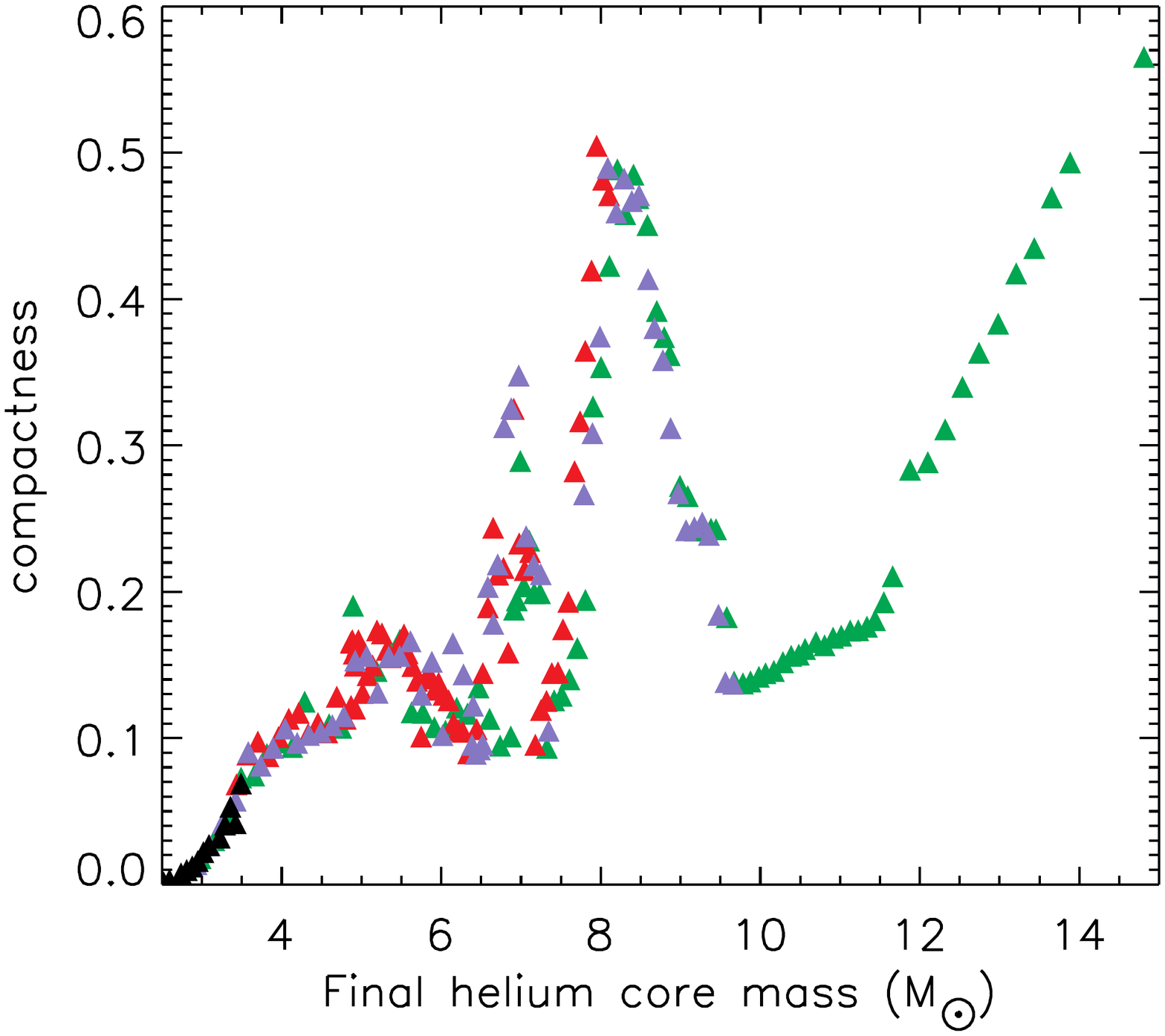}
\includegraphics[width=0.48\textwidth]{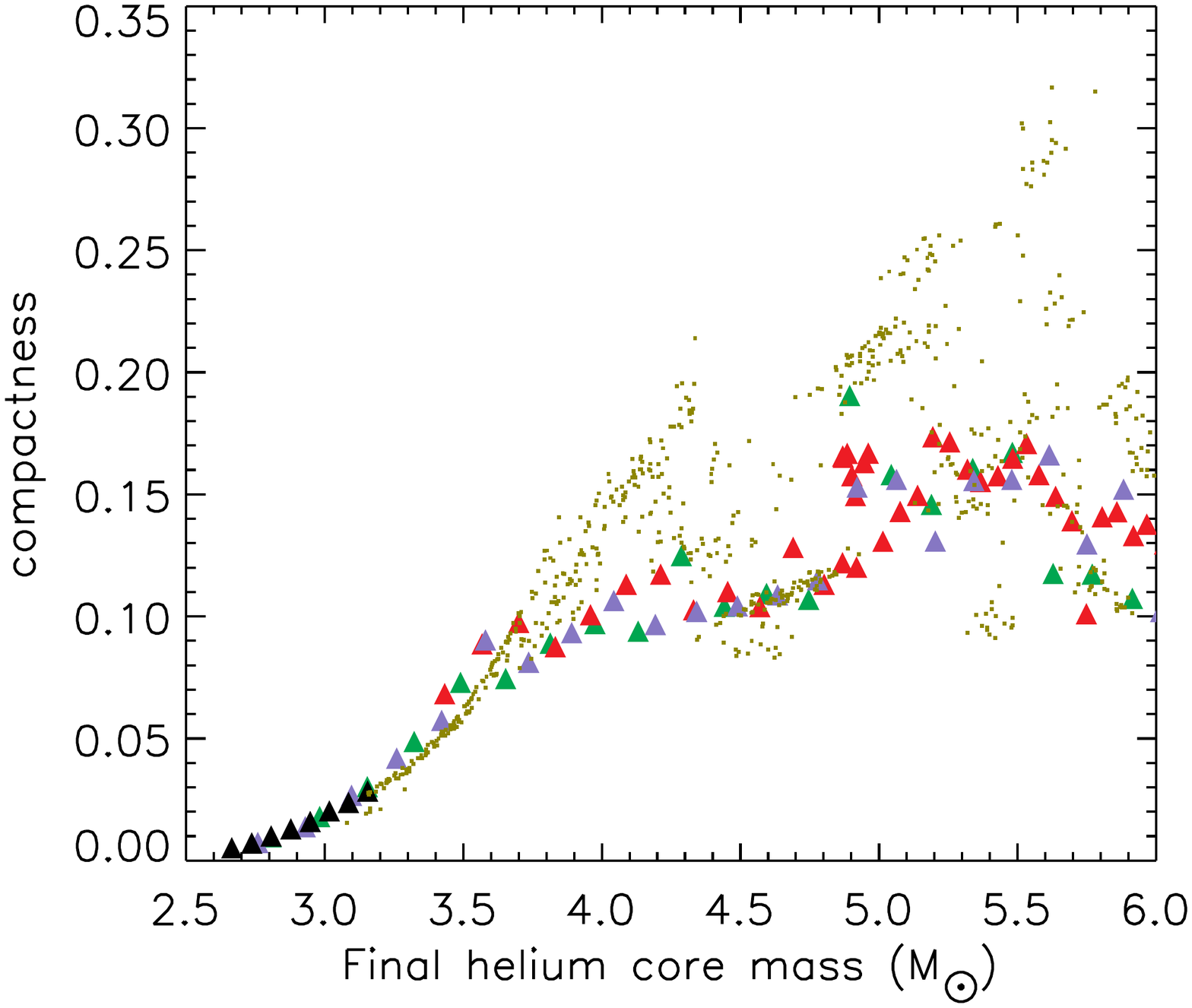}
\includegraphics[width=0.48\textwidth]{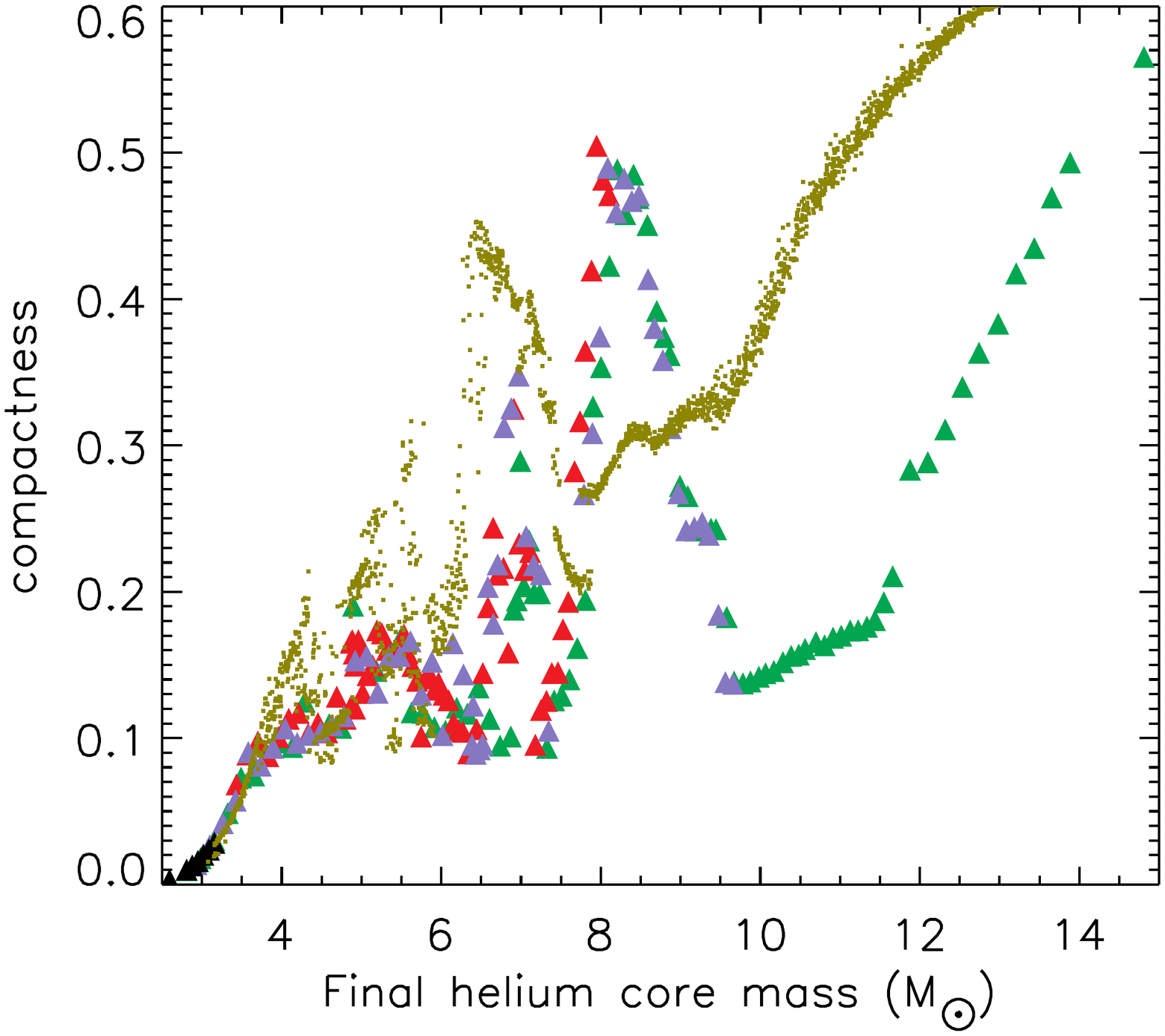}
\caption{Presupernova compactness as defined by \citet{Oco11} as a
  function of final core mass. (Top two panels:) The different color
  points indicate variations in the stellar physics and mass loss rate
  used in the models. Green is for the new models with the fiducial
  mass loss rate (\Tab{models}) and standard nuclear physics. Blue
  points used a reduced surface boundary pressure.  Black points used
  a large nuclear reaction network and finer zoning. Red points used
  1.5 times the fiducial mass loss rate. Note the smooth variation
  below 6 \Msun \ and between 10 and 12 \Msun \ and the large values
  between 7 and 9 \Msun \ and above 11 \Msun. (Bottom panels:) The new
  results are compared to the $\dot M/2$ case of \citet{Suk18} (gold
  points), see their Fig. 8. Note the large number of high compactness
  parameters below 6 \Msun. The peak in compactness for the single
  stars was at 6.5 \Msun. For the new set approximating binary
  evolution it is at 8 \Msun. The minimum above the first peak now has
  smaller compactness and is shifted to high presupernova masses.
  \lFig{compactfig}}
\end{figure*}

\begin{figure}
\includegraphics[width=0.48\textwidth]{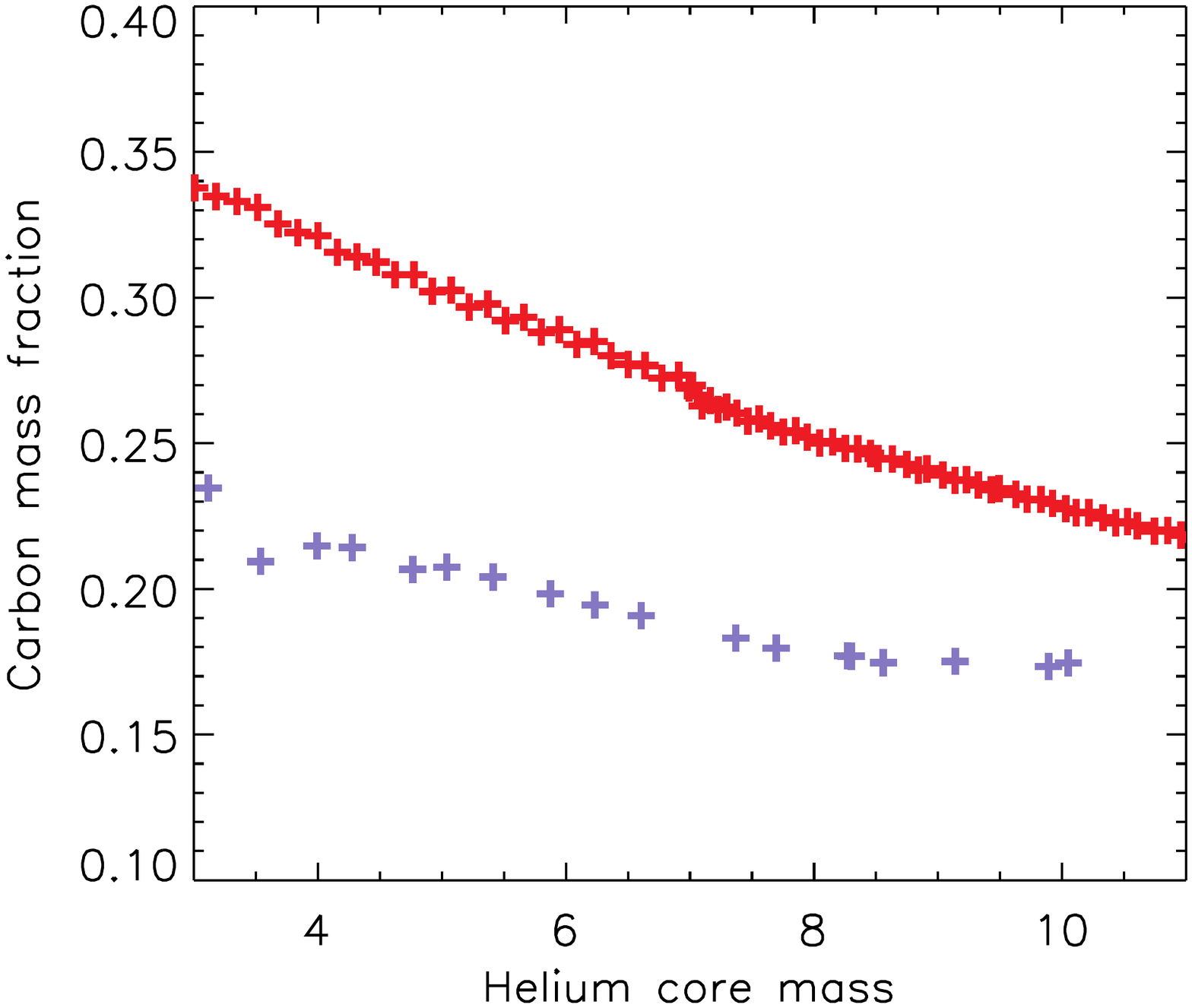}
\caption{Central carbon mass fraction at carbon ignition for the
  present set of models (red points) and a corresponding set of
  massive stars with hydrogen envelopes \citep{Suk18}. The x-axis
  gives the current value of the helium core mass which is very close
  to its final value. The mass losing helium stars studied here have a
  substantially higher carbon abundance. \lFig{carbon}}
\end{figure}

\subsection{Compactness}
\lSect{compact}

\Fig{compactfig} shows the compactness parameter \citep{Oco11} 
\begin{equation}
\xi_{M}=\frac{M/\Msun}{R(M)/1000\, {\rm
    km}}\Big|_{t_{{\rm bounce}}}, 
    \lEq{compacteq}
\end{equation}
as a function of final core mass for a variety of presupernova
models. $\xi$ is customarily evaluated at M = 2.5 \Msun \ and denoted
$\xi_{2.5}$.  In the figure, black points are for models with initial
masses 3.2 to 4.5 \Msun \ which have final masses 2.59 to 3.49 \Msun
\ (\Tab{models}). These are in turn derived from stars whose original
main sequence masses were 13.8 to 15.6 \Msun. For lighter stars, the
final mass is less than 2.5 \Msun \ and $\xi_{2.5}$ is undefined,
though it approaches zero. The lighter models (above 2.5 \Msun) were
evolved to the presupernova state using the large nuclear reaction
network to calculate energy generation, neutrino losses, and
neutronization. Normal mass loss was assumed. Fine surface zoning was
employed, down to $5 \times 10^{-9}$ \Msun, and a relatively small
boundary pressure, 10$^4$ dyne cm$^{-2}$, was employed.

Green points in \Fig{compactfig} also assumed the standard mass loss
rate, but used a combination of the approximation network and
quasiequilibrium network. Surface zoning was coarser than for the
black points with finest zones near 10$^{-7}$ \Msun. A surface
boundary pressure 10$^8$ dyne cm$^{-2}$ was used.  Blue points are for
a more limited survey that used similar resolution, but slightly
reduced boundary pressure, 10$^7$ dyne cm$^{-2}$.  Red points came
from a similar survey to the green points, but with 1.5 times the mass
loss rate (see \Tab{models}). Good agreement among different sets
evolved with different physics where they overlap in presupernova mass
suggests the robustness of the pattern.

The overall behavior is less chaotic than found by \citet{Suk14} and
\citet{Suk18} for stars in this mass range (see Fig. 8 of
\citet{Suk18} for final helium cores from 3.5 to 6 \Msun). Though 
clearer evidence might emerge with a greater number of models, there
is less indication here of multiple solutions for a given final mass,
especially branches with high compactness parameters. In general, the
models, especially the lower mass ones, look more likely to
explode. Helium cores up to 6.0 \Msun \ in final mass have $\xi_{2.5}
\ltaprx 0.15$. A final helium core mass of 6.2 \Msun \ corresponds to
an initial mass of 9.0 \Msun \ (\Tab{models}) which in turn
corresponds to a main sequence mass of about 30 \Msun.  The helium
core at death for a 30 \Msun \ star that did not lose its entire
envelope would have been 10 \Msun \ (\Fig{cores}) and would have had
$\xi_{2.5} \approx 0.40$ \citep{Suk18}. Such a large difference will
very likely affect whether the star explodes or not. Stars evolved in
mass exchanging binaries have very different deaths than single stars
of the same initial mass.

Most striking is the absence of models with compactness $\xi_{2.5}
\gtaprx 0.17$ for presupernova masses less than 6 \Msun. In
\citet{Suk18}, there were many models with $\xi_{2.5}$ between 0.17
and 0.3 for helium core masses around 4.5 to 6.0 \Msun \ (see their
Fig. 6). These might have provided a population of low mass black
holes. While a single parameter characterization cannot substitute for
a full study of core collapse, $\xi_{2.5}$ is known to correlate well
with other more physical measures of explodability \citep{Ert16,Suk18},
as well as with the results of simple 1D neutrino transport
calculations \citep{Suk16}. It thus seems likely that the distribution of
black hole masses will be different for binaries and single
stars. Since all mass measurements of black hole masses are made in
binaries, usually with a history of interaction, this has important
implications.  The peak in compactness where a large number of black
holes are likely to be born is also shifted slightly upwards. in
\cite{Suk18} the peak was centered at helium core masses of 7
\Msun. Here it is at 8 \Msun.

Why are the results so different, even when compared for the same
final helium core masses? Several factors contribute. First, a helium
core inside a massive star has different boundary conditions since the
former is bounded by a dense, hot hydrogen burning shell and the
latter is not. The helium core in a star that still has its hydrogen
envelope is arbitrarily defined as where the hydrogen mass fraction
first goes below 0.01 moving inwards. Typical temperatures for that
shell during helium burning inside a massive star are near $4 \times
10^7$, and the pressure is a few times 10$^{16}$ dyne cm$^{-2}$. In
bare helium stars, such conditions are usually achieved a few tenths
\Msun \ beneath the surface. Helium cores in massive stars, so
defined, evolve like slightly heavier bare helium cores.

Second, and more importantly, the carbon mass fraction at carbon
ignition is substantially higher in the mass-losing helium stars than
in in the mass-gaining helium cores of massive hydrogenic stars
(\Fig{carbon}). Inside a massive star, the helium core grows because
of mass processed through the hydrogen burning shell. As it grows, the
extent of the central convective core also increases \citep[Fig. 3
  of][]{Woo02}. Thus, at late times, extra helium is mixed into the
burning region, turning more carbon to oxygen. In a mass-losing star,
the opposite occurs. The convective core recedes (second panel of
\Fig{convection1}) leaving behind a gradient in the carbon abundance
with a larger abundance resulting from less destruction by
$^{12}$C($\alpha,\gamma)^{16}$O.  This larger carbon abundance has a
substantial impact on the presupernova core structure. As previously
noted \citep{Bar94,Tim96,Suk14}, a major change in core structure
happens when the star transitions from burning carbon in its center
convectively to burning it radiatively. For single stars, this occurs
near 20 \Msun, or for a final helium core mass of 6.2 \Msun. For the
mass losing helium stars, the transition is pushed up to 7.2 \Msun. A
similar shift would occur for a smaller rate for the
$^{12}$C($\alpha,\gamma)^{16}$O reaction \citep[see Fig 24
  of][]{Suk14}. Here we use an S-factor at 300 keV of 165 keV barns,
which is within the current experimental error bar, but could be
slightly high \citep{DeB17}.

\begin{figure*}
\includegraphics[width=0.48\textwidth]{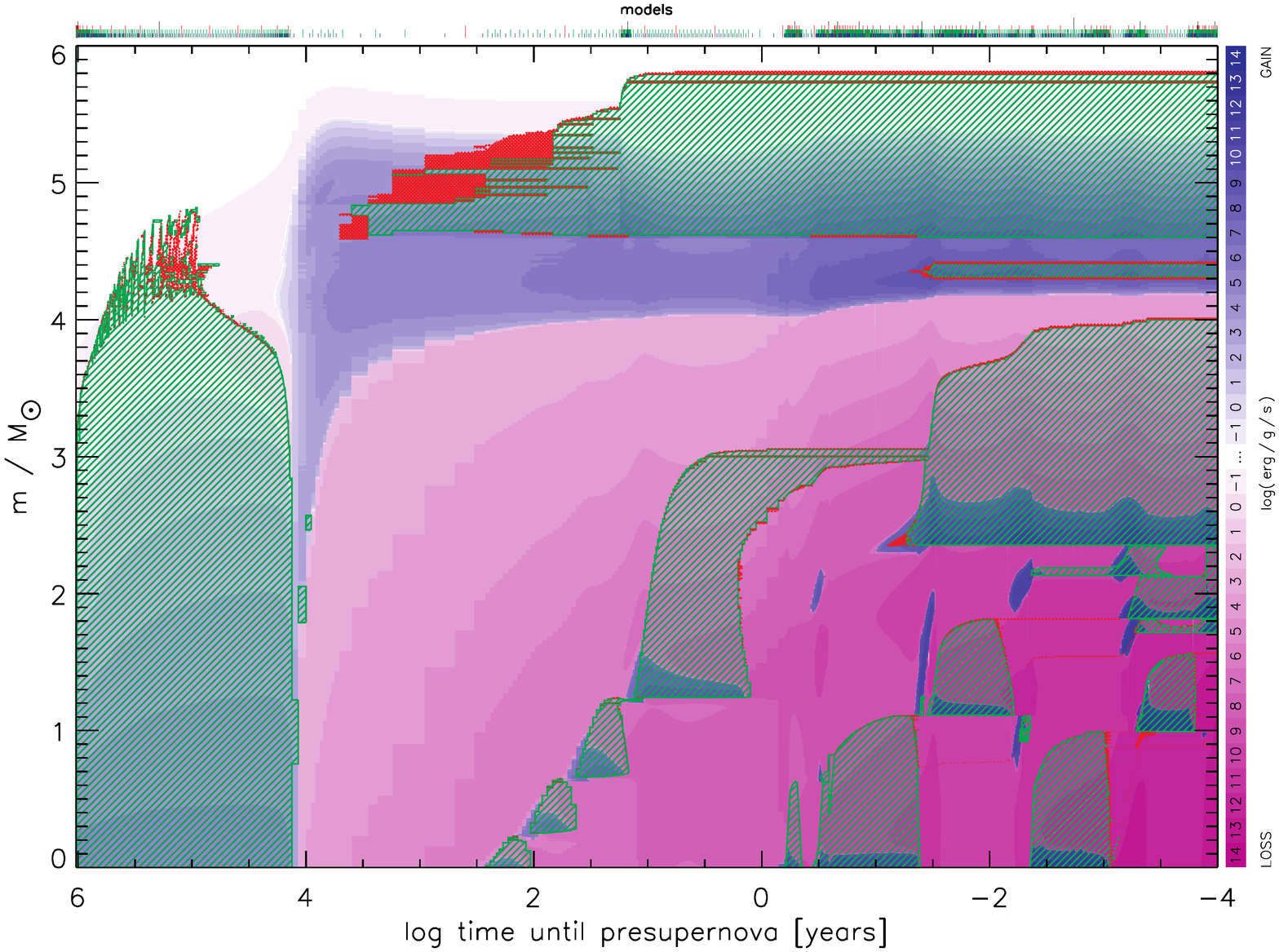}
\includegraphics[width=0.48\textwidth]{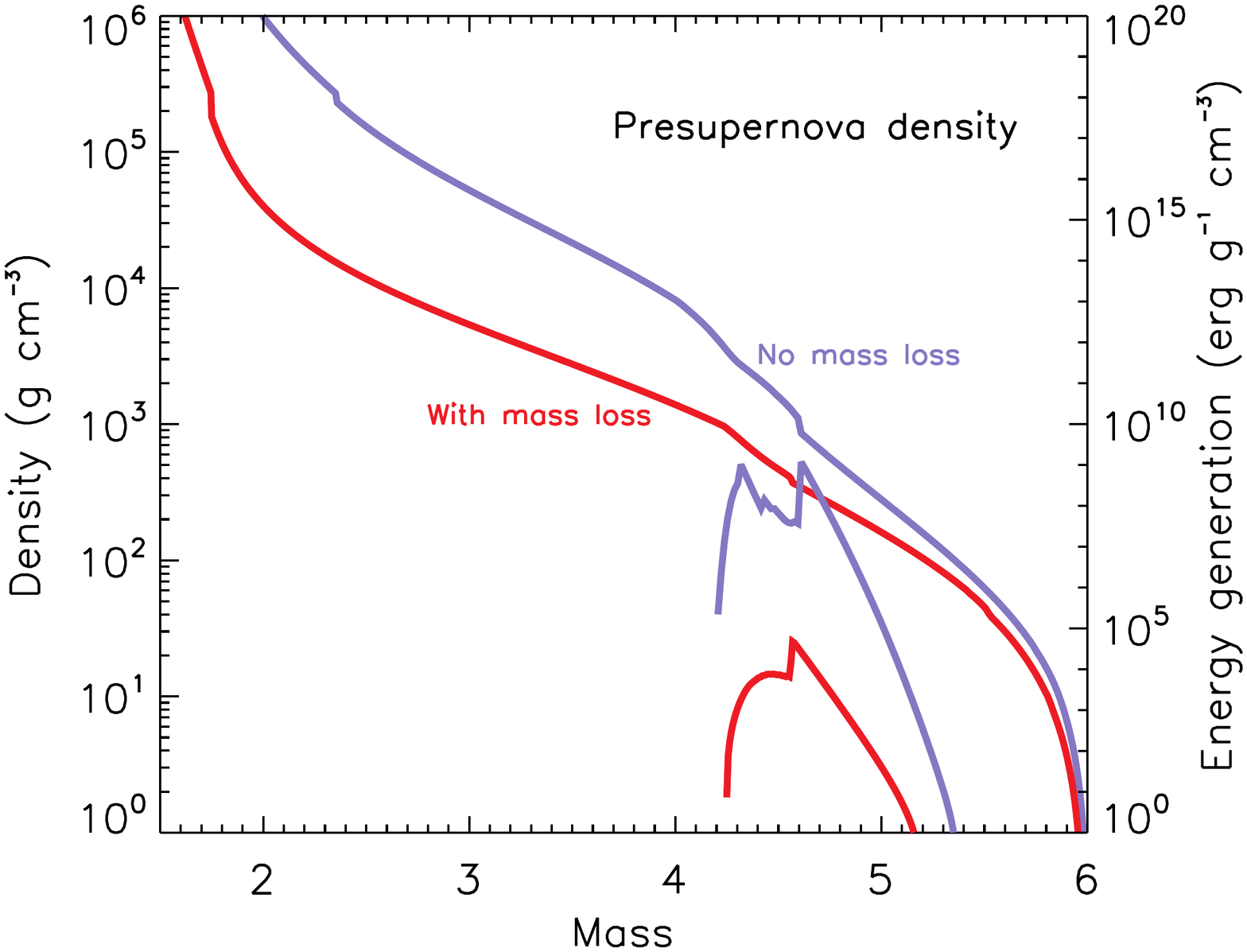}
\includegraphics[width=0.48\textwidth]{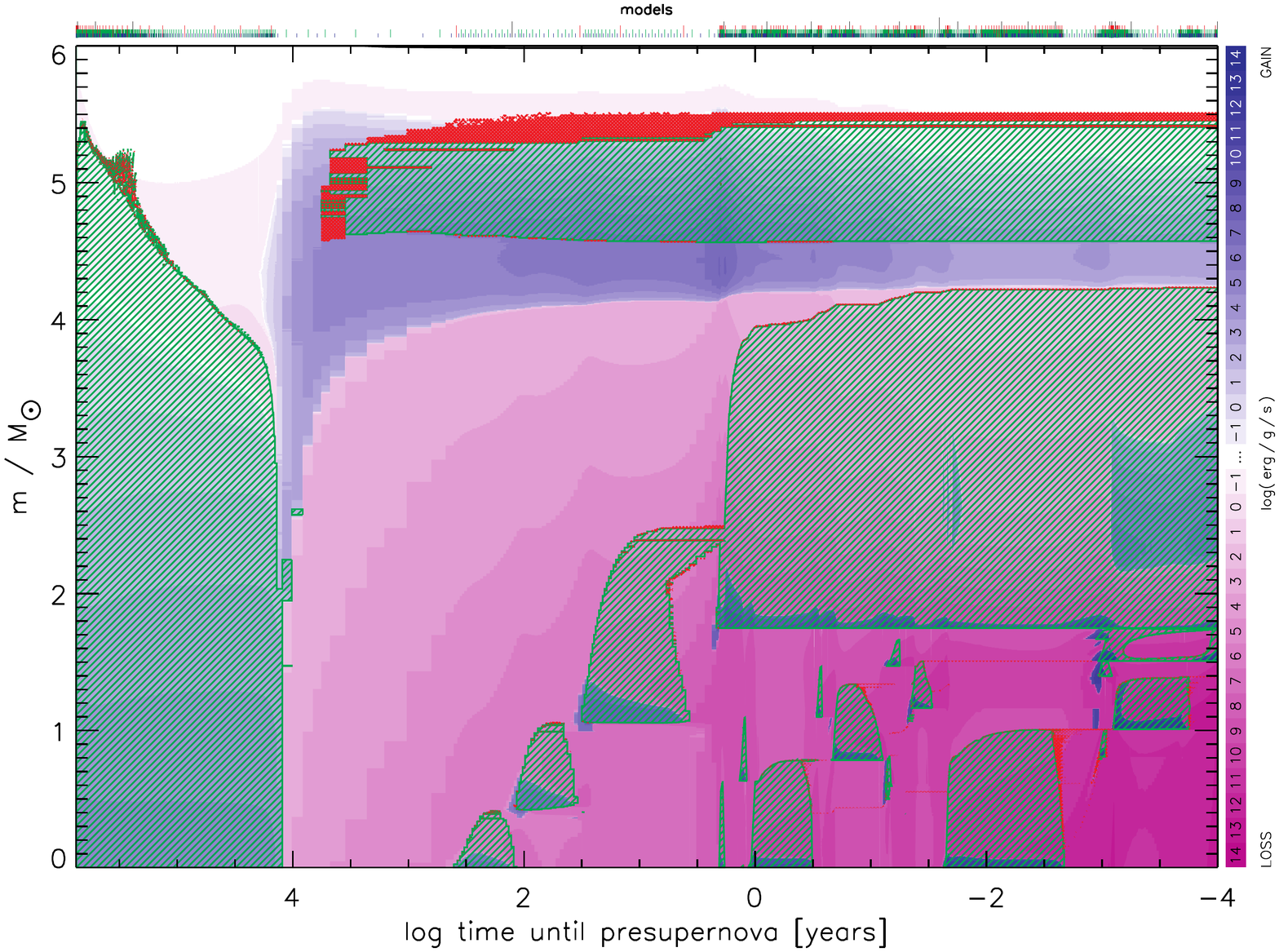}
\includegraphics[width=0.48\textwidth]{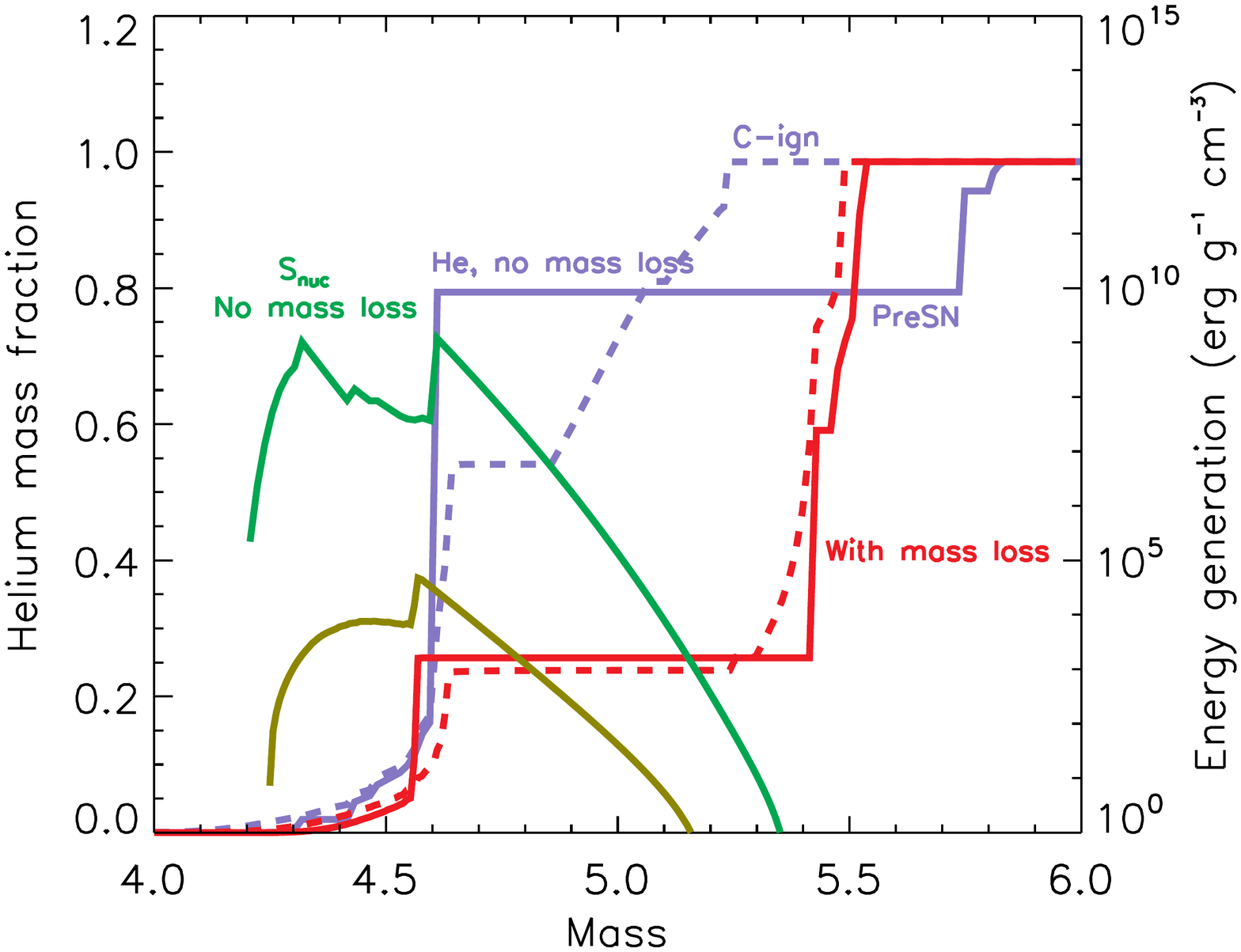}
\caption{Convective history in two helium stars that both ended
  their lives with a mass close to 6.0 \Msun. The top left panel is a
  star evolved at constant mass. The bottom left panel is for a star
  that began as an 8.625 \Msun \ helium star, but ended its life with
  5.98 \Msun \ (see \Tab{nomdot}). The right top panel gives the
  presupernova density profiles and helium shell energy generation for
  the same two stars. The bottom right panel shows the helium mass
  fraction at carbon ignition (dashed lines) and presupernova stars
  (solid lines) and the corresponding energy generation for the
  presupernova stars. Note the very energetic helium burning shell in
  the case without mass loss is altering the structure of the outer
  core.\lFig{convection1}}
\end{figure*}

Finally, the outer structure of the helium core, and thus the strength
of its helium burning shell are different when the core loses mass. A
set of 22 helium stars from 3.5 to 14 \Msun \ was evolved without mass
loss. The compactness plot (not shown) in \Fig{compactfig} would be
about half way between the \citet{Suk18} set and the present study. In
\Tab{nomdot}, a subset of these is compared with their counterparts in
the mass losing set that ended up with the same final helium core
mass.  The final mass of the carbon-depleted core ($M_{\rm ONe}$) and
compactness parameter, $\xi_{2.5}$, are given. For comparison, the
ONe-core masses and compactness parameters are given for seven
equivalent mass-losing stars from the standard set. These stars began
their lives as much heavier helium stars, $M_{\rm init}$, but ended
with final masses, $M_{\rm fin}$, very close to those of the stars
evolved at constant mass. The compactness parameters are
systematically lower and less variable than for the constant mass
stars.

\Fig{convection1} shows the convective history and presupernova
structure for two of the models in \Tab{nomdot}, both of which ended
their lives with a mass very near 6.0 \Msun. Neither star changed its
total mass appreciably after carbon ignition. In the mass losing 8.63
\Msun \ model, a strong, coupled carbon, oxygen, and neon burning
shell ignites at 1.75 \Msun \ about a year before the explosion. This
burning expands the overlying material making the radius at 2.5 \Msun
\ larger, i.e., the compactness parameter smaller. In the constant
mass star, a similar shell ignites at 2.36 \Msun, but only about a
week before the explosion. This burning is too late and too
far out to greatly affect the radius at 2.5 \Msun \ and the star dies
with a large compactness parameter.

As to why this difference exists, it may have to do with the
modulation of the helium burning shell which sets the boundary
condition for the core. \Fig{convection1} shows that while the
mass-losing cores initially produce larger CO cores, the helium stars
evolved without mass loss develop deeper, stronger helium burning
shells in their later evolution that modulate the presupernova density
structure.

\begin{deluxetable}{ccccccc} 
\tablecaption{Compactness in Models Without Mass Loss} 
\tablehead{ \colhead{${M_{\rm init,0}}$}     & 
            \colhead{${M_{\rm ONe}}$}        &
            \colhead{${\xi_{2.5}}$}         &
            \colhead{${M_{\rm init}}$}       & 
            \colhead{${M_{\rm fin}}$}        & 
            \colhead{${M_{\rm ONe}}$}         &
            \colhead{${\xi_{2.5}}$}         
            \\
            \colhead{[\Msun]}              &  
            \colhead{[\Msun]}              &
            \colhead{No $\dot {\rm M}$}    &
            \colhead{[\Msun]}              &
            \colhead{[\Msun]}              &
            \colhead{[\Msun]}              & 
            \colhead{$\dot {\rm M}$}                     
            }
 4.0 &  1.78 & 0.132 & 5.25  & 3.98 & 1.76 & 0.097   \\
 4.5 &  1.86 & 0.162 & 6.125 & 4.52 & 1.71 & 0.109  \\
 5   &  1.81 & 0.129 & 7.00  & 5.04 & 1.83 & 0.159  \\
 5.5 &  2.00 & 0.200 & 7.875 & 5.56 & 1.87 & 0.167  \\
 6   &  2.35 & 0.265 & 8.625 & 5.98 & 1.75 & 0.102  \\
 6.5 &  2.18 & 0.180 & 9.625 & 6.54 & 2.02 & 0.127  \\ 
 7   &  2.51 & 0.285 & 10.75 & 6.95 & 2.31 & 0.194         
\enddata
\tablecomments{The first three columns give the constant mass of the
  models evolved without mass loss, the location of the combined neon
  and carbon burning shells in the presupernova star, and the
  compactness parameter. The next four columns give similar
  information for the models evolved with mass loss. $M_{\rm fin}$ is
  the presupernova mass, chosen to be close to that of the models with
  no mass loss.}  \lTab{nomdot}
\end{deluxetable}

\section{Light Curves}
\lSect{lite}

\subsection{Type Ib and Ic}
\lSect{sn1bc}

Given the broad range of masses explored, a great variety of supernova
light curves are possible. These include essentially all Type I
supernovae that do not have rotationally powered light curves and are
Type Ia.  Some models resemble closely those already in the literature
and will have similar observable properties. Based upon
\Fig{compactfig}, it is expected that most models with final masses up
to 6 \Msun \ will explode and leave neutron star remnants. This
includes stars with initial helium core masses up to 9 \Msun
\ (\Tab{models}) and thus main sequence masses up to 30 \Msun. For
final masses below about 2.0 \Msun, the core is so compact that both
the explosion energy and $^{56}$Ni production probably decline below
what is needed for typical Type Ib and Ic supernovae
\citep{Suk16}. This leaves the 2 to 6 \Msun \ as the likely
presupernova mass for common Ibc supernovae. Previous studies of
models in this mass range have shown good agreement with observations
\citep{Des12,Des16}. The transition from Ib to Ic for progenitor
masses around 5 \Msun \ with residual helium masses of $\sim0.3$ \Msun
\ found by \citet{Des15} is in reasonable accord with \Tab{models}
which shows the CO-core being uncovered for final masses above 5 \Msun
\ for the 1.5 times standard mass loss cases. Residual helium in the
present models is $\gtaprx0.2$ \Msun. Pending further study, this
might be supportive of the higher mass loss rate.

The progenitor of Type Ib supernovae iPTF13bvn was inferred to have a
mass at death near 3.5 \Msun \ \citep{Ber14,Eld15}.  A final mass of
3.5 \Msun \ corresponds to present models with initial helium cores
masses 4.5 - 5 \Msun \ (\Tab{models}). These would have come from main
sequence stars of about 20 \Msun \ and would have log luminosity and
temperature 4.85 and 4.75 respectively. This is within the allowed
observational range \citep[Fig. 1 of][]{Eld15}.

\begin{figure}
\includegraphics[width=0.48\textwidth]{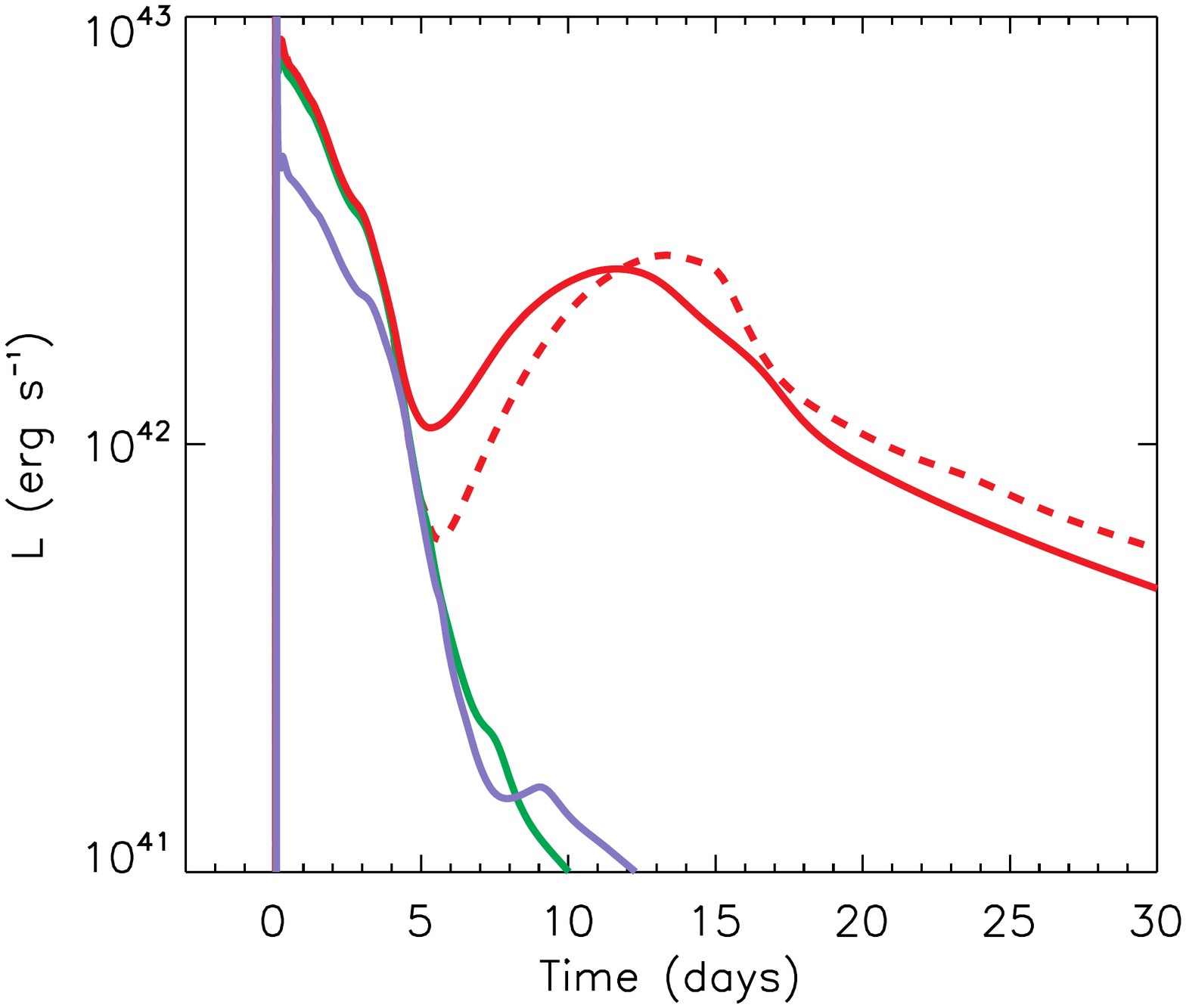}
\includegraphics[width=0.48\textwidth]{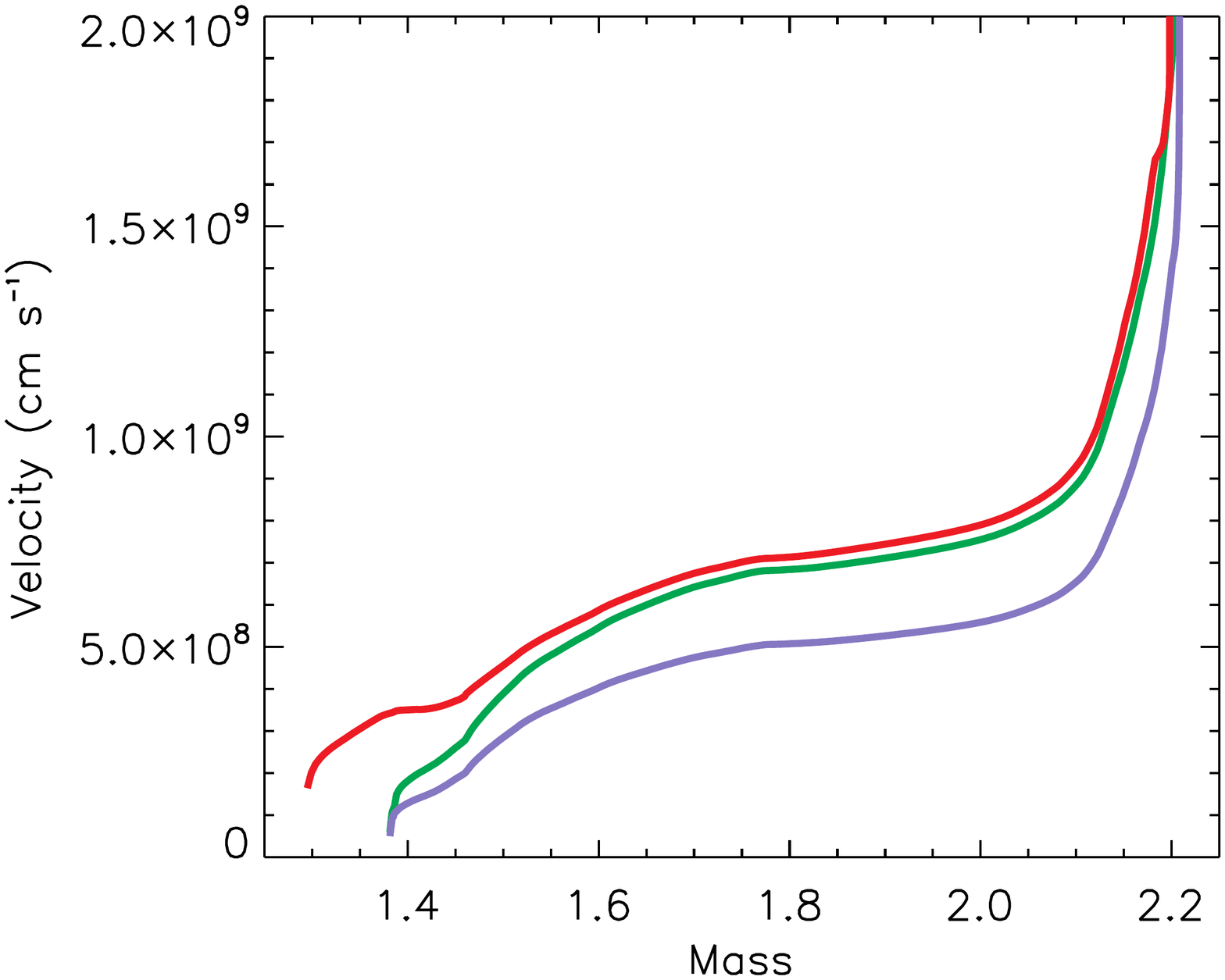}
\caption{(Top:) Light curves for the 2.7 \Msun \ models
  (\Tab{sisn}). These results are characteristic of stars that
  experience radius expansion, but lack a strong silicon flash. The
  radius of this model $6.5 \times 10^{12}$ cm.  The green and blue
  curves correspond to two different explosion energies (0.26 and 0.47
  $\times 10^{51}$ erg; Models 2.7A and 2.7B respectively). The light
  curves with secondary maxima (red; Models 2.7C and Cm) had pistons
  situated deep in the star and produced 0.071 \Msun \ of $^{56}$Ni.
  The dashed red line shows the effect of moderate mixing. (Bottom:)
  The terminal velocity profiles for the three models.
  explosions. \lFig{lite2.7}}
\end{figure}

An island of possible Ib or Ic supernovae with progenitor masses near
10 - 12 \Msun \ is also noted, but these will be rarer and probably
have broader fainter light curves due to their slower expansion speed.
Most of the stars with final masses between 7 and 30 \Msun \ will
produce black holes.

For final masses above 30 \Msun, and especially above 37 \Msun, PPISN
result. For a given final mass, the light curves should be 
similar to those calculated for helium stars evolved without mass loss
by \citet{Woo17}.

\subsection{Low mass models with radius expansion}
\lSect{bigr}

Some models are sufficiently different from those already published to
be worthy of immediate attention. Chief among these are the low mass
stars with extended envelopes and the stars that experience silicon
flashes \citep[see also][for more realistic transport studies
  of similar parametrized models]{Kle18}.

Typical of stars that have large radii as presupernovae, but have {\sl
  not} experienced a strong silicon flash, is the 2.7 \Msun \ model
(\Tab{siign}) which has a presupernova mass of only 2.21
\Msun. Assuming that its envelope was not lost to the binary companion
during the second phase of expansion, the presupernova star had a
radius of $6.5 \times 10^{12}$ cm. Two explosions were introduced in
this model using pistons at 1.29 \Msun \ and 1.38 \Msun.  The smaller
piston mass was at the edge of the iron core; the larger, where the
entropy per baryon was $S/N_A k$ = 4.0. Explosion energies of 2.6 and
4.7 $\times 10^{50}$ erg, as measured by the kinetic energy of ejecta
at infinity, were generated for the mass cut at 1.38 \Msun \ and one
with explosion energy $5.1 \times 10^{50}$ erg for the mass cut at
1.29. These limits on piston mass were chosen to approximate the
minimum and maximum $^{56}$Ni production. The explosions with the
piston farther out produced very little $^{56}$Ni, 0.0025 \Msun \ and
0.0029 \Msun \ for the low and high explosion energy respectively. The
deeper piston produced 0.071 \Msun.  This is an upper bound since
ejecting neutronized iron from deeper in would not increase $^{56}$Ni
production. More realistic explosion models usually produce less
\citep{Suk16}. A version of the high nickel yield model was also
calculated in which the ejected composition was moderately, but
artificially mixed.

\Fig{lite2.7} shows the results. All explosions exhibit an initial
peak lasting about 4 days. More energetic explosions are brighter, as
expected from scaling laws for Type IIP supernovae \citep{Suk16}. A
larger radius would also give a brighter initial explosion, but
larger radii were not found for models that did not experience silicon
flashes (\Fig{density}). The envelope would also be in greater danger
of being lost to the companion.  Effective temperatures on
days 0.5, 1, 3 and 5 for the higher energy model were 33,000, 23,000,
13,000, and 10,500 K, so this would be a very blue
transient. Velocities are typically around 5000 km s$^{-1}$ for the
less energetic model and 7000 km s$^{-1}$ for the more energetic one
(\Fig{lite2.7}). Most of the mass exterior to 1.46 \Msun \ is helium,
with about 10\% by mass of carbon and oxygen from a prior helium
burning shell extending to 1.82 \Msun. Matter interior to 1.46 \Msun
\ is mostly oxygen with traces of heavier freshly synthesized
elements, notably silicon, magnesium, and $^{56}$Ni. There is also, of
course, a primordial abundance of all elements characteristic of solar
metallicity in the ejected helium.

After the first week, the display, in those models that make
appreciable $^{56}$Ni, is dominated by its decay. In models without
mixing this decay produces a pronounced secondary maximum with
luminosity roughly proportional to nickel mass.  The effective
temperature at these later times is not well determined in the KEPLER
code, but is estimated to be about 7000 K after the first week.  With
a moderate amount of mixing (helium mixed to a mass fraction of 0.3 at
the bottom of the ejecta and $^{56}$Ni exceeding 0.01 out to 1.95
\Msun), the minimum between the two peaks is eroded.

Recently \citet{De18} have reported observations of SN 2014ft, a Type
Ic supernova that they attribute to an ``ultra-stripped'' helium star
in a binary system. This will be discussed further in \Sect{sidefl},
but the models here also bear a superficial resemblance to that event.
The observed light curve is initially very blue, evolves rapidly, and
has two peaks. The initial peak was brighter than 10$^{43}$ erg with a
temperature greater than 32,000 K, but declined within a day to 10,000
K and luminosity of approximately $1 \times 10^{42}$ erg
s$^{-1}$. Over the next 6 days the supernova rebrightened to over $2
\times 10^{42}$ erg s$^{-1}$ while the temperature remained near
10,000 K. Velocities in the second peak were $\sim10,000$ km s$^{-1}$
and the spectrum showed lines of helium and carbon.

Compared with the 2.7 \Msun \ model (\Fig{lite2.7}), temperatures,
velocities, composition, and even the bolometric light curve agree
qualitatively. The initial peak in SN 2014ft was much briefer though,
and brighter than the model. The secondary peak also occurred
earlier. A model with a larger radius and greater explosion energy
would have had a brighter first peak and one with a smaller ejected
mass would have evolved more quickly. For the opacities, mass
loss, and zoning employed here though, a radius much greater than 100
\Rsun \ is unlikely for models that avoid becoming SAGB stars
(\Fig{density}). Explosion energies much greater than $5 \times
10^{50}$ erg are also not expected for stars in this mass range
\citep{Suk16}. A larger explosion energy would also give greater
velocities than observed. The observed initial luminosity is thus
problematic, though perhaps only by a factor of two. A smaller ejected
mass is certainly possible and could be obtained by modestly altering
the uncertain mass loss rate or removing only part of the envelope in
a final binary interaction. All in all, the model seems promising and
worth further exploration. It is similar in properties to the one
suggested by \citet{De18}, but somewhat simpler in that it is the
first supernova in the binary that makes SN 2014ft, not the second
(see their Fig. 6), and all that has been assumed is that a main
sequence star of about 14 \Msun \ lost its envelope to a companion
when it ignited helium burning.

\subsection{Silicon deflagration}
\lSect{sidefl}

Some low mass models may have experienced explosive mass ejection due
to silicon deflagration or detonation a few weeks prior to their final
explosion (\Sect{deflag}). Here the major uncertainty is how much
silicon burns promptly (on a sonic crossing time for the dense core)
to iron in the runaway.  The greater the speed of the mass ejection,
the farther this matter travels before the iron core finally collapses
and launches a second, more powerful shock. Ejected envelopes that
expand to 10$^{14}$ - 10$^{15}$ cm before core collapse give very
luminous, long lasting secondary explosions, even for supernovae with
low kinetic energies. Those that eject less mass and spread to
$\ltaprx10^{14}$ cm give fainter, briefer ones. If very small mass of
silicon burns, less than $\sim$0.1 \Msun, too little mass is ejected
too slowly to greatly modify the light curve of the terminal
explosion. The small mass falls back or is overtaken by the terminal
shock long before the supernova becomes bright.

\Tab{sisn} shows some results for specific cases. Four presupernova
models based on the 2.5 \Msun \ progenitor each had a total mass of
2.07 \Msun \ at the time of their initial instability. Since a remnant
of $\sim1.4$ \Msun \ is left by the explosion and additional matter is
ejected by the silicon flash, the amount left to be ejected in the
final explosion is small. Silicon deflagrations are parametrized by
the amount of silicon that burns to iron in the runaway, an adjustable
parameter in this study. Since the deflagration ignites far off center
in this model (at 0.49 \Msun), and only matter external to the
ignition point burns in the initial flash, the amount of silicon that
can burn is limited. The most energetic model considered burned 0.294
\Msun.  On the other hand, about 0.1 \Msun \ of silicon must burn to
cause a large amplitude oscillation of bound core and launch an
appreciable shock. This determined the least energetic case
considered. Typical ejection speeds were 100 km s$^{-1}$ to 800 km
s$^{-1}$ (Model 2.5D) though higher speeds were present in both
(\Fig{lite2.5}). In Model 2.5A, the ejecta had not reached terminal
speed, and some would have fallen back had the supernova not exploded
first. For Models 2.5C and 2.5D the entire envelope was ejected.

\Tab{sisn} also shows that, usually, the more powerful the silicon
flash, the longer the wait until iron-core collapse. Times varied from 19
to 62 days.  During this time the ejected or partly ejected envelope
expanded to a few $\times 10^{13}$ cm (Model 2.5A) to 10$^{15}$ cm
(Model 2.5D).


\begin{deluxetable*}{ccccccccc} 
\tablecaption{Low Mass Supernovae} 
\tablehead{ \colhead{${M_{\rm init}}$}           & 
            \colhead{${M_{\rm fin}}$}            & 
            \colhead{$\Delta{M_{\rm Si}}$}       &
            \colhead{${M_{\rm ej}}$}             & 
            \colhead{${t_{\rm delay}}$}           &
            \colhead{${M_{\rm Fe}}$}             &
             \colhead{${M_{\rm pist}}$}           &
             \colhead{${KE_{\rm SN}}$}           &
             \colhead{${M(^{56}Ni)}$}             
            \\
            \colhead{[\Msun]}                  &  
            \colhead{[\Msun]}                  &  
            \colhead{[\Msun]}                  &  
            \colhead{[\Msun]}                  &  
            \colhead{[day]}                    &  
            \colhead{[\Msun]}                  &  
            \colhead{[\Msun]}                  &  
            \colhead{[10$^{51}$ \, erg]}        &       
            \colhead{[\Msun]}                    
            }
\startdata
2.5A  & 2.07 & 0.102 & 0.11 & 27 & 1.29 & 1.29 & 0.34 & 0.04    \\
2.5B  & 2.07 & 0.124 & 0.25 & 19 & 1.29 & 1.29 & 0.32 & 0.04    \\
2.5C  & 2.07 & 0.171 & 0.70 & 20 & 1.29 & 1.34 & 0.13 & $<$0.01 \\
2.5D  & 2.07 & 0.294 & 0.70 & 50 & 1.22 & 1.30 & 0.22 & $<$0.01 \\
2.7A  & 2.21 &   -   & -    & -  & 1.29 & 1.38 & 0.26 & 0.0027 \\
2.7B  & 2.21 &   -   &  -   & -  & 1.29 & 1.38 & 0.47 & 0.0024 \\
2.7C  & 2.21 &  -   &  -    & -  & 1.29 & 1.29 & 0.52 & 0.071 \\
2.7Cm & 2.21 &   -   &  -   & -  & 1.29 & 1.29 & 0.52 & 0.071 \\
3.0A  & 2.45 & 0.446 & 0.73 & 62 & 1.32 & 1.40 & 0.25 & $<$0.01 \\
3.0B  & 2.45 & 0.446 & 0.73 & 62 & 1.32 & 1.40 & 0.42 & $<$0.01 \\
3.0C  & 2.45 & 0.446 & 0.73 & 62 & 1.32 & 1.32 & 0.86 & 0.06    \\
3.2A  & 2.59 & 0.138 & 0.016 & 15 & 1.36 & 1.36 & 0.97 & 0.10    \\
3.2B   & 2.59 & 0.138 & 0.016 & 15 & 1.36 & 1.36 & 0.97 & 0.05    \\
3.2Bm  & 2.59 & 0.138 & 0.016 & 15 & 1.36 & 1.36 & 0.97 & 0.05    \\
\enddata
\tablecomments{A suffix ``m'' indicates a model that was artificially
  mixed. All models except the 2.7 experienced a silicon deflagration
  with mass ejection as indicated.  .}  \lTab{sisn}
\end{deluxetable*}

Following iron-core collapse, terminal explosions were introduced in
each model using a piston, $M_{\rm pist}$, to generate a final kinetic
energy of $KE_{\rm SN}$ (\Tab{sisn}). In those cases with low energy
flashes where radioactivity might contribute substantially to the
light curve, the piston was situated as deeply as possible, at the
edge of the iron core, and a moderately high explosion energy was
invoked. This resulted in the synthesis of 0.05 to 0.1 \Msun \ of
$^{56}$Ni, a rather standard amount of $^{56}$Ni for Type Ib and Ic
supernovae. For other models (2.5C, 2.5D, 3.0A, 3.0B) the piston was
placed where the entropy rose to $S/N_Ak$ = 4, a standard value often
used to describe the mass cut in core collapse supernovae. This was
well outside the iron core where the density was lower. As a result,
only a small amount of $^{56}$Ni was synthesized (\Tab{sisn}).

The impact of this piston-ejected mass with the previously ejected
envelope had dramatic consequences for the light curve that were most
sensitive to the mass of the previously ejected shell and its radius.
For the 2.5 \Msun \ models (\Fig{lite2.5}), the time the light curve
stayed near its peak value varied from days to weeks. While the
approximate nature of the effective temperature is to be emphasized
for a single-temperature code, these are all very blue transients. The
brightest events which peaked later had the cooler temperatures,
43,000 K and 23,000 K for Models 2.5C and 2.5D respectively. Still
cooler temperatures might have been obtained had the ejected shell
been at larger radius. 

The sharp falloff in luminosity around 25 days in Models 2.5C and 2.5D
occurs as the shock reaches the edge of the shell ejected by the
silicon flash and the supernova becomes optically thin. In a sense,
these very luminous displays are simply an extension of supernova
shock break out as the high velocity ejecta interacts with the
circumstellar medium. The photosphere remains outside of the shock
until the shock exits the shell, and light diffuses ahead of it. The
main source of energy is circumstellar interaction and the radiating
region piles up, in the 1D code, in a very thin shell. These light
curves are thus also similar to what has been seen for CSM interaction
before, e.g., in PPISN \citep{Woo17}.

The faster evolving Models 2.5A and 2.5B, on the other hand, have
light curves that, even by day 1, are dominated by diffusion from the
shock-heated layers well inside the shell ejected in the flash. The
decline rate thus reflects their large initial radius, which even
before the silicon flash, was $7.3 \times 10^{12}$ cm, and not so much
the mass or energy of the previously ejected shell. A much higher
energy explosion would be required for Models 2.5A or 2.5B to decline
more quickly. Even before the silicon flash, these models had large
radii.

\begin{figure}
\includegraphics[width=0.48\textwidth]{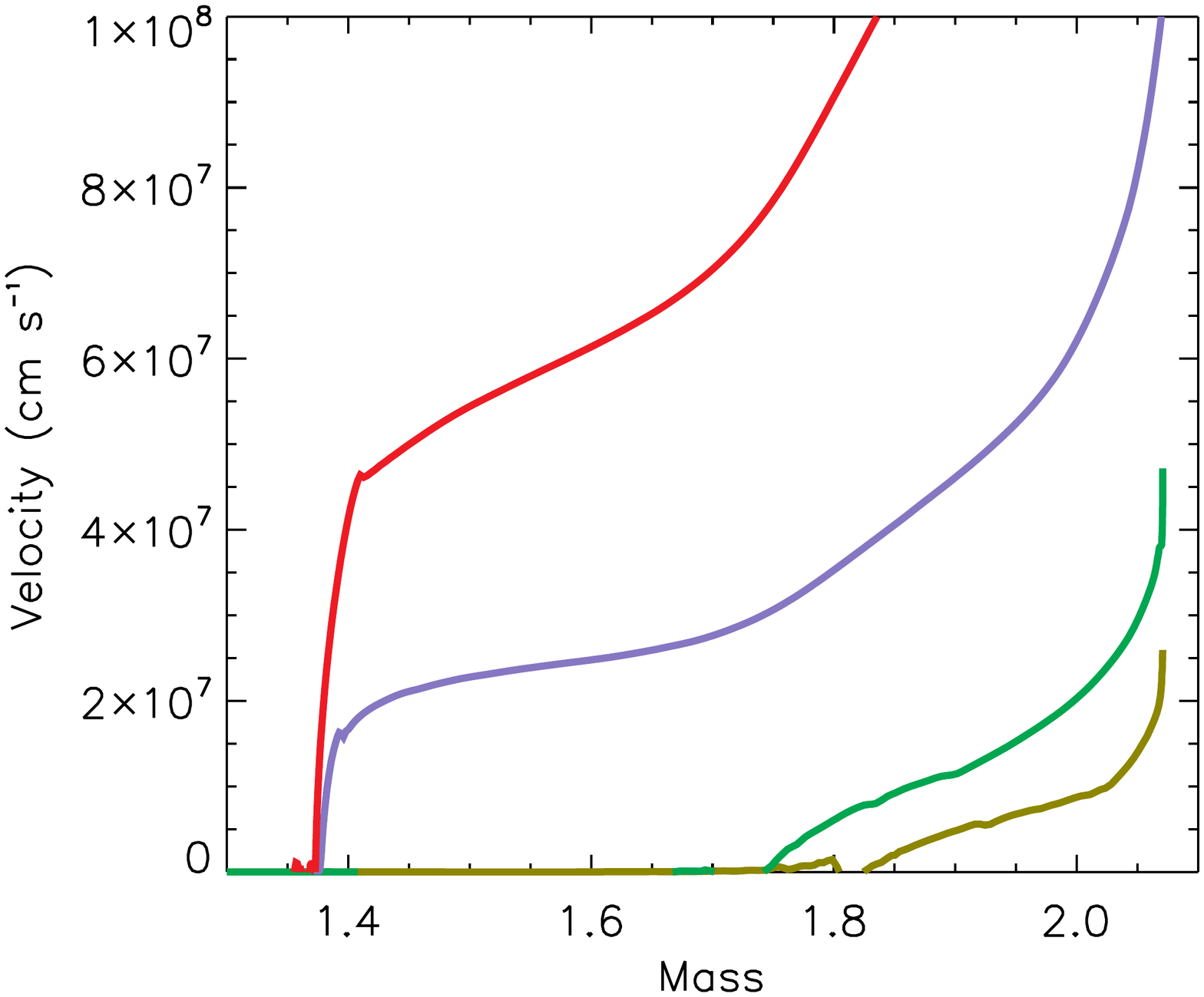}
\includegraphics[width=0.48\textwidth]{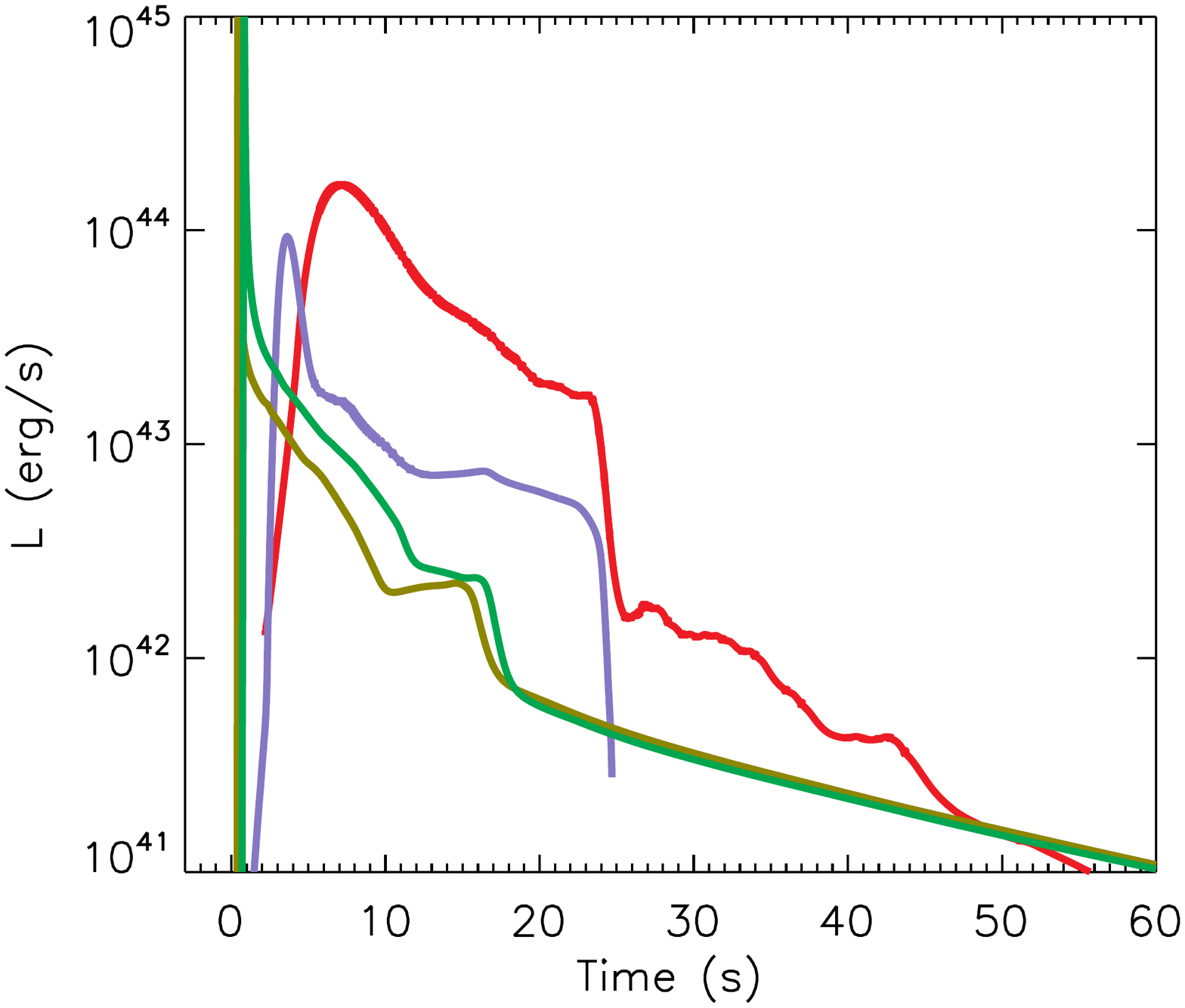}
\includegraphics[width=0.48\textwidth]{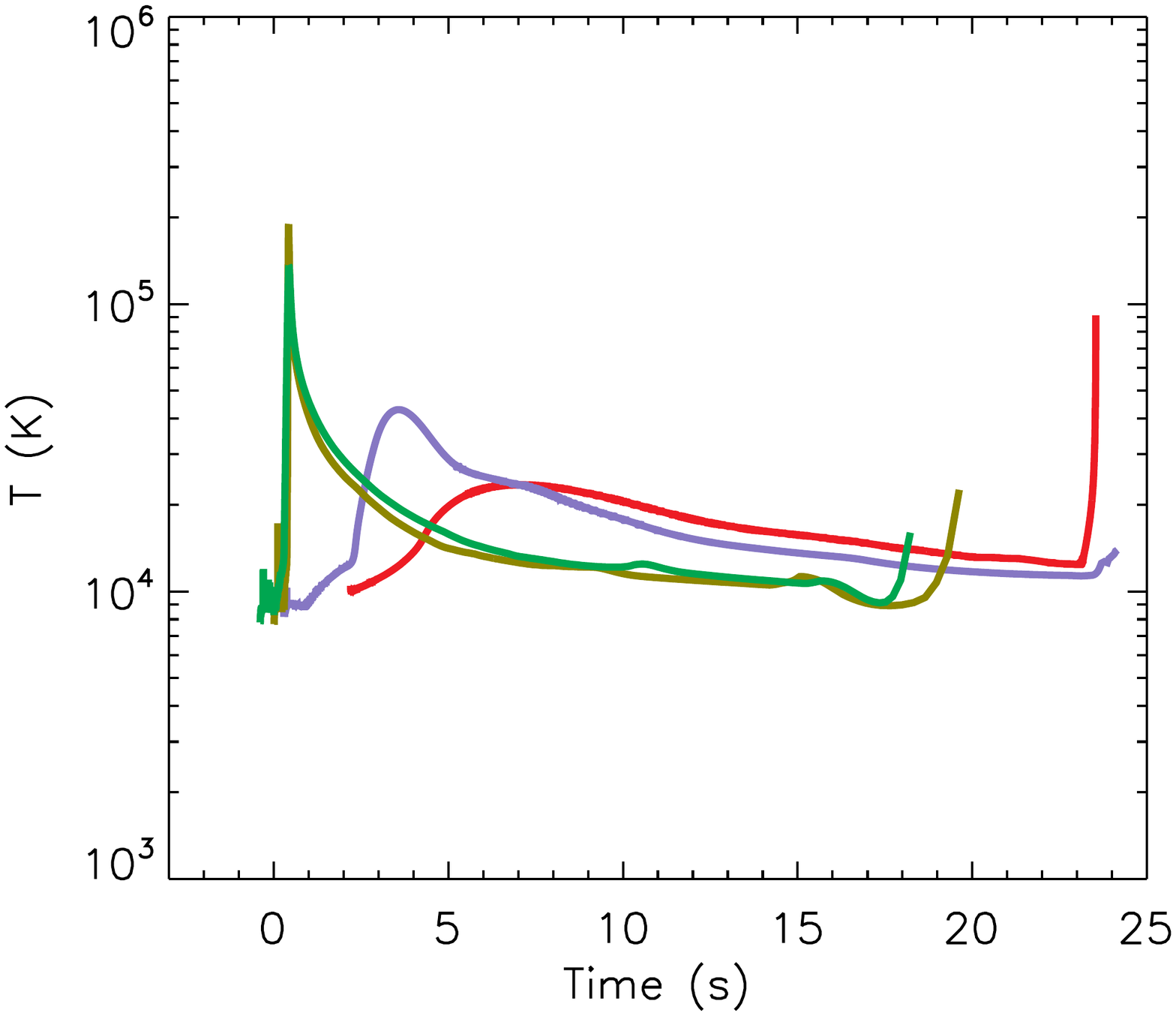}
\caption{Light curves for the 2.5 \Msun \ models in \Tab{sisn}. The
  red, blue, green and gold curves are for Models 2.5D, 2.5C, 2.5B, and
  2.5A respectively. These results are characteristic of stars that
  eject their envelope following a silicon deflagration and core
  collapse several weeks later. (Top:) Velocities of the matter
  ejected by the silicon flash at the time of core collapse
  (\Tab{sisn}). (Middle:) Light curves for terminal explosions
  impacting these distributions of matter. (Bottom:) Effective
  temperatures. The temperature at the well defined peaks for Model
  2.5C (blue) and Model 2.5D (red) are 43,000 K and 23,000 K
  respectively. The rapid rise in T$_{\rm eff}$ at the end is
  artificial and is due to the supernova becoming optically
  thin. \lFig{lite2.5}}
\end{figure}

Two other models explored the consequences of silicon deflagration in
more massive stars (\Tab{siign}).  A low energy deflagration in the
3.2 \Msun \ model ejected only 0.02 \Msun \ with energy $5 \times
10^{46}$ erg. When the iron core collapsed 15 days later, this matter,
with characteristic speeds less than 1000 km s$^{-1}$ was still within
10$^{14}$ cm.  A simulated explosion with the piston located just
outside the iron core at 1.36 \Msun \ with kinetic energy $0.97 \times
10^{51}$ erg gave the light curve in \Fig{lite3.2}.  The initial
transient is again essentially shock break out with circumstellar
interaction. Matter is piles up in a thin shell beneath ionized matter
that is optically thick. Radiation diffuses ahead. The bright emission
ceases when the shock reaches the edge of the shell. Substantial
radioactivity was produced in Model 3.2A (\Tab{sisn}) and this powers
a bright secondary peak at about 10 days. Due to the uncertainty in
where to put the mass cut, the nickel mass was varied. Model 3.2B (red
line in the figure) shows the effect of moderately mixing the ejecta
and dividing the $^{56}$Ni mass by two.

The similarity of this model to the observed properties of SN 2014ft
\citep[iPTF14gqr][]{De18} is striking and better than for the 2.7
\Msun \ models. Indeed its discoverers proposed a silicon flash as a
possible explanation in their paper.  At the time of their first
observation, the supernova had cooled to 32,000 K with a luminosity of
$\sim 2 \times 10^{43}$ erg s$^{-1}$. Here that luminosity occurs on
the fading tail of a much brighter hotter event when the age of the
supernova is 1.6 days and the temperature is 26,000 K. The temperature
at the second peak, 10,000 K, and its luminosity, about $2 \times
10^{42}$ erg s$^{-1}$ also agree with SN 2014ft, but the maximum
occurs substantially later, at about 11 days post explosion {\sl vs} 7
days in the observations. This is because of the larger mass of
ejected helium and heavy elements in the present model, 1.21 \Msun
\ as opposed to the 0.3 \Msun \ estimated by \citet{De18}.  One could
vary parameters in order to obtain a better match, but until an actual
explosion model is available, this would just duplicate what
\citet{De18} have already done.

In the 3.0 \Msun \ models, the effect of more energetic silicon
flashes was explored. More mass is ejected to larger radii and the
final light curves are brighter.  This is perhaps more natural than in
the 2.5 \Msun \ models since the flash occurs closer to the center of
the star and has the potential of burning more silicon to
iron. Explosions with even mild kinetic energies produce very bright
light curves in these models (\Fig{lite3.0}). Varying the explosion
energy and the location of the piston did not alter the light curve
qualitatively, though larger explosion energies did modestly increase
the peak luminosity. Because of the very strong circumstellar
interaction, the production of even 0.06 \Msun \ of $^{56}$Ni in
Model 3.0C had no discernible effect on the light curve. A broader
light curve, more consistent with common superluminous Type I
supernovae \citep{Gal19}, would be produced had the matter ejected by
the silicon flash expanded farther. The amount of silicon burned in
the flash is already maximal, but if the mass of the ejected envelope
had been smaller, it would have acquired a larger velocity and
expanded farther before core collapse. The time scale for silicon
shell burning and flame propagation might possibly be lengthened by a
factor of two.  Models like 3.2 that burn a lot of silicon in the
flash are thus worth further exploration as SLSN-I prototypes.

\begin{figure}
\includegraphics[width=0.48\textwidth]{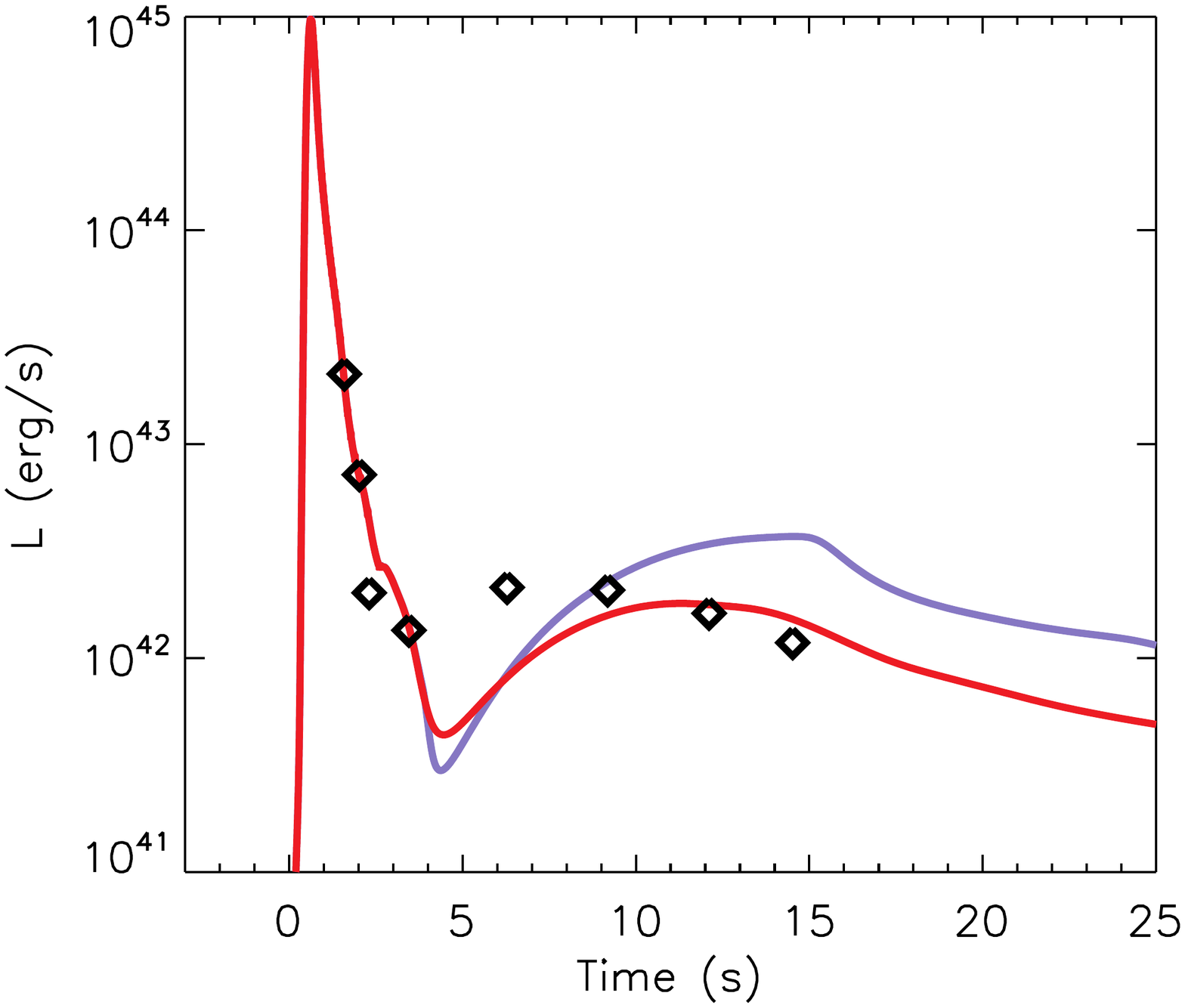}
\includegraphics[width=0.48\textwidth]{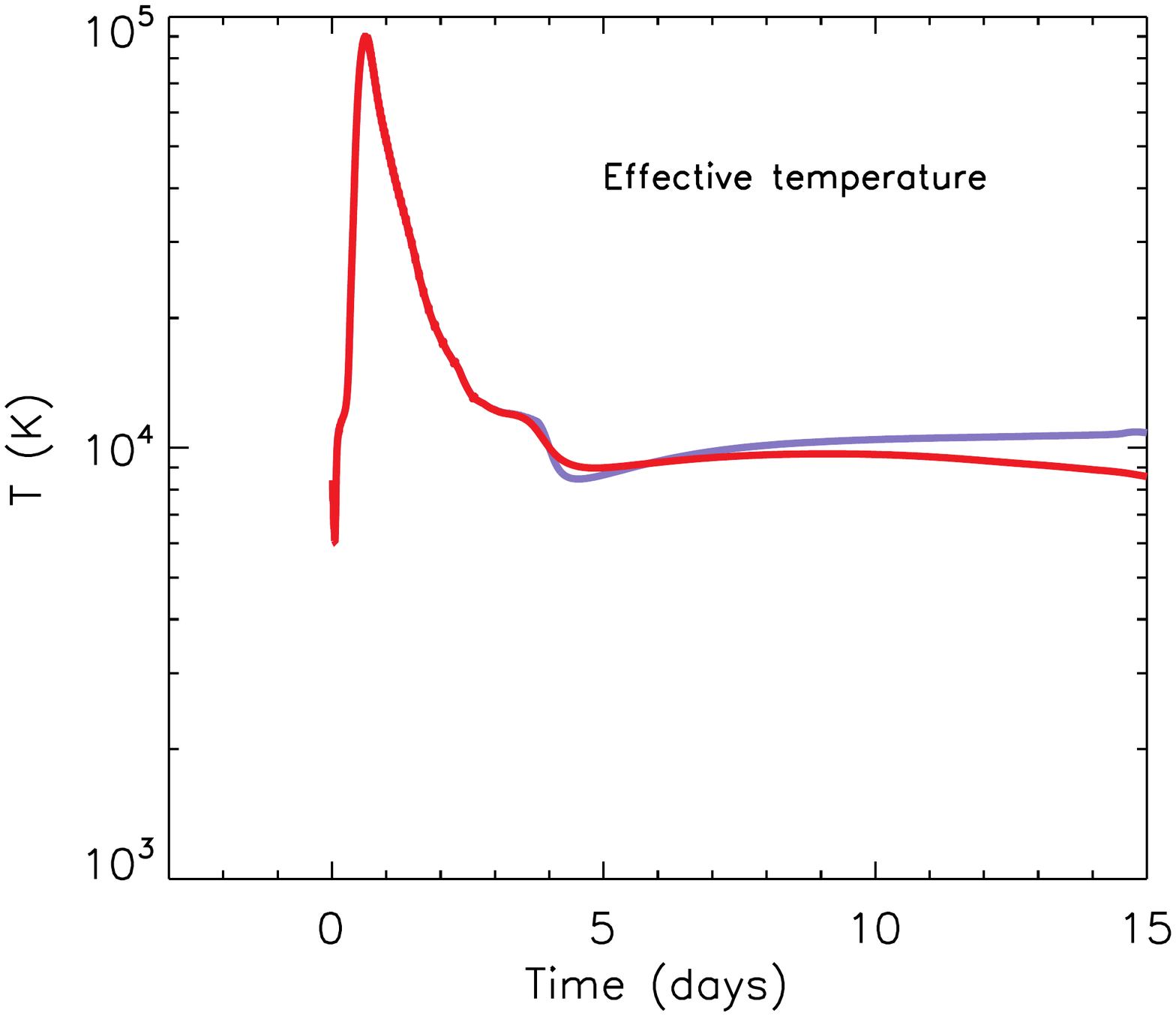}
\caption{(Top:) Light curves Models 3.2B (blue) and 3.2Bm (red).  See
  \Tab{sisn}.  These models experienced a weak silicon deflagration
  several weeks prior to core collapse. The density distribution at
  the time of iron-core collapse is given in \Fig{lite3.0}. Both
  explosions had a kinetic energy of 0.97 $\times 10^{51}$ erg and
  ejected 0.05 \Msun \ of $^{56}$Ni. (Bottom:) The effective
  temperature for the same two models. The similarity to the
  bolometric light curve of SN 2014ft \citep{De18}, plotted as black
  points in the top figure, is striking. \lFig{lite3.2}}
\end{figure}

\begin{figure}
\includegraphics[width=0.48\textwidth]{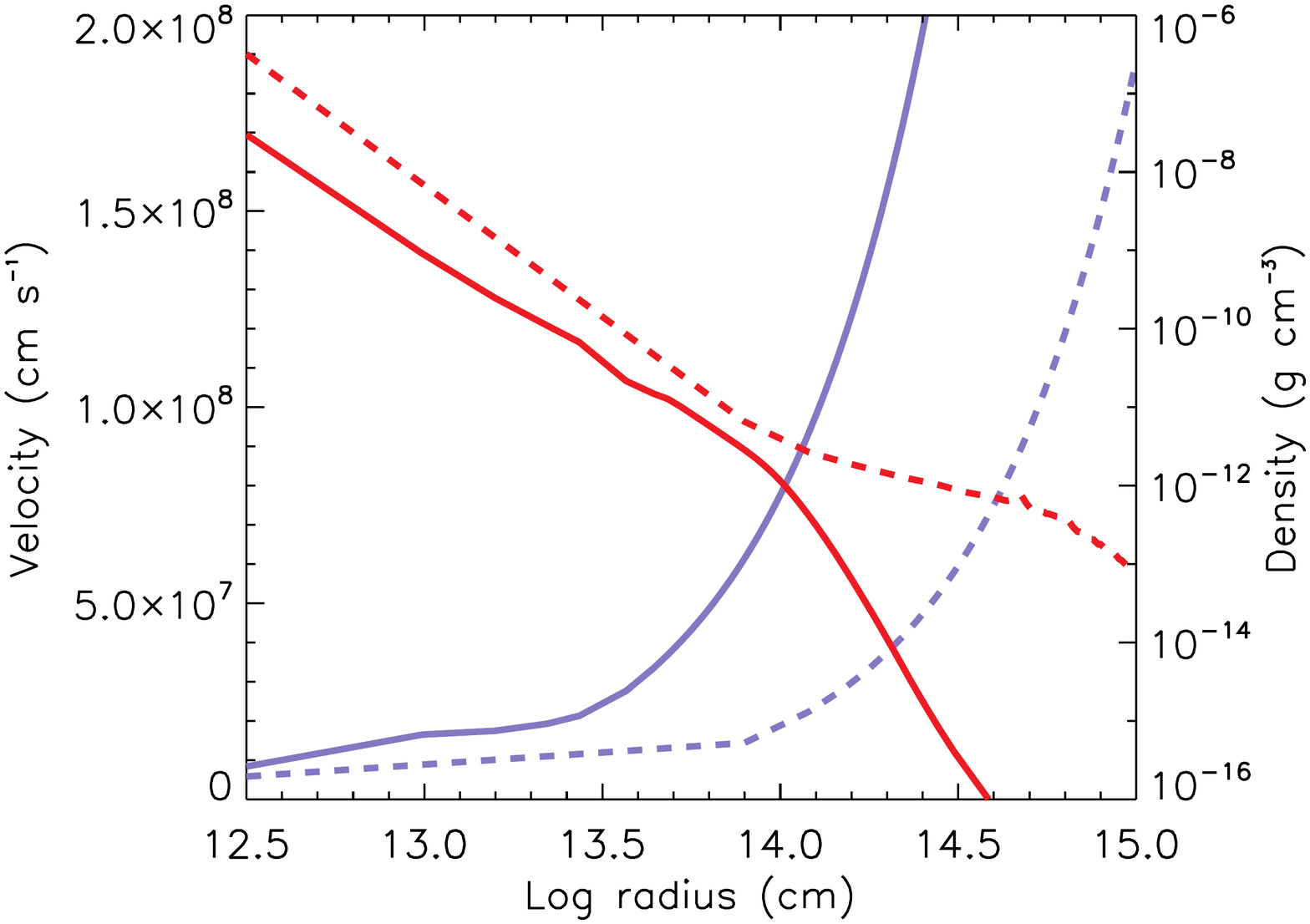}
\includegraphics[width=0.48\textwidth]{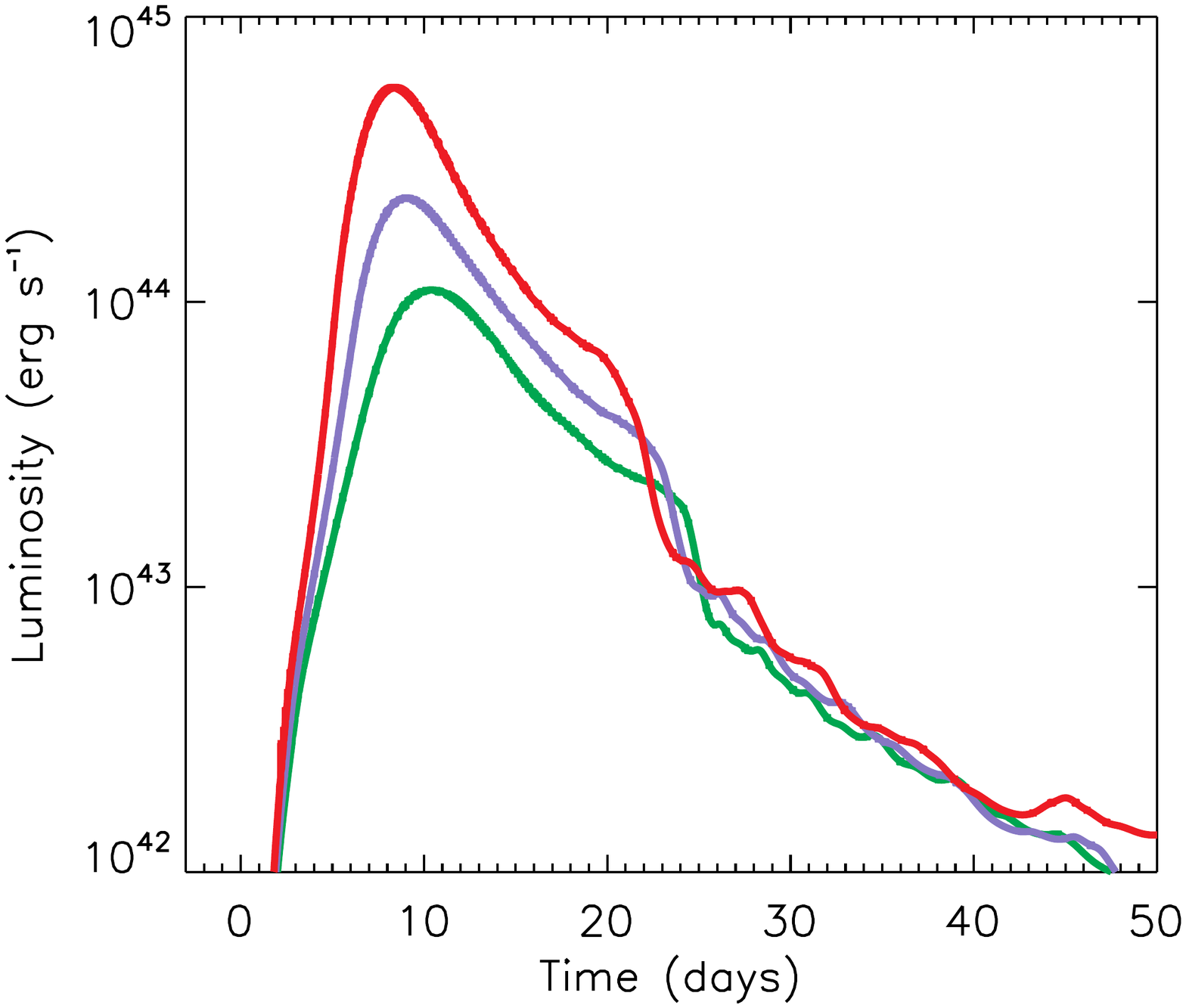}
\caption{ (Top:) Distribution of density (red) and velocity (blue) in
  the mass ejected by silicon flashes in two models evaluated at the
  time of iron-core collapse. The solid lines are for Model He3.0,
  which burned a lot of silicon explosively (\Tab{sisn}) and ejected
  0.73 \Msun \ of helium envelope and core prior to collapse, and the
  dashed lines are for Model He3.2, which ejected only 0.02 \Msun \ of
  envelope. The latter model was used to calculate the light curves
  in \Fig{lite3.2}.  (Bottom:) The light curves produced when
  terminal explosions having kinetic energies 0.25 (Model 3.0A;
  green), 0.42 (Model 3.0B; red), and 0.86 (Model 3.0C; blue) $\times
  10^{51}$ erg produced as it impacts the density distribution shown
  as the solid line in the top panel. Model He3.0C, the red line, had
  a piston deeper in the star and also ejected 0.06 \Msun \ of
  $^{56}$Ni, though this radioactivity had no impact on the peak of
  the light curve. \lFig{lite3.0}}
\end{figure}

\section{Conclusions}
\lSect{conclude}

The evolution of mass losing helium stars with initial masses 1.6 to
120 \Msun \ has been surveyed with a sufficient number of models to
determine the systematics of the supernovae and compact remnants they
produce. The mass loss rates are those of \citet{Yoo17}, with some
variation to test the sensitivity of results. These stars are taken to
represent the outcomes of mass exchanging binaries in which the helium
core is uncovered near the time of helium ignition. That is, early
Case B mass transfer is assumed to completely remove the hydrogen
envelope. Alternatively such stars could be formed by chemically
homogeneous evolution with only a small modification of composition,
but not essential hydrodynamics. 

\begin{deluxetable}{cccl} 
\tablecaption{Critical masses in Close Binary Systems} 
\tablehead{ \colhead{ZAMS}                     & 
            \colhead{Initial}                  & 
            \colhead{Pre-SN}                   &
            \colhead{}                    
            \\ 
            \colhead{Star}                     &
            \colhead{He star}                  &
             \colhead{Mass}                 &
             \colhead{Characteristics}               
            \\
            \colhead{[\Msun]}                  &  
            \colhead{[\Msun]}                  &  
            \colhead{[\Msun]}                  &  
            \colhead{}                           
           }
\startdata
\vspace{0.05 in}
   $<$13  & $<$2.4    &     -     &  SAGB star, WD              \\
13 - 13.5 & 2.4 - 2.5 & 2.0 - 2.1 & SAGB star, rad-expansion  \\
\vspace{0.05 in}
           &           &           &ECSN, fast SN Ib, little $^{56}$Ni\\
13.5  - 16 & 2.5 - 3.2 & 2.1 - 2.6 &  Si Flash, rad-expansion,       \\ 
\vspace{0.05 in}
           &           &           &     peculiar SN Ib              \\
\vspace{0.05 in}
16  - 30   & 3.2 - 10  & 2.6 - 7   &    Ordinary SN Ib, Ic           \\
\vspace{0.05 in}
30 - 120   &  10 - 60  & 7  - 30   &  Mostly BH, Massive SN Ic        \\
\vspace{0.05 in}
120 - 140  &  60 - 70  &  30 - 35  &   Weak PPISN, BH                 \\
\vspace{0.05 in}
140 - 250  & 70 - 125  &  35 - 62  &   Strong PPISN, BH               \\
\vspace{0.05 in}
250 - 500  & 125 - 250 & 62 - 133 &   PISN, no remnant                \\
\vspace{0.05 in} $>$500 & $>$250 & $>$ 133 & Black holes \enddata
\tablecomments{These are for non-rotating solar metallicity stars
  using the standard mass loss rate. The ``Initial He star masses''
  correspond to section headers in \Sect{evolution} and
  \Sect{conclude}. Equivalent main sequence masses are particularly
  uncertain at very high mass and crude estimates are given. The
  transition mass between NeO white dwarfs and electron-capture
  supernovae, shown here as initial helium core mass = 2.4 \Msun, is
  also very uncertain. } \lTab{summary}
\end{deluxetable}

The evolution of such stars is qualitatively different from that of
the helium cores inside single stars
\citep[\Tab{summary};][]{Pol02}. For comparison, for single stars the
main sequence mass range for ECSN and silicon flash supernovae would
extend to 9 and 10.5 \Msun respectively \citep{Woo15}. Most normal
Type IIP supernovae would come from stars lighter than 20 \Msun
\ \citep{Suk16}; substantial black hole formation would start at 20
\Msun, followed by the possibility of black hole formation or Type Ic
supernovae at higher masses, depending on mass loss and uncertain
explosion physics.

These changes reflect chiefly the different relation between initial
(main sequence) mass and final (presupernova) mass, and the larger
fraction of the the presupernova star that is carbon and oxygen in the
mass-losing models. Approximations are given relating the presupernova
mass to both the initial helium core mass (\Eq{mfinhe}) and the
original main sequence mass (\Eq{mfinzams}). For example, a 15 \Msun
\ single star, when it dies, has a helium core of 4.3 \Msun. A 15
\Msun \ star in a binary that loses its envelope at helium ignition
and experiences mass loss to a wind has a final helium core mass of
only 2.4 \Msun. These lighter presupernova stars develop more compact
structures (i.e., small compactness parameters;
  \Fig{compactfig}) and are probably easier to explode.  A large
fraction will have the right masses and structures to make Type Ib and
Ic supernovae (\Sect{lite}). Others have unusual light curves that
may or may not yet have been observed.

Particularly interesting are the light curves of low mass stars, those
here with initial helium core masses between 2.5 and 3.2 (\Tab{sisn}).
These stars experience significant radius expansion after helium
depletion (\Fig{density}). A subset also experiences a violent silicon
flash similar to that in the 10 \Msun \ single stars studied by
\citet{Woo15}. When these stars explode, the larger radius leads to a
bright, brief, blue display (\Fig{lite2.7}) that lasts a few days to a
week.  The emission arises from the diffusion of shock deposited
energy out of the ejected helium envelope. If the explosion makes
substantial $^{56}$Ni, then the light curve may have two peaks, the
second resembling a Type Ib supernova, but occurring somewhat earlier
because of the low ejected mass. For those events that experience
silicon deflagration a wide variety of light curves are possible
(\Fig{lite2.5}, \Fig{lite3.2}, and \Fig{lite3.0}), depending chiefly
on the mass and radius of the shell ejected by the silicon flash
(\Tab{sisn}). Even a weak terminal explosion can give supernovae that
would be classified as superluminous (SLSN-I; \Fig{lite3.0}). A weaker
flash can still cause radius expansion. Coupled with the production of
$^{56}$Ni, an event resembling SN 2014ft \citep{De18} might be
produced (\Fig{lite3.2}, \Sect{sidefl}).

Several light curves might agree better with observations had the mass
loss rate been higher. A lower mass helium envelope, given the same
energy and time, would have expanded to larger radii and produced a
longer, cooler transient more consistent with observed SLSN-I. A lower
mass envelope would also decrease the wait for the second peak in
Model 3.2. This would be more consistent with observations of SN
2014ft. The mass loss has to occur at the right time though, after the
core structure has already been determined, or more mass loss will
just shift the evolution downwards in mass and change its outcome.  As
\citet{Yoo17} has noted, a somewhat larger mass loss rate might also
be more consistent with observations of WO stars (\Fig{he}), producing
them at lower mass. More mass loss would also decrease the mass of
Type Ic supernova progenitors (\Sect{sn1bc}).

Other conclusions are summarized by mass range (see also
\Tab{summary}).

\subsection{$M_{\rm He,i}$ = 1.6 - 2.5 \Msun}

Stars in this mass range develop SAGB-like structures with low density
helium envelopes surrounding dense cores that slowly grow as mass
accretes through a stable, thin, helium-burning shell.  Helium stars
with initial masses below 2.5 \Msun \ also fail to ignite silicon
burning (\Tab{finallow}) and thus avoid iron-core collapse. Below 1.8
\Msun, even carbon burning does not ignite promptly following central
helium burning.  From 1.9 \Msun \ to 2.4 \Msun \ stars ignite
carbon-burning off center, forming flames that eventually burn to the
center of the star (\Fig{convection}).  The subsequent evolution
depends upon how much low density envelope is retained and burns
though the thin shell.

If the envelope is lost to a companion when the star first expands and
develops an SAGB-like structure, helium stars with initial masses
below 1.9 \Msun \ end up as CO white dwarfs with masses up to
$\sim$1.05 \Msun \ (\Tab{cign}). Stars from 1.9 to 2.4 \Msun \ 
become NeO white dwarfs, but can retain up to $\sim$0.1 \Msun \ of
unburned carbon on their outsides (less for higher masses). This might
be important if the star ever becomes a classical nova.

If the envelope is lost only to winds and not the companion, the core
continues to grow until the envelope is gone or the star
explodes. Carbon eventually ignites when the core mass approaches 1.26
\Msun, i.e., for all models with initial masses greater than 1.7 \Msun
\ (\Tab{finallow}).  If this is the first time carbon has ignited, a
flame might move to the center of the star producing a NeO white
dwarf, or it may stall \citep{Den13,Far15,Bro16}. If carbon has
already burned in the interior, then it ignites again, sometimes
repeatedly as thick layers of carbon-rich ash accumulate beneath the
thin helium shell. The critical mass for these carbon flashes
decreases as the mass of the core grows and eventually, for the
highest masses in this range, carbon burns steadily in a thin shell.

For this wind-dominated case, the 1.6 \Msun \ model is the heaviest to
unambiguously make a CO white dwarf.  Higher mass loss rates or late
stage mass transfer would raise this limit. In principle, a CO white
dwarf as massive as 1.26 \Msun \ could be created in a model that had
not yet ignited carbon (e.g., the 1.7 \Msun \ model) and lost its
envelope just as the core reached 1.26 \Msun. If the envelope is 
not lost in heavier models, the core ignites carbon at 1.26 \Msun
\ and eventually grows to the Chandrasekhar mass, about 1.38 \Msun,
producing an electron-capture supernova. This explosion would
transpire in an extended red-giant like structure composed entirely of
helium, and these Type I supernovae would have different properties
from both Type IIP or Ib events (\Sect{lite}; \Fig{lite2.5}). With no
mass loss except winds, this would happen for all models between 1.9
and 2.5 \Msun, a broad range.  In practice, the wind dominated case
may only be appropriate for the more massive stars in this range. The
radii of the 2.4 and 2.5 \Msun \ models at carbon ignition are
relatively small (\Tab{cign}) and might not readily interact with a
companion.

In \Tab{summary} it is assumed that all models below 2.4
\Msun \ lose their envelopes to their companions and become white
dwarfs. There may be a narrow range where the radius is not large
enough for this second stage of mass transfer and electron-capture
supernovae result \citep[see also][]{Tau15}.

\subsection{$M_{\rm He,i}$ = 2.5 - 3.2 \Msun}

Off-center silicon ignition characterizes all models in this mass
range. The 2.5 \Msun \ model is the lightest to experience iron-core
collapse. It ignites silicon 0.504 \Msun \ off center (\Tab{siign})
under very degenerate circumstances. Depending upon the degeneracy and
and offset of the ignition, the silicon flash develops into a violent
runaway with properties that cannot be accurately represented in a
one-dimensional study. Similar behavior was observed for the 3.1, 3.2,
and 3.3 \Msun \ models where the temperature ran away to a peak value
near $6 \times 10^{9}$ K and initiated a deflagration or detonation
(\Sect{deflag}).

The explosive silicon burning does not unbind the core, but initiates a
large amplitude oscillation that launches a shock into the helium
envelope.  Depending upon how much silicon burns, a parameter in these
calculations, either a small fraction of the envelope may be ejected
at a few hundred km s$^{-1}$, or the entire envelope, with a speed of
about 1,000 - 2,000 km s$^{-1}$.  After the oscillations damp, the
remaining core evolves as in the lower mass stars. A silicon flame is
kindled that burns to the center of the star. After a bit more silicon
burning in a shell, the core collapses to a neutron star. Even the
neutrino-powered wind from the proto-neutron star has the potential of
creating a bright display when it runs into the ejected envelope in
some of these models (\Sect{sidefl}; \Fig{lite3.0}).

\subsection{$M_{\rm He,i}$ = 3.2 - 9 \Msun}

Common Type Ib and Ic supernovae should come from these
stars. Presupernova masses range from 2.6 to 6 \Msun
\ (\Tab{models}). This is also the region where the greatest number of
successful explosions might make appreciable $^{56}$Ni. Even should
helium stars below 2.5 \Msun \ keep their envelopes until they die,
the steep density gradient at their edges (\Fig{density}) would
preclude making much nickel. Stars in this mass range should be
comparatively easy to explode (\Fig{compactfig}). Thus it is
reassuring that this is also the mass range that produces explosions
whose light curves and spectra agree with observations
\cite{Des12,Des15,Des16}. A scarcity of Type Ib and Ic supernovae with
final masses above 6 or 7 \Msun \ would be consistent with the lack of
Type IIP progenitors with helium cores above $\sim$6 \Msun \, i.e.,
main sequence masses above $\sim$20 \Msun \ \citep{Sma09}. Perhaps
both have a common cause - the difficulty exploding more massive cores
(\Fig{compactfig}).

Of course, the compactness parameter, $\xi_{2.5}$ is, by itself, an
overly simple representation of core structure, but it is known to
correlate with other measures of explodability
\citep[e..g.,][]{Ert16,Suk18}. Somewhat surprisingly, in addition to
being smaller for presupernova masses less than 6 \Msun, the
compactness parameter is less chaotic for stars in binaries where the
hydrogen envelope is lost than in single stars
(\Fig{compactfig};\Sect{compact}).  Pending further study, fewer low
mass black holes are expected in this range.

The small mass of the exploding star compared to its companion, which
now has its original mass plus, perhaps, the hydrogen envelope of the primary,
suggests that, absent strong kicks, the system will remain
bound. These are thus the likely progenitor systems of massive x-ray
sources that contain neutron stars.

\subsection{$M_{\rm He,i}$ = 9 to 70 \Msun}

Most of the helium stars in this range seem likely to make black
holes, with a pronounced peak between 7 and 9 \Msun
\ (\Fig{compactfig}).  There is also a narrow range of models with
final masses around 10 - 12 \Msun \ that might explode. Given their
large masses, these (neutrino-powered) explosions might produce
broader, fainter light curves \citep{Ens88}, inconsistent with common
Type Ic supernovae. Because of their high mass, they may also be
relatively rare. The black hole peak at 8 \Msun \ is derived from
helium stars with initial masses near 14 \Msun, which, in turn, come
from main sequence masses near 40 \Msun.  The 11 \Msun \ supernova
would be derived from a star with original helium core mass 22 \Msun
\ and main sequence mass near 55 \Msun.  If these models do explode,
then a new potentially detectable mass gap in black hole masses would
exist around 10 - 12 \Msun. This is a less robust gap than the one
above 46 \Msun \ (see below), and more sensitive to uncertain physics
like the rates for $^{12}$C($\alpha,\gamma)^{16}$O and mass loss.

The new results for compactness differ substantially from those for
single stars \citep[\Fig{compactfig};][]{Suk18} and suggest the
distribution of black hole masses will be different in mass exchanging
binaries and single stars.

The relation between helium core mass and main sequence mass used here
applies to non-rotating stars only, which is probably an unrealistic
approximation.  With substantial rotation, Gamma-ray bursts and
magnetars might come from these sorts of stars. With rotation, the
main sequence mass required to produce a given helium core mass would
be reduced, especially in the case of chemically homogeneous
evolution.

\subsection{$M_{\rm He,i}$ = 30 - 120 \Msun}

A new limit is derived for the maximum black hole mass implied by the
pulsational pair instability. In a binary system, pulses will eject
all mass external to 46 \Msun \ (\Tab{ppisntab}). This is smaller than
the previous limit, 52 \Msun, given by \citet{Woo17} because: a) the
limit is smaller for helium cores that lack a hydrogen envelope to
tamp the explosion \citep[see also][]{Woo17}; and b) the CO-core is
larger in presupernova stars for helium stars that lose their hydrogen
envelope early on. This limit is consistent with all LIGO detections
thus far \citep{Abo18}

Helium cores with solar metallicity and initial masses greater than 60
\Msun \ encounter a weak instability in the oxygen burning shell
(\Tab{ppisntab}) that results in the low energy ejection of a small
amount of mass. The stronger instability, with ejected kinetic
energies greater that 10$^{49}$ erg, begins for helium core masses
greater than 75 \Msun. These corresponds to main sequence masses of
roughly 130 and 160 \Msun (\Eq{mzamsh}) for the weak and strong
pulsational pair-instability respectively. If the mass loss rate is
reduced by a factor of two as might occur for lower metallicity, these
numbers are about 70\% as great (\Fig{mloss}), i.e., a main sequence
mass of only 110 \Msun \ could produce violent PPISN.  Of course these
numbers can be reduced if the core is a product of chemically
homogeneous evolution. Then a 75 \Msun \ ``helium star'' could be
produced by a rapidly rotating main sequence star with just a slightly
larger mass.

\subsection{Subsequent interactions with the other star}

While the mass loss history of the first star in the binary to fill
its Roche lobe is uncertain, the evolution of the second, the mass
acceptor, is even more so. If the mass exchange is conservative, the
``effective'' main sequence mass for the secondary in
\eq{mzams} and \eq{mzamsh} is its initial mass plus the mass lost by
the primary which is the main sequence mass minus the final mass
(\Eq{mfinzams}). This is an upper bound since some mass will be lost,
especially if there is a common envelope phase. 

Several caveats should be noted. First, the 46 \Msun \ limit derived
here for black holes remaining following pulsational pair-instability
supernovae (\Sect{ppisn}) strictly applies only to the mass of the
primary immediately after the first explosion. An uncertain amount of
matter could accrete onto the black hole, especially if it experiences
``hypercritical'' accretion \citep[e.g.][]{Bel02,Bet07} during a
second common envelope stage.  \citet{Osh05} estimate the fractional
increase in black hole mass to be 5 to 25\%. When the second explosion
happens, assuming the envelope has been lost, that helium star too
will be subject to the 46 \Msun \ limit, but the original primary
could have grown to over 46 \Msun.

Second, in addition to increasing the mass of the secondary, accretion
may also increase its spin. If the star rotates rapidly enough, it
will evolve chemically homogeneously. Very massive helium stars could
result, even at solar metallicity and a second common envelope might
be avoided, though the orbit would remain wide.

Finally, because the more massive stars have a constant hydrogen
burning lifetime given by the Eddington limit, about 3 million years,
and a constant helium burning lifetime $\sim10$\% of that, it may not
not be uncommon for very massive stars born at the same time to die at
nearly the same time, even if they have different masses. The delay
time between supernovae could be short and the first mass exchange
might begin when both stars have already developed helium cores.

\subsection{Type II supernovae in binaries}

While this paper is about stars that lose their hydrogen envelopes
near helium ignition, binary mass transfer is certainly capable of
producing supernovae that retain at some of their
hydrogen. Interacting binaries are the probable origin of Type IIb
supernovae like SN 1993J. Mergers can also produce supernovae like
1987A. For a recent discussion including many light curves, see
\citet{Eld18}. In general, a presupernova star that retains even a
small hydrogen envelope in an interacting binary will resemble much
more closely the outcome of single star evolution. The small mass of
the envelope will affect the optical display, but the helium core,
explosion physics, remnant masses, and nucleosynthesis will be more
like single stars of the same initial mass.

\section*{Acknowledgments}

This work has been partly supported by NASA NNX14AH34G. The author
acknowledges valuable advice from Sung-Chul Yoon on the mass loss
rates and helpful discussions with Thomas Ertl, Alexander Heger,
Thomas Janka, and Tuguldur Sukhbold. Sukhbold also provided data for
\Fig{compactfig}.

\end{document}